\documentclass[11pt,doublespacing]{article}

\usepackage{fullpage}
\usepackage{setspace}
\usepackage{authblk}
\usepackage{cite}
\usepackage{hyperref}

\usepackage{graphicx}

\usepackage{amsmath}
\usepackage{amssymb}

\usepackage{float}
\usepackage{subfigure}
\usepackage{array}
\usepackage{multirow}
\usepackage{makecell}
\usepackage{rotating}

\usepackage{enumitem}
\allowdisplaybreaks

\usepackage[title]{appendix}

\begin{document}

\title{A Thin Sheet Volume Integral Equation Solver for Simulation of Bianisotropic Metasurfaces}

\author[1]{Sebastian Celis Sierra}
\author[1]{Meruyert Khamitova}
\author[2]{Ran Zhao}
\author[1]{Sadeed Bin Sayed}
\author[1]{Hakan Bagci\vspace{0.5cm}}

\affil[1]{Electrical and Computer Engineering (ECE) Program, Computer, Electrical, and Mathematical Science and Engineering (CEMSE) Division
\authorcr King Abdullah University of Science and Technology (KAUST)
\authorcr Thuwal, 23955-6900, Saudi Arabia
\authorcr e-mail: sebastian.celissierra@kaust.edu.sa; meruyert.khamitova@kaust.edu.sa; sadeed.sayed@kaust.edu.sa; hakan.bagci@kaust.edu.sa\vspace{0.5cm}}

\affil[2]{Electronic Science and Engineering
\authorcr University of Electronic Science and Technology of China (UESTC)
\authorcr Chengdu, 61173, China
\authorcr e-mail: ran.zhao@uestc.edu.cn\vspace{0.5cm}}


\date{}
\maketitle
\newpage

\begin{abstract}
A thin-sheet (TS) volume integral equation (VIE) formulation incorporating generalized sheet transition conditions (GSTCs) is presented for the simulation of three-dimensional (3D) bianisotropic metasurfaces. The metasurface is represented as an equivalent thin sheet, with its constitutive tensors derived from the GSTC susceptibility tensors. Invoking the thin-sheet approximation, the governing VIEs are reduced to surface integral equations (SIEs), in which tangential and normal flux density components are treated as distinct sets of unknowns and discretized using Rao-Wilton-Glisson and pulse basis functions, respectively. In contrast to conventional GSTC approaches based on conventional SIEs, which represent only tangential fields, the proposed framework rigorously enforces the bianisotropic GSTCs, including normal field interactions,  while retaining the flux-based VIE character of the formulation. Numerical examples, including a comparison with an SIE-GSTC solver that neglects the normal susceptibility components, demonstrate the accuracy and robustness of the proposed TS-VIE-GSTC solver for polarization rotation on planar and curved sheets, perfect reflection, beam steering with spatially varying susceptibilities, multi-directional attenuation, and oblique phase-shift transformation.

\end{abstract}

\section{Introduction}\label{sec:Intro}
Metasurfaces are transforming the way communication, sensing, and imaging platforms operate by enabling precise manipulation of electromagnetic field polarization, phase, and amplitude~\cite{pan2016, maci2017, liaskos2018, dai2021}. This capability has driven a broad range of applications, including meta-lens design~\cite{MS_Lenses,MS_Lense_MIMO}, beam steering~\cite{MS_Antennas}, holography~\cite{MS_Holograms}, and radar cross-section (RCS) reduction~\cite{MS_RCS_Reduction}. As metasurface research transitions from unit-cell design to electrically large system-level deployments, the need for accurate computational modeling of their electromagnetic response becomes indispensable. 

Existing full-wave solvers, while widely used in the design of individual metasurface unit cells, are not well-suited for simulating metasurfaces deployed in realistic system-level scenarios, such as when the metasurface forms part of an aircraft structure for radar cross-section reduction~\cite{refmeta1,refmeta2} or is deployed on a building surface to enhance wireless signal coverage~\cite{refmeta3}. This limitation stems from the requirement to resolve subwavelength unit-cell features within an electrically large domain, which results in large, ill-conditioned matrix equations that are computationally prohibitive to assemble and solve~\cite{Caloz_Book}.

A widely adopted strategy to address this bottleneck is to replace the metasurface with an equivalent thin sheet, across which generalized sheet transition conditions (GSTCs) are enforced~\cite{Caloz_Book, idemen2011discontinuities, Kuester}. GSTCs capture the field discontinuities induced by the actual metasurface and relate the fields on either side of the sheet through four surface susceptibility tensors $\bar{\bar{\chi}}_\mathrm{\mathrm{ab}}(\mathbf{r})$, $\mathrm{a,b} \in \{\mathrm{e,m}\}$~\cite{Caloz_Book, idemen2011discontinuities, Kuester}, collectively forming a bianisotropic model of the metasurface response. In practice, however, simplified monoanisotropic approximations that neglect cross-coupling interactions are often employed for computational convenience, resulting in inaccurate solutions. 

GSTCs have been incorporated into electromagnetic solvers for metasurface simulation, including finite-difference, finite-element, and discontinuous Galerkin formulations, in both the frequency and time domains~\cite{FDFD, FEM_Caloz, FDTD_Jia_Xiao, FDTD_Caloz, FDTD_Wu, DG1,chen_2023}. These solvers rely on sparse, local discretizations and can be highly efficient. In open-region scattering problems, however, they require both the scatterers and the surrounding background region to be discretized up to a truncation boundary, on which absorbing boundary conditions, perfectly matched layers, or exact nonlocal boundary conditions are imposed, and they are susceptible to numerical phase dispersion~\cite{jin2015theory}. For electrically large problems, such as the system-level scenarios that motivate the use of GSTCs~\cite{costas2007, costas2009, zhekov2018, backer1997, conor2019, cclu2001, jiang2024}, this region extends over the entire domain of interest, so that the number of unknowns and the accumulated dispersion error grow with its electrical size.

In contrast, integral equation (IE) solvers, discretized using the method of moments (MoM)~\cite{Harrington1993}, are particularly attractive for open-region scattering problems, since they require discretization only of the scatterers and inherently satisfy the radiation condition without a truncation boundary~\cite{jin2015theory}. These advantages come at the cost of a dense system matrix. However, fast solvers based on the fast multipole method and its multilevel extension~\cite{Rokhlin1990, Coifman1993, Chew, Chew2001, Yucel2018}, or on fast Fourier transform (FFT)-based schemes such as the adaptive integral method~\cite{Bleszynski1996, Sharma2021}, can reduce the computational and memory costs to near-linear scaling with the number of unknowns, thereby enabling the simulation of electrically large structures. Building on these properties, surface integral equation (SIE) formulations incorporating GSTCs have been developed for metasurface simulation in both two-dimensional (2D)~\cite{Gupta_Coordinates,MoM_Sandeep,gupta2023p2, celis_2023a} and three-dimensional (3D) problems~\cite{budhu_2024, Kim2025, celis_2023b, celis_2024a, Journal1_Celis_2025, CelisSierra2026_ST}.  

Despite this progress, incorporating bianisotropic GSTCs with normal susceptibility components into 3D SIE frameworks remains a challenge, since such models require the field components normal to the metasurface, whereas the unknowns of the 3D SIE-GSTC formulations above represent only tangential fields. SIE formulations with normal field components as unknowns do exist, but they have been developed for other purposes, such as low-frequency stabilization and charge extraction~\cite{R1_Reviewer3, R2_Reviewer3, R3_Reviewer3, R4_Reviewer3}, rather than for enforcing GSTCs. Normal susceptibility components are not specific to bianisotropic media: the normal entries of $\bar{\bar{\chi}}_{\mathrm{ee}}$ and $\bar{\bar{\chi}}_{\mathrm{mm}}$ govern the angular scattering response and arise in unit cells supporting dipole moments along the surface normal, such as capacitively loaded loops, which support normal magnetic dipoles~\cite{Gupta_Coordinates}, and particles containing rods oriented along the surface normal, which support normal electric dipoles~\cite{Achouri_Angular_Scattering}. Neglecting either the bianisotropic or the normal components can lead to inaccurate solutions~\cite{celis_2024a, Journal1_Celis_2025, CelisSierra2026_ST, Gupta_Coordinates, Achouri_Angular_Scattering, ScottNew}. To the best of the authors' knowledge, SIE formulations that account for the normal components have so far been reported only for 2D problems~\cite{Gupta_Coordinates, gupta2023p2}, where they have also been used to verify ray-optical solutions for curved metasurfaces~\cite{ScottNew}. Their extension to 3D, however, is not straightforward.

This challenge is addressed here by representing the metasurface as an equivalent thin sheet (TS) using volume integral equations (VIEs) formulated in terms of the unknown electric and magnetic flux densities. Invoking the thin-sheet approximation, the volume integrals reduce to surface integrals, in which tangential and normal components of the flux densities are treated as distinct sets of unknowns~\cite{TS}. The constitutive tensors in this formulation are derived directly from the four surface susceptibility tensors $\bar{\bar{\chi}}_\mathrm{\mathrm{ab}}$, $\mathrm{a,b} \in \{\mathrm{e,m}\}$ of the bianisotropic GSTCs representing the metasurface response. This framework enables rigorous enforcement of the GSTCs, including normal field interactions, within a surface integral equation setting, while retaining the flux-based VIE character of the formulation. Consequently, the proposed TS-VIE-GSTC formulation overcomes the limitations of conventional SIE-GSTC approaches, enabling accurate modeling of bianisotropic metasurface responses previously inaccessible to such solvers. A preliminary version of this formulation was presented in~\cite{celis_2024b}.

The primary contributions of this work are threefold. First, to the best of the authors' knowledge, the thin-sheet reduction is extended to bianisotropic media for the first time. While thin-sheet formulations have been developed for isotropic~\cite{TS,ren_accurate_2017}, anisotropic~\cite{Guan2019Anisotropic}, and bi-isotropic~\cite{deng_thin_2013} media, none of these explicitly addresses the volume bound-charge contributions arising from the tensor--vector products inherent to the bianisotropic VIE, which are rigorously treated here. This extension is independent of the GSTCs and applies to thin bianisotropic layers in general. Metasurfaces are a natural application, since the GSTC susceptibility tensors directly supply the bianisotropic constitutive tensors. Second, normal field components are systematically incorporated through the flux density-based formulation, enabling the enforcement of GSTCs with normal susceptibility components, which conventional 3D SIE-GSTC solvers cannot enforce. Third, the proposed TS-VIE-GSTC solver is validated through a parametric study of a monoisotropic thin shell against the analytical Mie-series solution and through five canonical wave transformations: polarization rotation on planar and curved sheets, perfect reflection, beam steering with spatially varying susceptibilities, multi-directional attenuation, and oblique phase-shift transformation, the last of which is also used for a comparison with an SIE-GSTC solver that neglects the normal susceptibility components. Collectively, the proposed TS-VIE-GSTC formulation provides a rigorous framework for modeling bianisotropic metasurfaces in electrically large open-domain environments.

The remainder of this paper is organized as follows. Section~\ref{sec:Formulation} presents the TS-VIE-GSTC formulation, including the VIE formulation for bianisotropic scatterers, the thin-sheet reduction of VIEs into SIEs, the equivalent thin-sheet representation of the GSTCs, and the discretization strategy, followed by comments on the treatment of inhomogeneous sheets, the placement of the normal unknown, the role of the sheet thickness, and the numerical aspects of the formulation. Section~\ref{sec:Numerical_Examples} provides numerical examples validating the accuracy and robustness of the proposed TS-VIE-GSTC solver. Finally, Section~\ref{sec:Conclusions} concludes the paper and discusses directions for future research.

\section{Formulation}\label{sec:Formulation}
\subsection{VIEs for Bianisotropic Scatterers}\label{sec:VIE}
Let $V$ denote the volume of a bianisotropic scatterer embedded in an unbounded homogeneous background medium characterized by permittivity $\varepsilon_0$ and permeability $\mu_0$. The scatterer is illuminated by time-harmonic incident electric and magnetic fields $\mathbf{E}^{\mathrm{inc}}(\mathbf{r})$ and $\mathbf{H}^{\mathrm{inc}}(\mathbf{r})$. The time dependence $e^{j\omega t}$, where $\omega$ is the angular frequency, is assumed and suppressed. The incident fields induce equivalent electric and magnetic volume currents $\mathbf{J}(\mathbf{r})$ and $\mathbf{M}(\mathbf{r})$ within $V$, which generate the scattered electric and magnetic fields $\mathbf{E}^{\mathrm{sca}}(\mathbf{r})$ and $\mathbf{H}^{\mathrm{sca}}(\mathbf{r})$. The scattered fields are expressed as~\cite{jin2015theory} 
\begin{equation}
\label{eq2}
\begin{aligned}
    &\mathbf{E}^\mathrm{sca}(\mathbf{r})=\frac{1}{j \omega \varepsilon_0} \mathcal{L}[\mathbf{J}](\mathbf{r})-\mathcal{K}[\mathbf{M}](\mathbf{r})\\
    &\mathbf{H}^\mathrm{sca}(\mathbf{r})=
    \frac{1}{j \omega \mu_0} \mathcal{L}[\mathbf{M}](\mathbf{r})+\mathcal{K}[\mathbf{J}](\mathbf{r}).
\end{aligned}
\end{equation}
Here, the volume integral operators $\mathcal{L}[\mathbf{X}](\mathbf{r})$ and $\mathcal{K}[\mathbf{X}](\mathbf{r})$ are defined as:
\begin{equation}
\begin{aligned}
\mathcal{L}[\mathbf{X}](\mathbf{r}) & =k^2_0\int_V G(\mathbf{r}, \mathbf{r}^{\prime}) \mathbf{X}(\mathbf{r}^{\prime}) d v^{\prime}
+\nabla\int_V  G(\mathbf{r}, \mathbf{r}^{\prime}) \nabla^{\prime}\cdot\mathbf{X}(\mathbf{r}^{\prime}) d v^{\prime}\\
\mathcal{K}[\mathbf{X}](\mathbf{r}) & =\nabla \times \int_V  G(\mathbf{r}, \mathbf{r}^{\prime})\mathbf{X}(\mathbf{r}^{\prime}) d v^{\prime}
\end{aligned}
\end{equation}
where $G(\mathbf{r}, \mathbf{r}^{\prime})= e^{-jk_0|\mathbf{r}-\mathbf{r}^{\prime}|}/(4\pi |\mathbf{r}-\mathbf{r}^{\prime}|)$ is the Green function of the background medium and $k_0 = \omega \sqrt{\varepsilon_0 \mu_0}$ is the corresponding wavenumber. The total fields are therefore expressed as
\begin{equation}
\begin{aligned}
\mathbf{E}(\mathbf{r}) &= \mathbf{E}^{\mathrm{inc}}(\mathbf{r}) + \mathbf{E}^{\mathrm{sca}}(\mathbf{r})\\
\mathbf{H}(\mathbf{r}) &= \mathbf{H}^{\mathrm{inc}}(\mathbf{r}) + \mathbf{H}^{\mathrm{sca}}(\mathbf{r}).
\end{aligned}
\label{eq:total_fields}
\end{equation}

VIE formulations in terms of electric and magnetic flux densities $\mathbf{D}(\mathbf{r})$ and $\mathbf{B}(\mathbf{r})$ are often preferred because they possess well-defined divergences and exhibit continuous normal components across material interfaces~\cite{TS,swg1984}. Within $V$, the bianisotropic constitutive relations are written as~\cite{Liu2021,Pasi_Anisotropy,Pasi_Bianisotropy}:
 \begin{equation}
 \label{eq3}
 \begin{aligned}
&\mathbf{E}(\mathbf{r}) =\bar{\bar{\alpha}}_1\cdot \mathbf{D}(\mathbf{r})+\bar{\bar{\alpha}}_2 \cdot \mathbf{B}(\mathbf{r}) \\
&\mathbf{H}(\mathbf{r})  =\bar{\bar{\alpha}}_3\cdot \mathbf{D}(\mathbf{r})+\bar{\bar{\alpha}}_4 \cdot \mathbf{B}(\mathbf{r}).
\end{aligned}    
 \end{equation}
Here, the tensors $\bar{\bar{\alpha}}_i$, $i=1,2,3,4$, characterize the response of the bianisotropic material within $V$. Specifically, $\bar{\bar{\alpha}}_1$ and $\bar{\bar{\alpha}}_4$ describe the purely electric and magnetic responses, respectively, while $\bar{\bar{\alpha}}_2$ and $\bar{\bar{\alpha}}_3$ account for the electric-magnetic cross-coupling. Invoking the volume equivalence principle~\cite{Liu2021,Pasi_Anisotropy,Pasi_Bianisotropy}, $\mathbf{J}(\mathbf{r})$ and $\mathbf{M}(\mathbf{r})$ are expressed as:
\begin{equation}
\label{eq4}
 \begin{aligned}
\mathbf{J}(\mathbf{r}) & =j \omega[\bar{\bar{\beta}}_1\cdot \mathbf{D}(\mathbf{r})+\bar{\bar{\beta}}_2 \cdot \mathbf{B}(\mathbf{r})] \\
\mathbf{M}(\mathbf{r}) & =j \omega[\bar{\bar{\beta}}_3\cdot \mathbf{B}(\mathbf{r})+\bar{\bar{\beta}}_4\cdot \mathbf{D}(\mathbf{r})].
\end{aligned}   
\end{equation}
Here, $\bar{\bar{\beta}}_{i}$, $i=1,2,3,4$, are the contrast tensors related to $\bar{\bar{\alpha}}_i$ as
\begin{equation}
\begin{bmatrix}
\bar{\bar{\beta}}_1 & \bar{\bar{\beta}}_2 \\
\bar{\bar{\beta}}_3 & \bar{\bar{\beta}}_4
\end{bmatrix}=\begin{bmatrix}
\bar{I}-\varepsilon_0 \bar{\bar{\alpha}}_1 & -\varepsilon_0 \bar{\bar{\alpha}}_2 \\
\bar{I}-\mu_0 \bar{\bar{\alpha}}_4 & -\mu_0 \bar{\bar{\alpha}}_3
\end{bmatrix}  
\end{equation}
where $\bar{I}$ is the $3 \times 3$ identity matrix.

Substituting~\eqref{eq2}-\eqref{eq4} into~\eqref{eq:total_fields} for $\mathbf{r} \in V$ yields two VIEs in the unknowns $\mathbf{D}(\mathbf{r})$ and $\mathbf{B}(\mathbf{r})$:
\begin{equation}
\label{eq5}
\begin{aligned}
 \mathbf{E}^{\mathrm{inc}}(\mathbf{r})&= \bar{\bar{\alpha}}_1\cdot \mathbf{D}(\mathbf{r})-\frac{1}{\varepsilon_0} \mathcal{L}[\bar{\bar{\beta}}_1\cdot \mathbf{D} ](\mathbf{r})+j \omega \mathcal{K}[\bar{\bar{\beta}}_4 \cdot \mathbf{D} ](\mathbf{r})
+\bar{\bar{\alpha}}_2 \cdot \mathbf{B}(\mathbf{r})\\
&-\frac{1}{\varepsilon_0} \mathcal{L}[\bar{\bar{\beta}}_2\cdot \mathbf{B} ](\mathbf{r})+j \omega \mathcal{K}[\bar{\bar{\beta}}_3 \cdot {\mathbf{B}}] (\mathbf{r})
\end{aligned}
\end{equation}
\begin{equation}
\label{eq6}
\begin{aligned}
 \mathbf{H}^{\mathrm{inc}}(\mathbf{r})&=\bar{\bar{\alpha}}_3 \cdot \mathbf{D}(\mathbf{r})-\frac{1}{\mu_0} \mathcal{L}[\bar{\bar{\beta}}_4\cdot \mathbf{D} ](\mathbf{r})-j \omega \mathcal{K}[\bar{\bar{\beta}}_1 \cdot \mathbf{D}](\mathbf{r})+\bar{\bar{\alpha}}_4 \cdot{\mathbf{B}}(\mathbf{r})\\
&-\frac{1}{\mu_0} \mathcal{L}[\bar{\bar{\beta}}_3\cdot {\mathbf{B}}] (\mathbf{r})-j \omega \mathcal{K}[\bar{\bar{\beta}}_2 \cdot {\mathbf{B}} ](\mathbf{r}).
\end{aligned}  
\end{equation}

\subsection{Bianisotropic GSTCs}\label{sec:GSTC}

Let $S$ denote the surface of an open metasurface of arbitrary shape embedded in the unbounded homogeneous background medium. The electromagnetic field discontinuities across the two sides of the metasurface are governed by the GSTCs~\cite{Susceptibility_Tensors}:
\begin{equation}
\label{eqGSTC}
\begin{aligned}
    \hat{\mathbf{n}}(\mathbf{r}) \times \Delta \mathbf{H}(\mathbf{r}) &=j\omega \mathbf{P}^{\mathrm{e}}_{\parallel}(\mathbf{r})-\hat{\mathbf{n}}(\mathbf{r}) \times\nabla_{\parallel}P^{\mathrm{m}}_{\bot}(\mathbf{r})\\
    \hat{\mathbf{n}}(\mathbf{r}) \times \Delta \mathbf{E}(\mathbf{r}) &=-j\omega \mu_0 \mathbf{P}^{\mathrm{m}}_{\parallel}(\mathbf{r})-\hat{\mathbf{n}}(\mathbf{r}) \times\frac{\nabla_{\parallel}P^{\mathrm{e}}_{\bot}(\mathbf{r})}{\varepsilon_0}.
\end{aligned}
\end{equation}
Here, $\Delta \mathbf{X}(\mathbf{r})=\mathbf{X}_1(\mathbf{r})-\mathbf{X}_2(\mathbf{r})$ denotes the discontinuity of field $\mathbf{X}(\mathbf{r})$ across $S$, and $\hat{\mathbf{n}}(\mathbf{r})$ is the unit normal vector to $S$.  Throughout the paper, the subscripts ``$\parallel$'' and ``$\perp$'' indicate components tangential and normal to $S$, respectively. The electric and magnetic polarizations $\mathbf{P}^{\mathrm{e}}(\mathbf{r})$ and $\mathbf{P}^{\mathrm{m}}(\mathbf{r})$ are expressed in terms of the fields as~\cite{Kuester}:
\begin{equation}
\label{eqPol}
\begin{aligned}
   \mathbf{P}^{\mathrm{e}}(\mathbf{r})&=\varepsilon_0 \bar {\bar {\chi}}_{\mathrm{ee}}\cdot \Sigma \mathbf{E}(\mathbf{r})+\sqrt{\mu_0 \varepsilon_0} \bar {\bar {\chi}}_{\mathrm{em}}\cdot \Sigma \mathbf{H}(\mathbf{r})\\
   \mathbf{P}^{\mathrm{m}}(\mathbf{r})&=\sqrt{\frac{\varepsilon_0}{\mu_0}} \bar {\bar {\chi}}_{\mathrm{me}}\cdot \Sigma \mathbf{E}(\mathbf{r})+\bar {\bar {\chi}}_{\mathrm{mm}}\cdot \Sigma \mathbf{H}(\mathbf{r})
\end{aligned}    
\end{equation}
where $\Sigma \mathbf{X}(\mathbf{r})=[\mathbf{X}_1(\mathbf{r})+\mathbf{X}_2(\mathbf{r})]/2$ denotes the average of field $\mathbf{X}(\mathbf{r})$ across $S$, and $\bar{\bar{\chi}}_\mathrm{\mathrm{ab}}$, $\mathrm{a,b} \in \{\mathrm{e,m}\}$ are the surface susceptibility tensors. 

Solving the GSTCs~\eqref{eqGSTC} for prescribed wave transformations, i.e., specified electromagnetic fields on both sides of the metasurface, yields a set of susceptibility tensors~\cite{Synthesis_Spherical,Susceptibility_Tensors,Journal1_Celis_2025,MoM_Sandeep}. In the most general case, $\bar{\bar{\chi}}_\mathrm{\mathrm{ab}}$ may exhibit up to $36$ independent degrees of freedom. Once determined, $\bar{\bar{\chi}}_\mathrm{\mathrm{ab}}$ serve as the targets for the design of corresponding physical unit-cell geometries~\cite{Synthesis_Spherical,Susceptibility_Tensors,Journal1_Celis_2025,MoM_Sandeep}.

\subsection{Thin-Sheet Formulation}\label{sec:TS-VIEs}
\begin{figure}[h]
\centering
\includegraphics[scale=0.6]{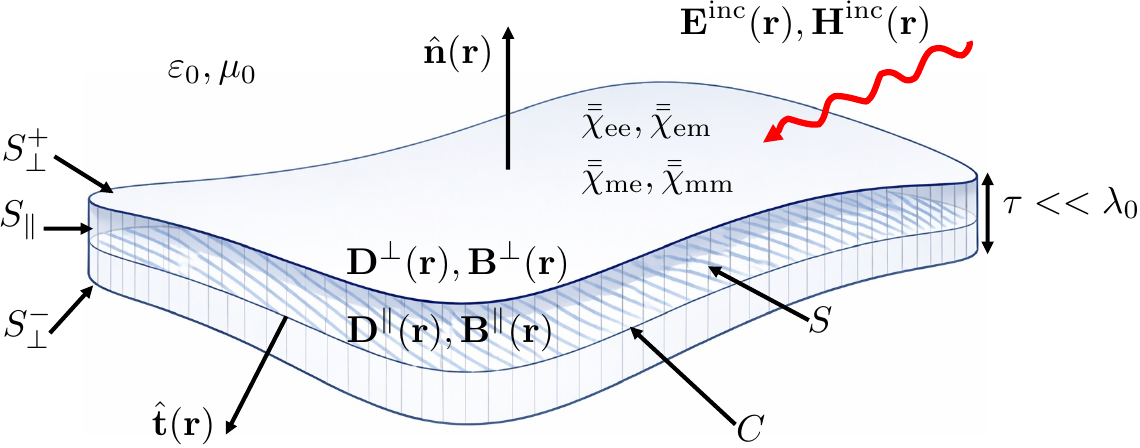}
\caption{Illustration of a thin bianisotropic sheet with boundary $S$ and incident source $\mathbf{E}^{\mathrm{inc}}(\mathbf{r}),\mathbf{H}^{\mathrm{inc}}(\mathbf{r})$.}
\label{fig:TS_Figure}
\end{figure}

Let $V$ denote the volume of a thin sheet (see Fig.~\ref{fig:TS_Figure}) of uniform thickness $\tau$, with $\tau \ll \lambda_0$, where $\lambda_0 = 2\pi/k_0$ is the wavelength in the background medium. The boundary of $V$ comprises three surfaces: the lateral surface $S_{\parallel}$, the top surface $S_{\perp}^{+}$, and the bottom surface $S_{\perp}^{-}$, where $S_{\perp}^{-}$ and $S_{\perp}^{+}$ are parallel. The outward-pointing unit normal vectors to these surfaces are denoted by $\hat{\mathbf{t}}(\mathbf{r})$, $\hat{\mathbf{n}}(\mathbf{r})$, and $-\hat{\mathbf{n}}(\mathbf{r})$, respectively. A mid-surface, located at a distance $\tau/2$ from the two parallel surfaces $S_{\perp}^{-}$ and $S_{\perp}^{+}$, coincides with $S$.

Under the thin-sheet approximation, the VIEs~\eqref{eq5} and~\eqref{eq6} are reformulated as SIEs~\cite{TS}. To this end, $\mathbf{D}(\mathbf{r})$ and $\mathbf{B}(\mathbf{r})$ are decomposed into tangential and normal components as:
\begin{equation}
\label{eqfield_decomp}
\begin{aligned}
    \mathbf{D}(\mathbf{r}) & =\mathbf{D}_{\parallel}(\mathbf{r})+\hat{\mathbf{n}}(\mathbf{r}){D}_{\bot}(\mathbf{r}) \\
    \mathbf{B}(\mathbf{r}) & =\mathbf{B}_{\parallel}(\mathbf{r})+\hat{\mathbf{n}}(\mathbf{r}){B}_{\bot}(\mathbf{r})
\end{aligned}    
\end{equation}
and the integral operators $\mathcal{L}[\bar{\bar{\beta}}_{i}\cdot \mathbf{X}](\mathbf{r})$ and $\mathcal{K}[\bar{\bar{\beta}}_{i}\cdot \mathbf{X}](\mathbf{r})$ are reformulated in thin-sheet form as detailed in the following subsections. 

\subsubsection{Thin-Sheet Reduction of $\mathcal{L}[\bar{\bar{\beta}}_{i}\cdot \mathbf{X}](\mathbf{r})$}\label{sec:operator}
The operator $\mathcal{L}[\bar{\bar{\beta}}_i \cdot \mathbf{X}](\mathbf{r})$, $i=1,2,3,4$, is decomposed into vector and scalar potential contributions:
\begin{equation}
\begin{aligned}
&\mathcal{L}[\bar{\bar{\beta}}_{i}\cdot\mathbf{X}](\mathbf{r})  = \mathcal{L}_{A}[\bar{\bar{\beta}}_{i}\cdot\mathbf{X}](\mathbf{r})+\nabla\mathcal{L}_{\phi}[\bar{\bar{\beta}}_{i}\cdot\mathbf{X}](\mathbf{r})
\end{aligned}
\end{equation}
where
\begin{equation}
\begin{aligned}
&\mathcal{L}_{A}[\bar{\bar{\beta}}_{i}\cdot\mathbf{X}](\mathbf{r})= k_0^2\int_V G(\mathbf{r}, \mathbf{r}^{\prime}) \bar{\bar{\beta}}_{i}\cdot\mathbf{X}(\mathbf{r}^{\prime}) d v^{\prime}\\
& \mathcal{L}_{\phi}[\bar{\bar{\beta}}_{i}\cdot \mathbf{X}](\mathbf{r})=  {\int_V G(\mathbf{r}, \mathbf{r}^{\prime})\nabla^{'}\cdot[\bar{\bar{\beta}}_{i} \cdot \mathbf{X}(\mathbf{r}^{\prime})]  d v^{\prime}} .
\end{aligned}
\end{equation}
Invoking $dv \approx \tau ds$ under the thin-sheet approximation~\cite{TS},
\begin{equation}
\label{eq:la_reduced}
\begin{aligned}
    \mathcal{L}_{A}[\bar{\bar{\beta}}_{i}\cdot\mathbf{X}](\mathbf{r})=\tau k_0^2\int_S G(\mathbf{r}, \mathbf{r}^{\prime}) \bar{\bar{\beta}}_{i}\cdot\mathbf{X}(\mathbf{r}^{\prime}) d s^{\prime}.
\end{aligned}
\end{equation}

The scalar potential contribution $\mathcal{L}_{\mathrm{\phi}}[\bar{\bar{\beta}}_{i}\cdot\mathbf{X}](\mathbf{r})$ requires careful treatment due to the divergence operator, which produces both volume and surface charge terms~\cite{TS}. To this end, the divergence is expanded as:
\begin{equation}
\label{eqdiv}
\nabla \cdot [\bar{\bar{\beta}}_{i} \cdot \mathbf{X}(\mathbf{r})]=[\nabla \cdot \bar{\bar{\beta}}_{i}] \cdot \mathbf{X}(\mathbf{r})+\bar{\bar{\beta}}_{i}:\nabla \mathbf{X}(\mathbf{r})
\end{equation}
where ``$:$" denotes a double inner product. Within $V$, $\nabla \cdot \bar{\bar{\beta}}_i = 0$; however, the discontinuity in $\bar{\bar{\beta}}_i$ across the surface of $V$ ($\bar{\bar{\beta}}_i=0$ in the background medium) introduces a delta function contribution~\cite{TS}. Since $V$ does not support free charges, $\nabla \cdot \mathbf{D}(\mathbf{r}) = 0$ and $\nabla \cdot \mathbf{B}(\mathbf{r}) = 0$, i.e., $\nabla\cdot\mathbf{X}(\mathbf{r})=0$. Consequently, for a scalar contrast $\beta_i$, the second term in~\eqref{eqdiv} reduces to $\beta_i\nabla\cdot\mathbf{X}(\mathbf{r})$ and vanishes, whereas for a tensor contrast, $\bar{\bar{\beta}}_i:\nabla\mathbf{X}(\mathbf{r})$ does not vanish in general and yields a volume contribution. As a result, $\mathcal{L}_{\mathrm{\phi}}[\bar{\bar{\beta}}_i\cdot\mathbf{X}](\mathbf{r})$ is expressed as~\cite{TS}:
\begin{equation}
\label{eqTSAni1}
\begin{aligned}
 \mathcal{L}_{\mathrm{\phi}}[\bar{\bar{\beta}}_{i}\cdot\mathbf{X}](\mathbf{r})&=-\int_{S^{+}_{\bot}} G(\mathbf{r}, \mathbf{r}^{\prime}) \hat{\mathbf{n}}(\mathbf{r}^{\prime}) \cdot[\bar{\bar{\beta}}_{i}\cdot \mathbf{X}(\mathbf{r^{\prime}})] d s^{\prime}+\int_{S^{-}_{\bot}} G(\mathbf{r}, \mathbf{r}^{\prime}) \hat{\mathbf{n}}(\mathbf{r}^{\prime}) \cdot[\bar{\bar{\beta}}_{i}\cdot \mathbf{X}(\mathbf{r^{\prime}})] d s^{\prime}\\
  &-\int_{S_{\parallel}} G(\mathbf{r}, \mathbf{r}^{\prime}) \hat{\mathbf{t}}(\mathbf{r}^{\prime}) \cdot[\bar{\bar{\beta}}_{i}\cdot \mathbf{X}(\mathbf{r^{\prime}})] d s^{\prime}+\int_V  G(\mathbf{r}, \mathbf{r}^{\prime}) \bar{\bar{\beta}}_{i}: \nabla^\prime \mathbf{X}(\mathbf{r}^\prime) d v^{\prime}.
\end{aligned}
\end{equation}
The tensor $\bar{\bar{\beta}}_{i}$ is decomposed into tangential–tangential, tangential–normal, normal–tangential, and normal–normal blocks. For equivalent thin-sheet models of 2D materials such as graphene and metasurfaces, tangential–normal and normal-tangential coupling terms are typically negligible or identically zero~\cite{Zhao2022MultitraceSIE, Caloz_Book}. Accordingly, $\bar{\bar{\beta}}_{i}$ is simplified to: 
\begin{equation}
\label{eqbeta2}
\bar{\bar\beta}_i=
\begin{bmatrix}
\bar{\bar\beta}_{i,\parallel} & \bar{0} \\
\bar{0} & \beta_{i,\bot}
\end{bmatrix}.
\end{equation}
Here, $\bar{\bar{\beta}}_{i,{\parallel}}$ denotes the $2 \times 2$ tangential-tangential block of $\bar{\bar{\beta}}_i$, and $\beta_{i,\perp}$ is the normal-normal component. Substituting~\eqref{eqbeta2} into~\eqref{eqdiv} yields
\begin{equation}
\label{eqdd_inner}
\bar{\bar{\beta}}_{i}: \nabla \mathbf{X}(\mathbf{r}) =\bar{\bar{\beta}}_{ i,\parallel}: \nabla_{\parallel} \mathbf{X}_{\parallel}(\mathbf{r})+\beta_{ i,\bot}\partial_{\bot} X_{\bot}(\mathbf{r}). 
\end{equation}
Substituting~\eqref{eqdd_inner} into~\eqref{eqTSAni1}, using the relations $\hat{\mathbf{n}}(\mathbf{r}) \cdot [\bar{\bar{\beta}}_{i}\cdot \mathbf{X}(\mathbf{r})] = \beta_{i, \perp} X_{\perp}(\mathbf{r})$ and $\hat{\mathbf{t}}(\mathbf{r}) \cdot [\bar{\bar{\beta}}_{i}\cdot \mathbf{X}(\mathbf{r})]=\hat{\mathbf{t}}(\mathbf{r}) \cdot [\bar{\bar{\beta}}_{i,\parallel} \cdot \mathbf{X}_{\parallel}(\mathbf{r})]$, and invoking $dv\approx \tau ds$ for the integral over $V$ and $ds\approx \tau dl$ for the integral over $S_{\parallel}$ yield
\begin{equation}
\label{eqL2}
\begin{aligned}
    &\mathcal{L}_{\phi}[\bar{\bar{\beta}}_{i}\cdot \mathbf{X}](\mathbf{r})= - \int_{S^{+}_\bot} G(\mathbf{r},\mathbf{r}^{\prime}) \beta_{i,\bot}X^{+}_{\bot}(\mathbf{r}^{\prime}) d s^{\prime}+ \int_{S^{-}_\bot} G(\mathbf{r},\mathbf{r}^{\prime})\beta_{i,\bot}X^{-}_{\bot}(\mathbf{r}^{\prime}) d s^{\prime}\\
    &- \tau\int_{C}G(\mathbf{r},\mathbf{r}^{\prime}) \mathbf{\hat{t}}(\mathbf{r}^{\prime}) \cdot [\bar{\bar{\beta}}_{i,\parallel} \cdot \mathbf{X}_{\parallel}(\mathbf{r}^{\prime})]   d l^{\prime}+\tau\int_S G(\mathbf{r}, \mathbf{r}^{\prime})[\bar{\bar{\beta}}_{i, \parallel}: \nabla_{\parallel}^{\prime} \mathbf{X}_{\parallel}(\mathbf{r}^{\prime})]  d s^{\prime}\\
&+\tau\int_S G(\mathbf{r}, \mathbf{r}^{\prime})\beta_{i, \bot} \partial^{\prime}_{\bot} X_{\bot}(\mathbf{r}^{\prime}) d s^{\prime} 
\end{aligned}    
\end{equation}
where $X^{+}_{\bot}(\mathbf{r})$ and $X^{-}_{\bot}(\mathbf{r})$ are the normal components of $X(\mathbf{r})$ on $S^{+}_{\perp}$ and $S^{-}_{\perp}$, respectively.

To reduce the number of independent unknowns, $\nabla \cdot \mathbf{X}(\mathbf{r}) = 0$ is decomposed into its tangential and normal components as 
\begin{equation}
\label{eq:divfree}
\nabla_{\parallel} \cdot \mathbf{X}_{\parallel}(\mathbf{r})+\partial_{\perp} X_{\perp}(\mathbf{r})=0.
\end{equation}

Unlike in~\cite{TS}, where the independent normal unknown is the value on $S_{\perp}^{+}$, the normal unknown is defined here on the mid-surface (see Section~\ref{sec:comments}). Let $X_{\perp}(\mathbf{r})$, $\mathbf{r}\in S$, denote the normal component of $\mathbf{X}(\mathbf{r})$ on $S$, approximated as $X_{\perp}(\mathbf{r}) \approx [X_{\perp}^{+}(\mathbf{r})+X_{\perp}^{-}(\mathbf{r})]/2$, and let $\partial_{\perp}X_{\perp}(\mathbf{r}) \approx [X_{\perp}^{+}(\mathbf{r})-X_{\perp}^{-}(\mathbf{r})]/\tau$. Then~\eqref{eq:divfree} expresses the normal components on the two faces in terms of $X_{\perp}(\mathbf{r})$ as
\begin{equation}
X_{\perp}^{\pm}(\mathbf{r}) = X_{\perp}(\mathbf{r}) \mp \frac{\tau}{2}\nabla_{\parallel}\cdot\mathbf{X}_{\parallel}(\mathbf{r}),\quad \mathbf{r}\in S.
\label{eq:normal_der}
\end{equation}
Substituting~\eqref{eq:divfree} and~\eqref{eq:normal_der} into~\eqref{eqL2} and moving the integrations over $S^{+}_{\perp}$ and $S^{-}_{\perp}$ onto $S$, by shifting the source point in $G(\mathbf{r},\mathbf{r}^{\prime})$ by $\pm\tau/2$, lead to
\begin{equation}
\label{eqL2preFinal}
\begin{aligned}
&\mathcal{L}_{\phi}[\bar{\bar{\beta}}_{i}\cdot \mathbf{X}](\mathbf{r})=-\int_{S} G(\mathbf{r},\mathbf{r}^{\prime}+\frac{\tau\hat{\mathbf{n}}(\mathbf{r}^{\prime})}{2})\beta_{i,\bot}X_{\bot}(\mathbf{r}^{\prime}) d s^{\prime}\\
&+\frac{\tau}{2}\int_{S}  G(\mathbf{r},\mathbf{r}^{\prime}+\frac{\tau\hat{\mathbf{n}}(\mathbf{r}^{\prime})}{2}) \beta_{i,\bot}\nabla^{\prime}_{\parallel} \cdot \mathbf{X}_{\parallel}(\mathbf{r}^{\prime} )  ds^{\prime}+\int_{S} G(\mathbf{r},\mathbf{r}^{\prime}-\frac{\tau\hat{\mathbf{n}}(\mathbf{r}^{\prime})}{2}) \beta_{i,\bot}X_{\bot}(\mathbf{r}^{\prime}) d s^{\prime}\\
&+\frac{\tau}{2}\int_{S}  G(\mathbf{r},\mathbf{r}^{\prime}-\frac{\tau\hat{\mathbf{n}}(\mathbf{r}^{\prime})}{2}) \beta_{i,\bot}\nabla^{\prime}_{\parallel} \cdot \mathbf{X}_{\parallel}(\mathbf{r}^{\prime} )  ds^{\prime}- \tau\int_{C}G(\mathbf{r},\mathbf{r}^{\prime})\mathbf{\hat{t}}(\mathbf{r}^{\prime}) \cdot [\bar{\bar{\beta}}_{i,\parallel} \cdot \mathbf{X}_{\parallel}(\mathbf{r}^{\prime})]   d l^{\prime}\\
&+\tau\int_S G(\mathbf{r}, \mathbf{r}^{\prime}) [\bar{\bar{\beta}}_{i, \parallel}: \nabla_{\parallel}^{\prime} \mathbf{X}_{\parallel}(\mathbf{r}^{\prime})] d s^{\prime}\\
&-\tau\int_S G(\mathbf{r}, \mathbf{r}^{\prime})\beta_{i, \bot} \nabla^{\prime}_{\parallel}\cdot \mathbf{X}_{\parallel}(\mathbf{r}^{\prime}) d s^{\prime} .
\end{aligned}   
\end{equation}
Finally, $\nabla \mathcal{L}_{\phi}[\bar{\bar{\beta}}_{i} \cdot \mathbf{X}](\mathbf{r})$ is decomposed to two operators as 
\begin{equation}
\nabla \mathcal{L}_{\phi}[\bar{\bar{\beta}}_{i} \cdot \mathbf{X}](\mathbf{r}) = \mathcal{T}_{\phi}[\bar{\bar{\beta}}_i \cdot \mathbf{X}_{\parallel}](\mathbf{r}) + \mathcal{N}_{\phi}[\bar{\bar{\beta}}_i \cdot \hat{\mathbf{n}}{X}_{\bot}](\mathbf{r})
\end{equation}
where $\mathcal{T}_{\phi}[\bar{\bar{\beta}}_i \cdot \mathbf{X}_{\parallel}](\mathbf{r})$ collects the terms driven by $\mathbf{X}_{\parallel}(\mathbf{r})$, and $\mathcal{N}_{\phi}[\bar{\bar{\beta}}_i \cdot \hat{\mathbf{n}}{X}_{\bot}](\mathbf{r})$ collects the terms driven by $X_{\perp}(\mathbf{r})$. Their expressions are given as
\begin{equation}
\label{eq:t_phi_reduced}
\begin{aligned}
&\mathcal{T}_{\phi}[\bar{\bar{\beta}}_i \cdot \mathbf{X}_{\parallel}](\mathbf{r})=\frac{\tau}{2}\int_{S} \nabla G(\mathbf{r},\mathbf{r}^{\prime}+\frac{\tau\hat{\mathbf{n}}(\mathbf{r}^{\prime})}{2}) \beta_{i,\bot}\nabla^{\prime}_{\parallel} \cdot \mathbf{X}_{\parallel}(\mathbf{r}^{\prime} )  ds^{\prime}\\
&+\frac{\tau}{2}\int_{S} \nabla G(\mathbf{r},\mathbf{r}^{\prime}-\frac{\tau\hat{\mathbf{n}}(\mathbf{r}^{\prime})}{2}) \beta_{i,\bot}\nabla^{\prime}_{\parallel} \cdot \mathbf{X}_{\parallel}(\mathbf{r}^{\prime} )  ds^{\prime}- \tau\int_{C}\nabla G(\mathbf{r},\mathbf{r}^{\prime})\mathbf{\hat{t}}(\mathbf{r}^{\prime}) \cdot [\bar{\bar{\beta}}_{i,\parallel} \cdot \mathbf{X}_{\parallel}(\mathbf{r}^{\prime})]   d l^{\prime}\\
&+\tau\int_S \nabla G(\mathbf{r}, \mathbf{r}^{\prime}) [\bar{\bar{\beta}}_{i, \parallel}: \nabla_{\parallel}^{\prime} \mathbf{X}_{\parallel}(\mathbf{r}^{\prime})] d s^{\prime}-\tau\int_S \nabla G(\mathbf{r}, \mathbf{r}^{\prime})\beta_{i, \bot} \nabla^{\prime}_{\parallel}\cdot \mathbf{X}_{\parallel}(\mathbf{r}^{\prime}) d s^{\prime} 
\end{aligned}
\end{equation}

\begin{equation}
\label{eq:n_phi_reduced}
\begin{aligned}
&\mathcal{N}_{\phi}[\bar{\bar{\beta}}_i \cdot \hat{\mathbf{n}}{X}_{\bot}](\mathbf{r})=-\int_{S}\nabla G(\mathbf{r},\mathbf{r}^{\prime}+\frac{\tau\hat{\mathbf{n}}(\mathbf{r}^{\prime})}{2}) \beta_{i,\bot}{X}_{\bot}(\mathbf{r}^{\prime})  d s^{\prime}\\
 &+\int_{S} \nabla G(\mathbf{r},\mathbf{r}^{\prime}-\frac{\tau\hat{\mathbf{n}}(\mathbf{r}^{\prime})}{2})  \beta_{i,\bot}{X}_{\bot}(\mathbf{r}^{\prime}) d s^{\prime}.
\end{aligned}   
\end{equation}
\subsubsection{Thin-Sheet Reduction of $\mathcal{K}[\bar{\bar{\beta}}_{i}\cdot \mathbf{X}](\mathbf{r})$}
Invoking $dv \approx \tau ds$ under the thin-sheet approximation~\cite{TS}, 
\begin{equation}
\label{eq:k_reduced}
\begin{aligned}
   &\mathcal{K}[\bar{\bar{\beta}}_{i}\cdot\mathbf{X}](\mathbf{r})=\tau \int_S\nabla  G(\mathbf{r}, \mathbf{r}^{\prime}) \times \bar{\bar{\beta}}_{i}\cdot\mathbf{X}(\mathbf{r}^{\prime})\,d s^{\prime}.
\end{aligned}    
\end{equation}
\subsection{Equivalent Thin-Sheet Representation of GSTCs}\label{sec:GSTC_mapping}
To incorporate the GSTCs into the VIE formulation, the metasurface is represented by an equivalent thin sheet of thickness $\tau$ centered on $S$, whose constitutive tensors $\bar{\bar{\alpha}}_i(\mathbf{r})$, $i=1,2,3,4$, are chosen such that the sheet reproduces the field discontinuities prescribed by the GSTCs. To this end, the electric and magnetic volume polarizations of the sheet, $\mathbf{D}(\mathbf{r})-\varepsilon_0\mathbf{E}(\mathbf{r})$ and $[\mathbf{B}(\mathbf{r})-\mu_0\mathbf{H}(\mathbf{r})]/\mu_0$, which give rise to the equivalent currents in~\eqref{eq4}, are required, when integrated across the thickness of the sheet, to equal the surface polarizations $\mathbf{P}^{\mathrm{e}}(\mathbf{r})$ and $\mathbf{P}^{\mathrm{m}}(\mathbf{r})$ of the GSTCs~\cite{Kuester,idemen2011discontinuities}. Since these volume polarizations are constant across the sheet under the thin-sheet approximation, this condition reads
\begin{equation}
\label{eq:match}
\begin{aligned}
\tau[\mathbf{D}(\mathbf{r})-\varepsilon_0\mathbf{E}(\mathbf{r})] &= \mathbf{P}^{\mathrm{e}}(\mathbf{r})\\
\tau[\mathbf{B}(\mathbf{r})-\mu_0\mathbf{H}(\mathbf{r})] &= \mu_0\mathbf{P}^{\mathrm{m}}(\mathbf{r}).
\end{aligned}
\end{equation}
Here, the fields on the left-hand side are those inside the sheet, whereas $\mathbf{P}^{\mathrm{e}}(\mathbf{r})$ and $\mathbf{P}^{\mathrm{m}}(\mathbf{r})$ are defined in terms of the average fields $\Sigma\mathbf{E}(\mathbf{r})$ and $\Sigma\mathbf{H}(\mathbf{r})$ using~\eqref{eqPol}. Consistent with~\eqref{eqbeta2}, $\bar{\bar{\chi}}_{\mathrm{ab}}$ are assumed to have no tangential--normal coupling and, therefore, the tangential and normal components of~\eqref{eq:match} are treated separately.

Since the tangential fields are continuous across the faces of the sheet, their values on the two faces equal those on the two sides of the metasurface. Under the thin-sheet approximation, the tangential fields inside the sheet are represented by their values on $S$, which equal the averages of the face values, i.e., $\mathbf{E}_{\parallel}(\mathbf{r}) = \Sigma\mathbf{E}_{\parallel}(\mathbf{r})$ and $\mathbf{H}_{\parallel}(\mathbf{r}) = \Sigma\mathbf{H}_{\parallel}(\mathbf{r})$. Substituting~\eqref{eqPol} into the tangential components of~\eqref{eq:match} then yields
\begin{equation}\label{eq:DB_par}
\begin{bmatrix}
\mathbf{D}_{\parallel}(\mathbf{r}) \\ \mathbf{B}_{\parallel}(\mathbf{r})
\end{bmatrix}
=
\begin{bmatrix}
\varepsilon_0 ( \bar{\bar{I}} + \bar{\bar{\chi}}_{\mathrm{ee},\parallel}/\tau) 
&
\sqrt{\mu_0 \varepsilon_0}\,\bar{\bar{\chi}}_{\mathrm{em},\parallel}/\tau
\\
\sqrt{\mu_0 \varepsilon_0}\,\bar{\bar{\chi}}_{\mathrm{me},\parallel}/\tau
&
\mu_0 ( \bar{\bar{I}} + \bar{\bar{\chi}}_{\mathrm{mm},\parallel}/\tau)
\end{bmatrix}
\begin{bmatrix}
\mathbf{E}_{\parallel}(\mathbf{r}) \\ \mathbf{H}_{\parallel}(\mathbf{r})
\end{bmatrix}
\end{equation}
where $\bar{\bar{I}}$ is the $2\times2$ identity matrix. Since~\eqref{eq3} expresses $\mathbf{E}_{\parallel}(\mathbf{r})$ and $\mathbf{H}_{\parallel}(\mathbf{r})$ in terms of $\mathbf{D}_{\parallel}(\mathbf{r})$ and $\mathbf{B}_{\parallel}(\mathbf{r})$, the tangential blocks of $\bar{\bar{\alpha}}_i(\mathbf{r})$ are obtained by inverting~\eqref{eq:DB_par}:
\begin{equation}
\label{eq:alpha_chi}
\begin{bmatrix}
\bar{\bar{\alpha}}_{1,\parallel} & \bar{\bar{\alpha}}_{2,\parallel} \\
\bar{\bar{\alpha}}_{3,\parallel} & \bar{\bar{\alpha}}_{4,\parallel}
\end{bmatrix}
=
\begin{bmatrix}
\varepsilon_0 ( \bar{\bar{I}} + \bar{\bar{\chi}}_{\mathrm{ee},\parallel}/\tau)
&
\sqrt{\mu_0 \varepsilon_0}\,\bar{\bar{\chi}}_{\mathrm{em},\parallel}/\tau
\\
\sqrt{\mu_0 \varepsilon_0}\,\bar{\bar{\chi}}_{\mathrm{me},\parallel}/\tau
&
\mu_0 ( \bar{\bar{I}} + \bar{\bar{\chi}}_{\mathrm{mm},\parallel}/\tau)
\end{bmatrix}^{-1}.
\end{equation}
In the tangential directions, the sheet thus behaves as a bianisotropic medium whose volume susceptibilities are $\bar{\bar{\chi}}_{\mathrm{ab},\parallel}/\tau$, e.g., with relative permittivity $\bar{\bar{I}}+\bar{\bar{\chi}}_{\mathrm{ee},\parallel}/\tau$.

For the normal components, $D_{\perp}(\mathbf{r})$ and $B_{\perp}(\mathbf{r})$ are continuous across the faces instead. Since $D_{\perp}(\mathbf{r})$ and $B_{\perp}(\mathbf{r})$ on $S$ are defined as the averages of their values on the two faces (Section~\ref{sec:TS-VIEs}), it follows that $\Sigma E_{\perp}(\mathbf{r}) = D_{\perp}(\mathbf{r})/\varepsilon_0$ and $\Sigma H_{\perp}(\mathbf{r}) = B_{\perp}(\mathbf{r})/\mu_0$, while $E_{\perp}(\mathbf{r})$ and $H_{\perp}(\mathbf{r})$ inside the sheet differ from these averages. Substituting these relations into the normal components of~\eqref{eqPol} and~\eqref{eq:match} yields
\begin{equation}\label{eq:DB_perp}
\begin{aligned}
\tau[D_{\perp}(\mathbf{r})&-\varepsilon_0 E_{\perp}(\mathbf{r})] =\chi_{\mathrm{ee},\perp} D_{\perp}(\mathbf{r}) + \sqrt{\varepsilon_0/\mu_0}\,\chi_{\mathrm{em},\perp} B_{\perp}(\mathbf{r})\\
\tau[B_{\perp}(\mathbf{r})&-\mu_0 H_{\perp}(\mathbf{r})] =\sqrt{\mu_0/\varepsilon_0}\,\chi_{\mathrm{me},\perp} D_{\perp}(\mathbf{r})+ \chi_{\mathrm{mm},\perp} B_{\perp}(\mathbf{r}).
\end{aligned}
\end{equation}
Solving~\eqref{eq:DB_perp} for $E_{\perp}(\mathbf{r})$ and $H_{\perp}(\mathbf{r})$ puts it in the form of the normal components of~\eqref{eq3}, from which the normal--normal components $\alpha_{i,\perp}$ follow as
\begin{equation}\label{eq:alpha_perp}
\begin{aligned}
\alpha_{1,\perp} &= \frac{\tau-\chi_{\mathrm{ee},\perp}}{\varepsilon_0\,\tau}, &
\alpha_{2,\perp} &= -\frac{\chi_{\mathrm{em},\perp}}{\sqrt{\mu_0\varepsilon_0}\,\tau},\\
\alpha_{3,\perp} &= -\frac{\chi_{\mathrm{me},\perp}}{\sqrt{\mu_0\varepsilon_0}\,\tau}, &
\alpha_{4,\perp} &= \frac{\tau-\chi_{\mathrm{mm},\perp}}{\mu_0\,\tau}.
\end{aligned}
\end{equation}
Unlike for the tangential blocks, no matrix inversion is required, since the right-hand sides of~\eqref{eq:DB_perp} are already expressed in terms of $D_{\perp}(\mathbf{r})$ and $B_{\perp}(\mathbf{r})$.

\subsection{Discretization and Matrix System}\label{sec:Discretization}
To numerically solve the thin-sheet reduced form of~\eqref{eq5}–\eqref{eq6} using the MoM, the mid-surface $S$ is discretized using a triangular mesh. The tangential components $\mathbf{D}_{\parallel}(\mathbf{r})$ and $\mathbf{B}_{\parallel}(\mathbf{r})$ and the normal components ${D}_{\bot}(\mathbf{r})$ and ${B}_{\bot}(\mathbf{r})$ are expanded using RWG~\cite{rao_electromagnetic_1982} and pulse basis functions~\cite{TS}, respectively, as
\begin{equation}
\label{eqExp}
\begin{aligned}
    \mathbf{D}(\mathbf{r}) & =\mathbf{D}_{\parallel}(\mathbf{r})+\hat{\mathbf{n}}(\mathbf{r}){D}_{\bot}(\mathbf{r})  = \sum_{n=1}^{N_{\parallel}} \{\bar{I}_{\mathrm{D},\parallel}\}_n \mathbf{f}_n(\mathbf{r})+\sum_{q=1}^{N_{\bot}} \{\bar{I}_{\mathrm{D},\bot}\}_{q} p_q(\mathbf{r})\\
    \mathbf{B}(\mathbf{r}) & =\mathbf{B}_{\parallel}(\mathbf{r})+\hat{\mathbf{n}}(\mathbf{r}){B}_{\bot}(\mathbf{r})  = \sum_{n=1}^{N_{\parallel}} \{\bar{I}_{\mathrm{B},\parallel}\}_n \mathbf{f}_n(\mathbf{r})+\sum_{q=1}^{N_{\bot}} \{\bar{I}_{\mathrm{B},\bot}\}_{q}p_q(\mathbf{r}).
\end{aligned}    
\end{equation}
Here, $\mathbf{f}_n(\mathbf{r})$, $n=1,\ldots, N_{\parallel}$, are the RWG functions defined over pairs of adjacent triangular elements~\cite{rao_electromagnetic_1982} and $p_q(\mathbf{r})$, $q=1,\dots,N_{\perp}$, are the pulse functions defined on triangular elements~\cite{TS}. The vectors $\bar{I}_{\mathrm{D},\parallel}$ and $\bar{I}_{\mathrm{B},\parallel}$ store the unknown expansion coefficients associated with $\mathbf{f}_n(\mathbf{r})$, while $\bar{I}_{\mathrm{D},\perp}$ and $\bar{I}_{\mathrm{B},\perp}$ store those associated with $p_q(\mathbf{r})$. The total numbers of edges and elements in the mesh are denoted by $N_{\parallel}$ and $N_{\perp}$, which are equal to the total numbers of $\mathbf{f}_n(\mathbf{r})$ and $p_q(\mathbf{r})$, respectively. The dimensions of $\bar{I}_{D,\parallel}$ and $\bar{I}_{B,\parallel}$ are $N_{\parallel}$, while those of $\bar{I}_{D,\perp}$ and $\bar{I}_{B,\perp}$ are $N_{\perp}$. Note that at the boundary edges of $S$, $\mathbf{f}_n(\mathbf{r})$ are defined as half RWG basis functions.  

Substituting~\eqref{eqExp} into the thin-sheet reduced form of~\eqref{eq5}-\eqref{eq6} and applying Galerkin testing using $\mathbf{f}_m(\mathbf{r})$, $m = 1,\dots,N_{\parallel}$, and 
$\hat{\mathbf{n}}(\mathbf{r})p_k(\mathbf{r})$, $k = 1,\dots,N_{\perp}$, as testing functions yields a linear system of dimension $(2N_{\parallel}+2N_{\perp}) \times 
(2N_{\parallel}+2N_{\perp})$, which can be written compactly as
\begin{equation}\label{final_mom}
\bar{\bar{Z}}\bar{I} = \bar{V}.
\end{equation}
Here, $\bar{\bar{Z}}$ is composed of $16$ block matrices $\bar{\bar{Z}}_{kl}$, $k,l \in \{1,2,3,4\}$, where each block has dimensions $N_{\parallel}\times N_{\parallel}$, $N_{\parallel}\times N_{\perp}$, $N_{\perp}\times N_{\parallel}$, or $N_{\perp}\times N_{\perp}$ depending on the indices $k$ and $l$. The vector $\bar{I}$ collects all unknown expansion coefficients, i.e., $\bar{I}_{\mathrm{D},\parallel}$, $\bar{I}_{\mathrm{D},\perp}$, $\bar{I}_{\mathrm{B},\parallel}$, and $\bar{I}_{\mathrm{B},\perp}$. Similarly, the vector $\bar{V}$ is formed by cascading vectors $\bar{V}_{\mathrm{E},\parallel}$, $\bar{V}_{\mathrm{E},\perp}$, $\bar{V}_{\mathrm{H},\parallel}$, and $\bar{V}_{\mathrm{H},\perp}$, where $\bar{V}_{\mathrm{E},\parallel}$ and $\bar{V}_{\mathrm{E},\perp}$ store $\mathbf{E}^{\mathrm{inc}}(\mathbf{r})$ tested using $\mathbf{f}_m(\mathbf{r})$ and $\hat{\mathbf{n}}(\mathbf{r})p_q(\mathbf{r})$, respectively, and $\bar{V}_{\mathrm{H},\parallel}$ and $\bar{V}_{\mathrm{H},\perp}$ store $\mathbf{H}^{\mathrm{inc}}(\mathbf{r})$ tested using $\mathbf{f}_m(\mathbf{r})$ and $\hat{\mathbf{n}}(\mathbf{r})p_q(\mathbf{r})$, respectively. The dimensions of $\bar{V}_{\mathrm{E},\parallel}$ and $\bar{V}_{\mathrm{H},\parallel}$ are $N_{\parallel}$, while those of $\bar{V}_{\mathrm{E},\perp}$ and $\bar{V}_{\mathrm{H},\perp}$ are $N_{\perp}$.

The explicit expressions for the entries of the block matrices, $\bar{\bar{Z}}_{kl}$, $k,l \in \{1,2,3,4\}$, and the vectors $\bar{V}_{\mathrm{E},\parallel}$, $\bar{V}_{\mathrm{E},\perp}$, $\bar{V}_{\mathrm{H},\parallel}$, and $\bar{V}_{\mathrm{H},\perp}$ are provided in Appendix~\ref{sec:matrix_entries}. The singularity and hypersingularity treatment of the integrals arising in the matrix entries is described in Appendix~\ref{sec:inner_product}.

\subsection{Comments}\label{sec:comments}
Four comments on the formulation are in order. First, the thin-sheet reduction of Section~\ref{sec:TS-VIEs} is derived for a homogeneous sheet. After discretization, however, the constitutive tensors are evaluated at the centroid of each element and assumed constant over it, so that the sheet is piecewise homogeneous. The reduction applies without modification to this case, with each element treated as a homogeneous subregion, as detailed in Appendix~\ref{sec:inhomogeneous}.

Second, the normal unknown is placed on the mid-surface $S$ rather than on the top face of the sheet as in~\cite{TS} (Section~\ref{sec:TS-VIEs}). This placement is consistent with the equivalent thin-sheet representation of the GSTCs, as follows from Section~\ref{sec:GSTC_mapping}. An unknown on the top face would instead be offset by $\tau/2$ from $S$, where the local terms $\bar{\bar{\alpha}}_i\cdot\mathbf{X}(\mathbf{r})$ are enforced. In addition, the mid-surface placement renders the face charges in~\eqref{eqL2preFinal} symmetric with respect to $S$.

Third, $\tau$ is a normalization parameter that distributes the surface polarizations across the sheet (Section~\ref{sec:GSTC_mapping}). It does not need to match the physical thickness of the unit cells, since the GSTCs already represent their response as that of a zero-thickness sheet. For the examples of Section~\ref{sec:Bianisotropic}, the accuracy is insensitive to $\tau$ as long as $\tau \ll \lambda_0$, while the iteration count remains constant or grows moderately as $\tau$ decreases. That said, a true zero-thickness model may be preferable when fields must be evaluated within a distance comparable to $\tau$ from the sheet, or when other scatterers lie at such a distance.

\begin{figure}[h]
\centering
\includegraphics[width=0.6\linewidth]{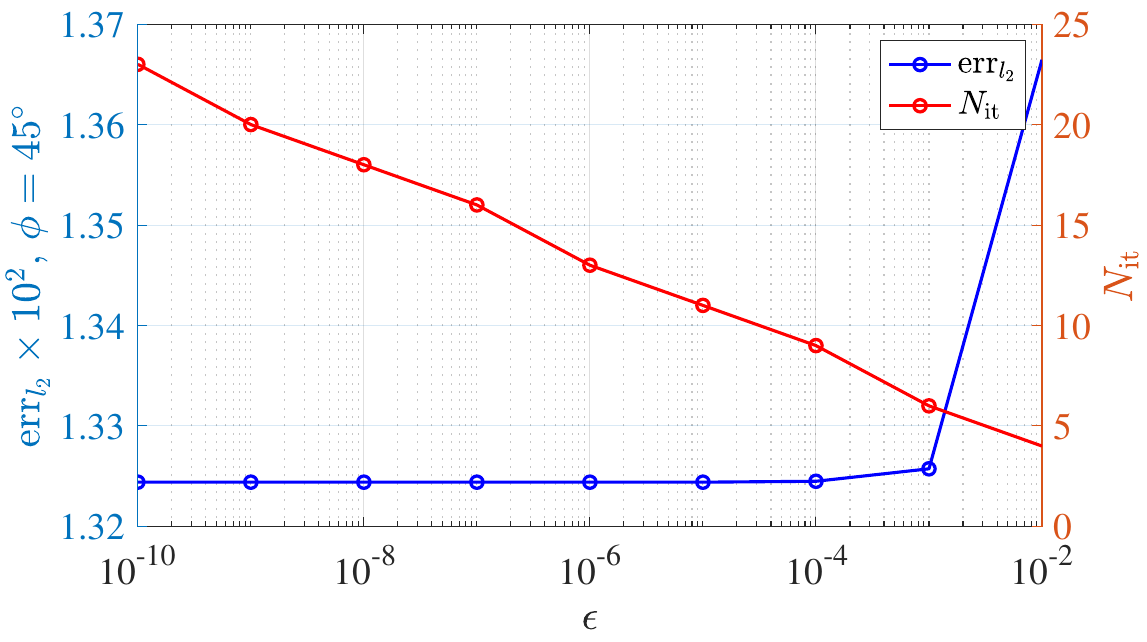}
\caption{Accuracy and convergence of the TS-VIE solver as a function of the convergence threshold $\epsilon$ for the spherical monoisotropic shell with $l_{\mathrm{avg}} = \lambda_0/10$ and $\tau = \lambda_0/30$. $\mathrm{err}_{\ell_2}$ and $N_{\mathrm{it}}$ are shown on the left and right axes, respectively.}
\label{Fig:Eps}
\end{figure}

\begin{figure}[h]
\centering
\subfigure[]{\includegraphics[width=0.6\linewidth]{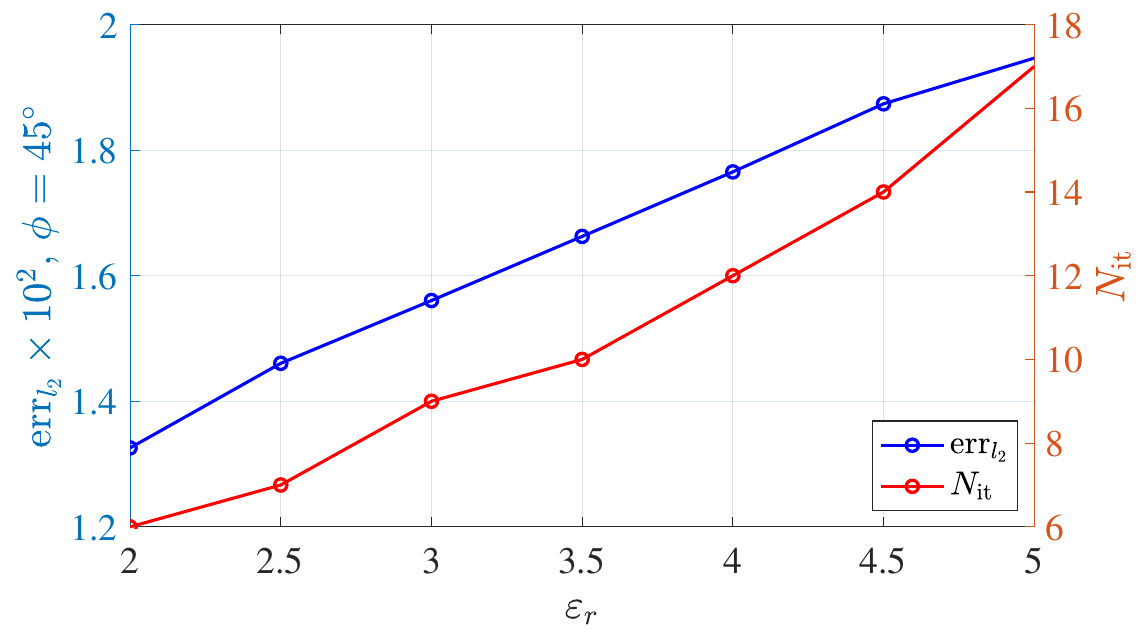}}\\
\subfigure[]{\includegraphics[width=0.6\linewidth]{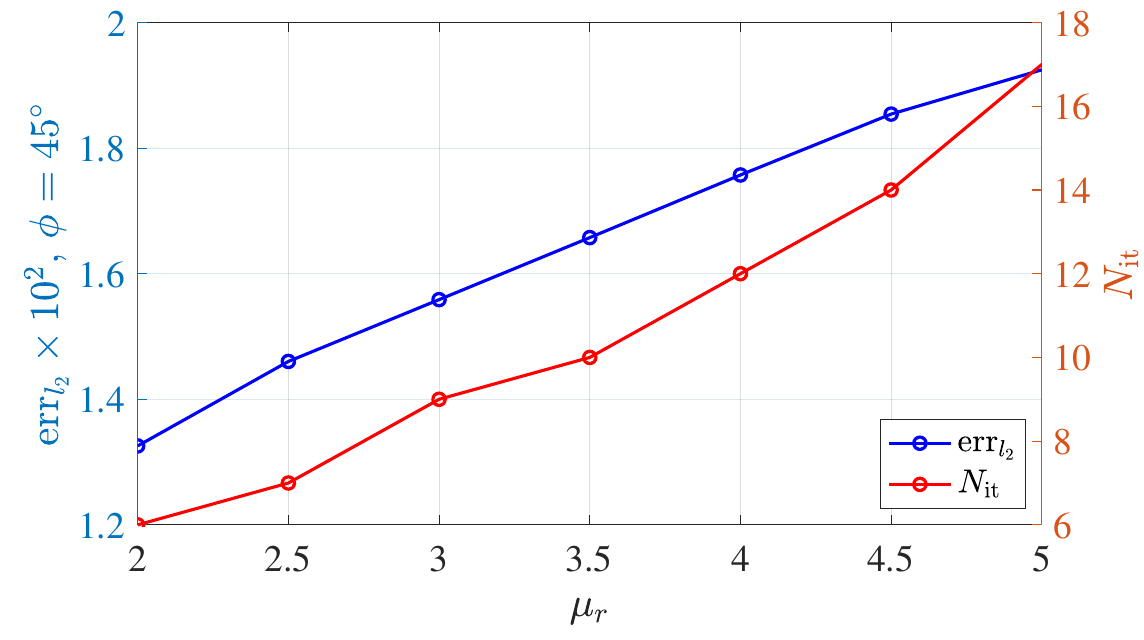}}
\caption{Accuracy and convergence of the TS-VIE solver as a function of the material parameters for the spherical monoisotropic shell with $\tau = \lambda_0/30$ and $l_{\mathrm{avg}} = \lambda_0/10$. $\mathrm{err}_{\ell_2}$ and $N_{\mathrm{it}}$ are shown on the left and right axes, respectively, (a) versus $\varepsilon_r$ for $\mu_r = 2$ and (b) versus $\mu_r$ for $\varepsilon_r = 2$.}
\label{Fig:Material}
\end{figure}

Fourth, the thin-sheet reduced equations are expected to inherit the numerical issues of conventional SIEs, such as the choice of basis and testing function spaces~\cite{CFIE_Pasi, Andriulli2008, Bagci2009, Cools2011, Bogaert2014}, the accuracy of the identity contribution arising from the principal-value treatment~\cite{CFIE_Pasi, Cools2011, Bogaert2014}, the first- or second-kind character of the operators~\cite{Andriulli2008, Bagci2009}, the low-frequency behavior~\cite{Vecchi1999, Andriulli2012, R4_Reviewer3}, and the dense-discretization breakdown~\cite{Andriulli2008, Bagci2009}, with the normally-tested operators adding further considerations~\cite{R1_Reviewer3, R2_Reviewer3, R3_Reviewer3, R4_Reviewer3}. The observed behavior is consistent with the second-kind structure of the flux-based VIE: the iteration count is insensitive to the discretization over the range of discretizations considered [Fig.~\ref{Fig:Tau_Anl}(b)] and grows only slowly as the convergence threshold is tightened (Fig.~\ref{Fig:Eps}), whereas it increases with the material contrast (Fig.~\ref{Fig:Material}) and with the magnitude of the susceptibilities (Section~\ref{sec:PEC}), as is typical of conventional VIEs. Neither the low-frequency nor the dense-discretization breakdown regime has been tested here, since neither arises for GSTC-modeled metasurfaces: the GSTCs homogenize the unit cells, so that the sheet is discretized at the scale of the wavelength rather than of the unit-cell features, and the metasurfaces of interest are not electrically small at their operating frequencies. A rigorous analysis of these operators, including alternative testing schemes for the normal contributions, is left for future work.

\section{Numerical Results}\label{sec:Numerical_Examples}

In this section, several numerical examples are presented to assess the accuracy, efficiency, and robustness of the proposed solver, hereafter referred to as the TS-VIE-GSTC solver. In all examples, the generalized minimal residual (GMRES) method~\cite{GMRES} is employed to solve the linear system~\eqref{final_mom}. GMRES is selected for its robust and stable convergence behavior for nonsymmetric linear systems. The GMRES iterations are terminated when the relative residual falls below the prescribed convergence threshold:
\begin{equation}
\dfrac{\left\|\bar{V}-\bar{\bar{Z}}\bar{I}^{(n)}\right\|}{\left\|\bar{V}\right\|} < \epsilon
\label{eq:conv}
\end{equation}
where $\bar{I}^{(n)}$ denotes the $n$-th iterate. The number of iterations required to reach convergence is denoted by $N_{\mathrm{it}}$. In some examples, a near-field preconditioner~\cite{Sertel2000,Saad1996} is used to improve the convergence of the iterative solution. The near-field matrix is formed from the entries of $\bar{\bar{Z}}$ associated with pairs of basis functions whose centers lie within a distance $R_{\mathrm{pre}}$ of each other. Its LU factorization, denoted by $\bar{\bar{P}}$, is used as the preconditioner: the system $\bar{\bar{P}}^{-1}\bar{\bar{Z}}\bar{I} = \bar{\bar{P}}^{-1}\bar{V}$ is solved, and $\bar{V}$ and $\bar{\bar{Z}}$ in~\eqref{eq:conv} are replaced by $\bar{\bar{P}}^{-1}\bar{V}$ and $\bar{\bar{P}}^{-1}\bar{\bar{Z}}$, respectively. In all examples, $E_u(\mathbf{r})$, $u\in \{x,y,z\}$ represents the component of the electric field in the $\hat{u}$ direction. All simulations are carried out on a workstation with two Intel Xeon Gold $6330$ processors ($56$ cores, $2.00\,\mathrm{GHz}$) and $1\,\mathrm{TB}$ of RAM.

\subsection{Monoisotropic Thin Shell}\label{Sec:Thin_Shell}
The accuracy and robustness of the TS-VIE solver (without GSTCs) are validated using a spherical monoisotropic shell embedded in free space and centered at the origin. The inner and outer radii of the shell are $a - \tau/2$ and $a + \tau/2$, respectively, where $a$ is the radius of the mid-surface $S$. The shell is characterized by a relative permittivity $\varepsilon_r$ and a relative permeability $\mu_r$, corresponding to
\begin{equation}
\bar{\bar{\alpha}}_1 = \frac{1}{\varepsilon_0\varepsilon_r} \bar{\bar{I}},\quad
\bar{\bar{\alpha}}_2 = \bar{\bar{0}},\quad
\bar{\bar{\alpha}}_3 = \bar{\bar{0}},\quad
\bar{\bar{\alpha}}_4 = \frac{1}{\mu_0\mu_r} \bar{\bar{I}}
\end{equation}
where $\bar{\bar{I}}$ is the $3 \times 3$ identity matrix.
The shell is illuminated by an $x$-polarized plane wave propagating in the $z$-direction with electric and magnetic fields
\begin{equation}
\mathbf{E}^{\mathrm{inc}}(\mathbf{r}) = \hat{\mathbf{x}} E_0 e^{-jk_0 z},
\quad \mathbf{H}^{\mathrm{inc}}(\mathbf{r}) = \hat{\mathbf{y}} E_0\sqrt{\varepsilon_0/\mu_0} e^{-jk_0 z}
\end{equation}
where $E_0 = 1\,\mathrm{V/m}$.

The accuracy is quantified using the relative $\ell_2$-norm error
\begin{equation}
\mathrm{err}_{\ell_2} =\sqrt{\frac{\sum_{p=1}^{N_p}\left|E^{\mathrm{far}}_{\mathrm{sim}}(\theta_p,\phi)
-E^{\mathrm{far}}_{\mathrm{ref}}(\theta_p, \phi)\right|^2}{\sum_{p=1}^{N_p}
\left|E^{\mathrm{far}}_{\mathrm{ref}}(\theta_p, \phi)\right|^2}}
\end{equation}
where $E^{\mathrm{far}}_{\mathrm{sim}}(\theta, \phi)$ and $E^{\mathrm{far}}_{\mathrm{ref}}(\theta, \phi)$ denote far-field electric fields computed using the TS-VIE solver and the analytical Mie-series solution, respectively. The error is evaluated at $N_p = 181$ uniformly spaced angles $\theta_p = (p-1)\Delta\theta$, $p = 1,\ldots,N_p$, with $\Delta\theta = 1^\circ$, at a fixed azimuthal angle $\phi = 45^\circ$.

First, the effect of the sheet thickness is examined for $a = 1\,\mathrm{m}$ and $\varepsilon_r = \mu_r = 2$. A total of $30$ simulations are performed for $\tau \in \{\lambda_0/n \mid n \in \{10,20,30,\ldots,100\}\}$ 
using three discretizations with average edge lengths $l_{\mathrm{avg}} = \lambda_0/10$, $l_{\mathrm{avg}}=\lambda_0/15$, and $l_{\mathrm{avg}}=\lambda_0/20$, at $f = 200\,\mathrm{MHz}$. The coarsest discretization ($l_{\mathrm{avg}} = \lambda_0/10$) results in $N_{\perp}=1\,372$ triangles and $N_{\parallel}=2\,058$ edges, the intermediate discretization ($l_{\mathrm{avg}} = \lambda_0/15$) in $N_{\perp}=3\,152$ triangles and $N_{\parallel}=4\,728$ edges, and the finest discretization ($l_{\mathrm{avg}} = \lambda_0/20$) in $N_{\perp}=5\,434$ triangles and $N_{\parallel}=8\,151$ edges. In all cases, \eqref{final_mom} is solved without a preconditioner, and $\epsilon = 10^{-3}$.

\begin{figure}[h]
\centering
\subfigure[]{\includegraphics[width=0.6\linewidth]{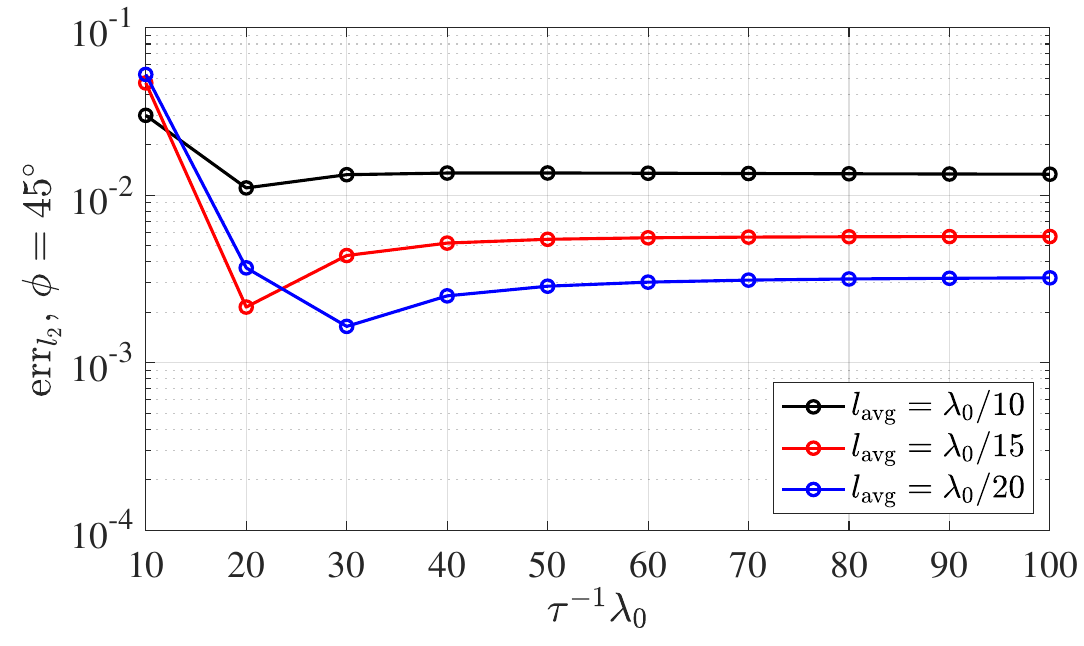}}\\
\subfigure[]{\includegraphics[width=0.6\linewidth]{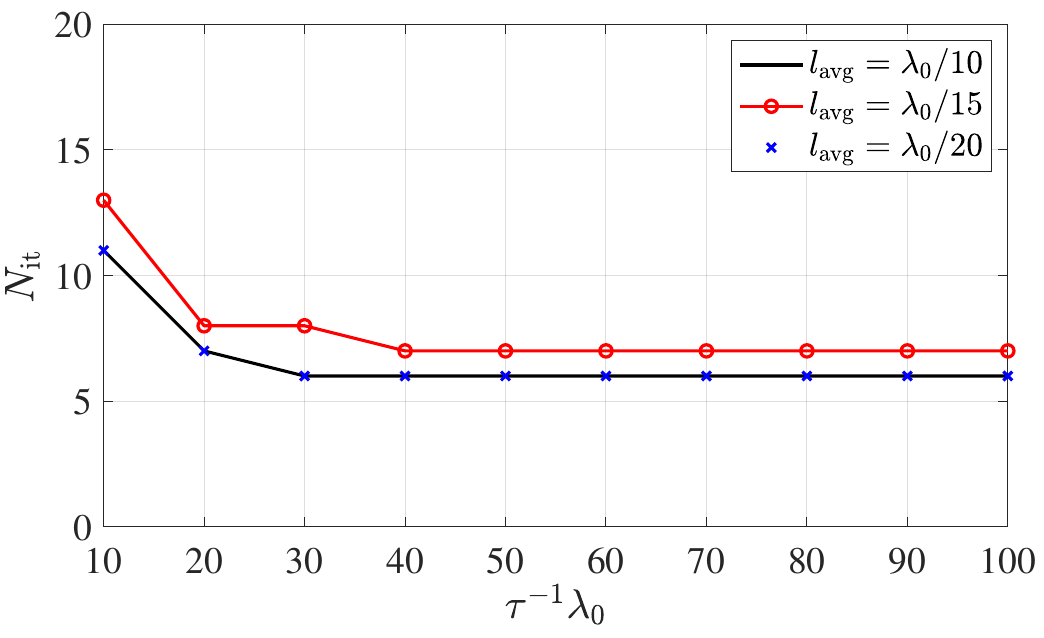}}
\caption{Accuracy and convergence of the TS-VIE solver as a function of $\tau^{-1}\lambda_0$ for the spherical monoisotropic shell. (a) $\mathrm{err}_{\ell_2}$ for three discretizations with $l_{\mathrm{avg}} = \lambda_0/10$, $\lambda_0/15$, and $\lambda_0/20$. (b) The corresponding $N_{\mathrm{it}}$ values.}
\label{Fig:Tau_Anl}
\end{figure}

Fig.~\ref{Fig:Tau_Anl}(a) reports $\mathrm{err}_{\ell_2}$ versus $\tau^{-1}\lambda_0$ for the three discretizations. As expected, the finer discretizations achieve lower errors. For all discretizations, the largest error occurs at $\tau=\lambda_0/10$, indicating that the thin-sheet condition $\tau \ll \lambda_0$ is no longer well satisfied at this thickness. As $\tau$ is reduced, the error decreases and reaches its minimum around $\tau=\lambda_0/20$ for $l_{\mathrm{avg}} = \lambda_0/10$ and $\lambda_0/15$, and around $\tau=\lambda_0/30$ for $l_{\mathrm{avg}} = \lambda_0/20$. Beyond this point, the error increases slightly and then saturates, consistent with the discretization error becoming dominant once the thin-sheet approximation is sufficiently accurate. The saturation level decreases with the mesh density, confirming this interpretation.

Fig.~\ref{Fig:Tau_Anl}(b) reports the corresponding $N_{\mathrm{it}}$ values. For all discretizations, the iteration count drops rapidly when decreasing $\tau$ from $\lambda_0/10$ and then remains constant for $\tau \leq \lambda_0 / 30$ for $l_{\mathrm{avg}} = \lambda_0/10$ and $\lambda_0/20$, and for $\tau \leq \lambda_0 / 40$ for $l_{\mathrm{avg}} = \lambda_0/15$. The former two stabilize at the same iteration count and the latter at one iteration more, indicating that the convergence is insensitive to $\tau$ in this regime and varies little with the mesh density over the range considered.

\begin{figure}[h]
\centering
\includegraphics[width=0.6\linewidth]{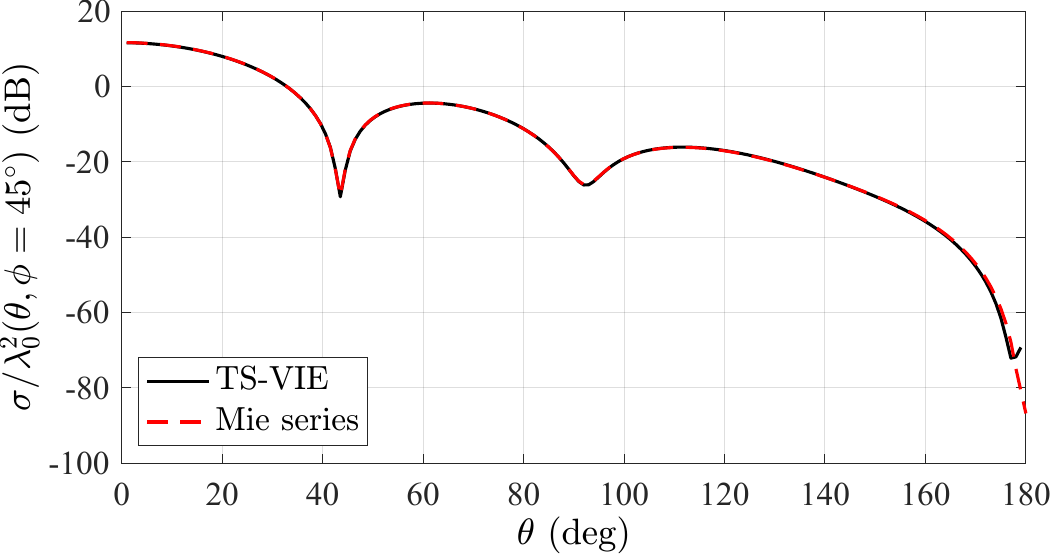}
\caption{Normalized monostatic RCS $\sigma/\lambda_0^2$ computed using the TS-VIE solver ($l_{\mathrm{avg}} = \lambda_0/10$, $\tau = \lambda_0/30$) compared against the analytical Mie-series solution for the monoisotropic spherical shell over $\theta \in [0^\circ, 180^\circ]$ at $\phi = 45^\circ$.}
\label{Fig:Mie_Tau}
\end{figure}

Fig.~\ref{Fig:Mie_Tau} compares the normalized monostatic RCS $\sigma/\lambda_0^2$ computed using the TS-VIE solver ($l_{\mathrm{avg}} = \lambda_0/10$, $\tau=\lambda_0/30$) against the analytical Mie-series solution over $\theta\in[0^\circ,180^\circ]$ at $\phi=45^\circ$. The numerical and analytical results are in excellent agreement across the entire angular range, including regions with pronounced dips and rapid angular variation, confirming the accuracy of the proposed solver.

Next, the effect of the convergence threshold is examined for $a = 1\,\mathrm{m}$ and $\varepsilon_r = \mu_r = 2$ at $f = 200\,\mathrm{MHz}$, with $l_{\mathrm{avg}} = \lambda_0/10$ and $\tau = \lambda_0/30$. A total of $9$ simulations are performed for $\epsilon \in \{10^{-n} \mid n \in \{2,3,\ldots,10\}\}$, , with \eqref{final_mom} solved without a preconditioner.

Fig.~\ref{Fig:Eps} reports $\mathrm{err}_{\ell_2}$ and $N_{\mathrm{it}}$ versus $\epsilon$. The error saturates at $\epsilon = 10^{-3}$ and does not change as $\epsilon$ is decreased further, indicating that the accuracy is limited by the discretization and the thin-sheet approximation rather than by the termination of the iterations.

Next, the effect of the material parameters is examined for $a = 1\,\mathrm{m}$ at $f = 200\,\mathrm{MHz}$, with $l_{\mathrm{avg}} = \lambda_0/10$ and $\tau = \lambda_0/30$. A total of $14$ simulations are performed, for $\varepsilon_r \in \{2, 2.5, 3,3.5,4,4.5, 5\}$ with $\mu_r = 2$, and for $\mu_r \in \{2, 2.5, 3,3.5,4,4.5, 5\}$ with $\varepsilon_r = 2$. In all cases, \eqref{final_mom} is solved without a preconditioner, and $\epsilon = 10^{-3}$.

Fig.~\ref{Fig:Material}(a)--(b) report $\mathrm{err}_{\ell_2}$ and $N_{\mathrm{it}}$ versus $\varepsilon_r$ and $\mu_r$, respectively. The two sweeps yield nearly identical results, as expected from the duality between the electric and magnetic problems for this configuration. In both cases, the error and the iteration count increase monotonically with the material parameter. The increase in the error follows from the shorter wavelength inside the sheet: increasing $\varepsilon_r$ or $\mu_r$ at fixed $\tau$ increases the electrical thickness of the sheet, so that the thin-sheet approximation degrades. The increase in the iteration count is attributed to the larger material contrast, which makes the matrix system more ill-conditioned.

Finally, the effect of the frequency is examined for $a = 0.5\,\mathrm{m}$ and $\varepsilon_r = \mu_r = 2$, with $\tau = \lambda_0/30$ at $f = 200\,\mathrm{MHz}$. A total of $6$ simulations are performed for $f \in \{200, 300, 400,500,600, 700\}\,\mathrm{MHz}$ using a single discretization with $l_{\mathrm{avg}} = \lambda_0/31$ at $f = 200\,\mathrm{MHz}$, which results in $N_{\perp} = 3\,152$ triangles and $N_{\parallel} = 4\,728$ edges. Since the mesh and $\tau$ are fixed while $f$ is varied, $l_{\mathrm{avg}}$ and $\tau$ increase relative to $\lambda_0$ with $f$, reaching $\lambda_0/8.9$ and $\lambda_0/8.6$, respectively, at $f = 700\,\mathrm{MHz}$. In all cases, \eqref{final_mom} is solved without a preconditioner, and $\epsilon = 10^{-3}$.

\begin{figure}[h]
\centering
\includegraphics[width=0.6\linewidth]{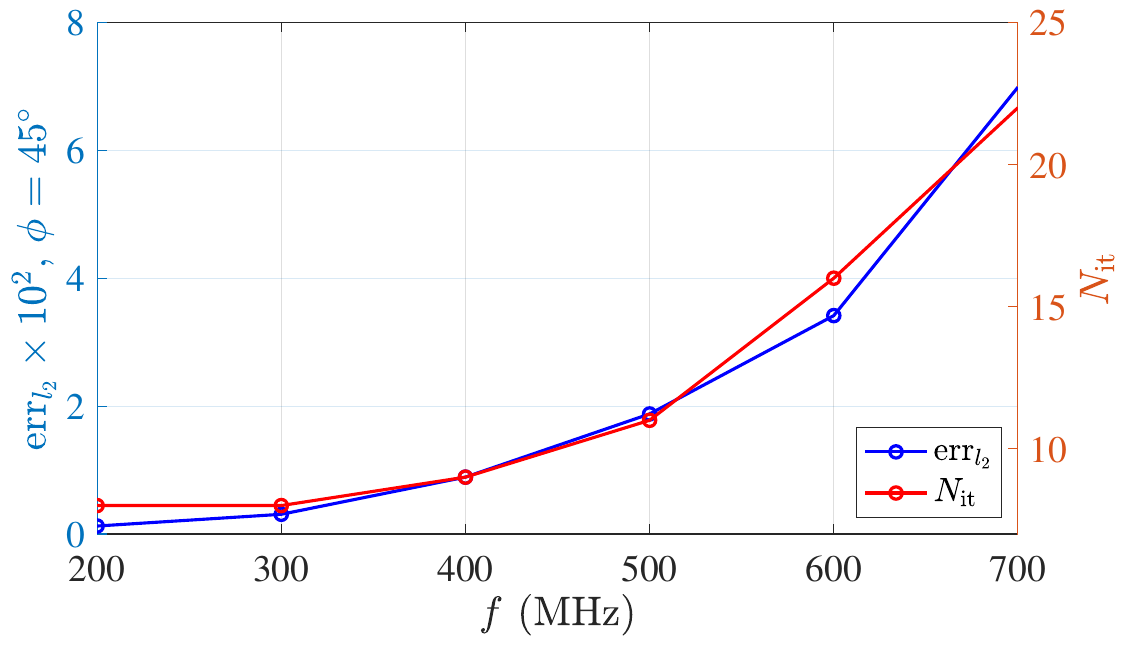}
\caption{Accuracy and convergence of the TS-VIE solver as a function of the frequency $f$ for the spherical monoisotropic shell with $\tau = \lambda_0/30$ and $l_{\mathrm{avg}} = \lambda_0/30$ at $f = 200\,\mathrm{MHz}$. $\mathrm{err}_{\ell_2}$ and $N_{\mathrm{it}}$ are shown on the left and right axes, respectively.}
\label{Fig:Freq}
\end{figure}

Fig.~\ref{Fig:Freq} reports $\mathrm{err}_{\ell_2}$ and $N_{\mathrm{it}}$ versus $f$. The error increases monotonically with $f$, as expected, since both the thin-sheet approximation and the discretization become less accurate as the shell becomes electrically thicker and the mesh coarser relative to the wavelength. The iteration count increases slightly, as the shell becomes electrically larger.

\subsection{Bianisotropic Metasurfaces}\label{sec:Bianisotropic}
Several metasurfaces designed to realize prescribed wave transformations are analyzed using the TS-VIE-GSTC solver. In all examples, the metasurface is embedded in free space and modeled as an equivalent thin cylindrical sheet of radius $2\,\mathrm{cm}$ centered at $(0,0,0.2\,\mathrm{cm})$.

The susceptibility tensors $\bar{\bar{\chi}}_\mathrm{ab}(\mathbf{r})$, $\mathrm{a,b} \in \{\mathrm{e,m}\}$, are obtained by solving the GSTCs~\eqref{eqGSTC} for the prescribed electromagnetic field transformation on both sides of the metasurface, using the synthesis procedure described in~\cite{MoM_Sandeep, Synthesis_Spherical, Susceptibility_Tensors, degruyter, Journal1_Celis_2025}. 

The metasurface is illuminated by an $x$-polarized Gaussian beam propagating in the direction
\begin{equation}
\hat{\mathbf{k}}^{\mathrm{inc}} =
\hat{\mathbf{z}}\cos\theta^{\mathrm{inc}}+\hat{\mathbf{y}}\sin\theta^{\mathrm{inc}}
\end{equation}
where $\theta^{\mathrm{inc}}$ denotes the angle of incidence with respect to the $z$-axis. The incident electric field is
\begin{equation}
\mathbf{E}^{\mathrm{inc}}(\mathbf{r})
=\hat{\mathbf{x}} E_0\left[{w_b}/{w(d)}\right]e^{-\frac{\rho^2}{w(d)^2}-j\Phi(d,\rho)} 
\end{equation}
where $E_0 = 1\,\mathrm{V/m}$, $d = \mathbf{r} \cdot \hat{\mathbf{k}}^{\mathrm{inc}}$ is the axial coordinate, $w_b$ is the beam waist radius at $d=0$, and
$\rho = \|\mathbf{r} - d \hat{\mathbf{k}}^{\mathrm{inc}}\|$ is the transverse radial distance. The beam radius is $w(d) = w_b \sqrt{1 + (d/{d_R})^2}$ where $d_R = \pi w_b^2/\lambda_0$ is the Rayleigh range. The phase term is
$\Phi(d,\rho)=k_0 d+[k_0\rho^2]/[2R(d)]-\Psi(d)$ where $R(d) = d [1 + (d_R/{d})^2]$ and $\Psi(d) = \tan^{-1}(d/d_R)$. The excitation frequency is $f = 60\,\mathrm{GHz}$. 

The surface $S$ is discretized with average edge length $l_{\mathrm{avg}} = \lambda_0/10$, resulting in $N_\perp = 11\,784$ triangles and $N_\parallel = 17\,802$ edges.
The accuracy is quantified using the relative $\ell_2$-norm error
\begin{equation}
\mathrm{err}_{\ell_2}=\sqrt{\frac{\sum_{p=1}^{N_p}\left|\mathbf{E}_{\mathrm{sim}}(\mathbf{r}_p)-\mathbf{E}_{\mathrm{ref}}(\mathbf{r}_p)
\right|^2}{\sum_{p=1}^{N_p}\left|\mathbf{E}_{\mathrm{ref}}(\mathbf{r}_p)\right|^2}}
\end{equation}
where $\mathbf{E}_{\mathrm{sim}}(\mathbf{r})$ and $\mathbf{E}_{\mathrm{ref}}(\mathbf{r})$ denote the electric fields computed using the TS-VIE-GSTC solver and by applying the prescribed wave transformation of the GSTCs to $\mathbf{E}^{\mathrm{inc}}(\mathbf{r})$, respectively.  

\subsubsection{Polarization Rotation}\label{sec:Pol_Rot}

Polarization rotation can be realized either through strictly monoanisotropic transformations, for which 
$\bar{\bar{\chi}}_{\mathrm{em}}(\mathbf{r}) = \bar{\bar{\chi}}_{\mathrm{me}}(\mathbf{r}) = \bar{\bar{0}}$, 
or through bianisotropic transformations satisfying 
$\bar{\bar{\chi}}_{\mathrm{ee}}(\mathbf{r}) = \bar{\bar{\chi}}_{\mathrm{mm}}(\mathbf{r}) = \bar{\bar{0}}$, 
as demonstrated in~\cite{degruyter,Journal1_Celis_2025}.

\begin{figure}[h]
\centering
\subfigure[]{\includegraphics[width=0.45\linewidth]{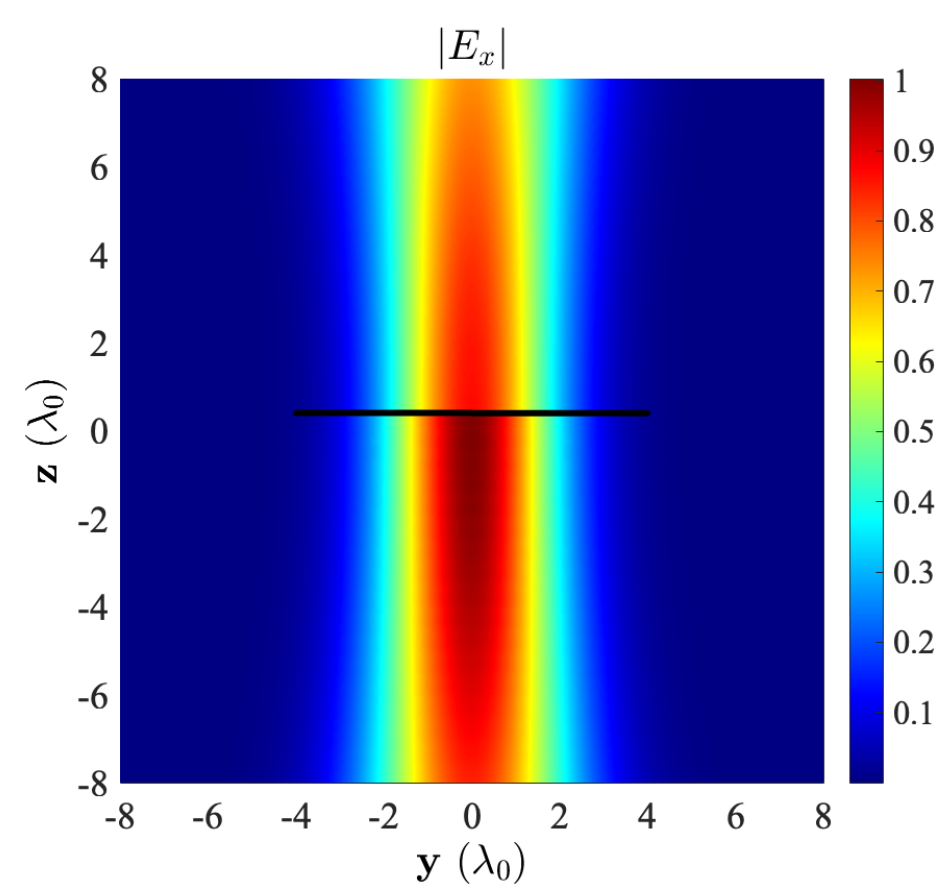}}\hfill
\subfigure[]{\includegraphics[width=0.45\linewidth]{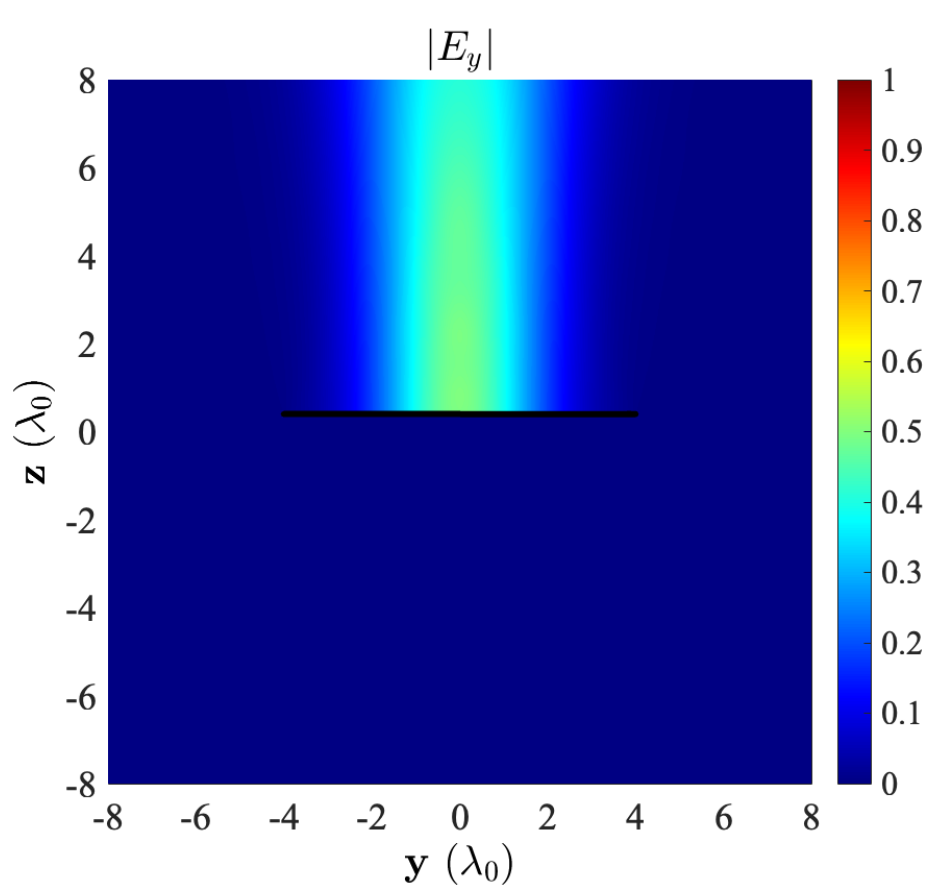}}\\
\subfigure[]{\includegraphics[width=0.6\linewidth]{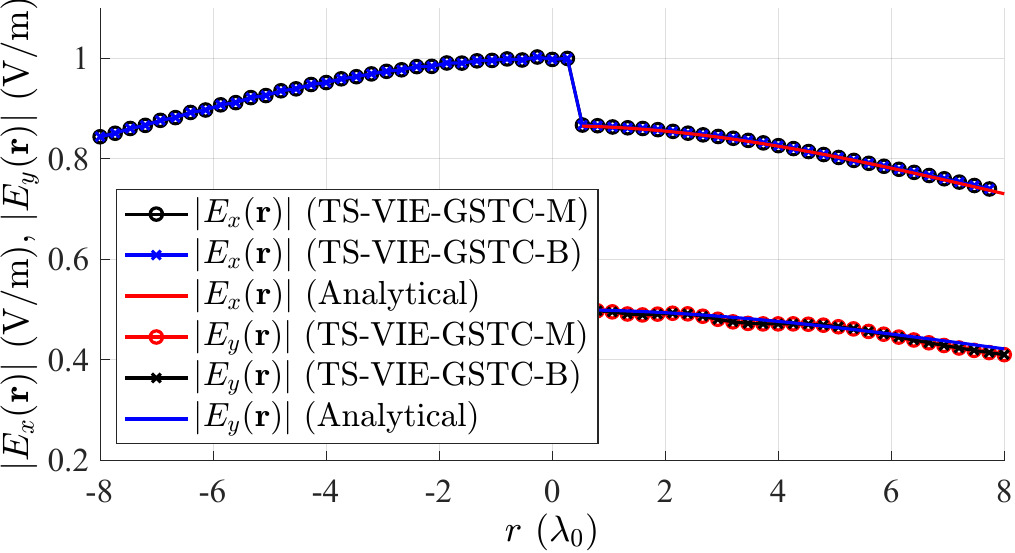}}
\caption{Polarization rotator.  
(a) $|E_x(\mathbf{r})|$ and (b) $|E_y(\mathbf{r})|$ on the $yz$-plane computed using the TS-VIE-GSTC solver for the monoanisotropic realization for $\tau=\lambda_0/30$. (c) $|E_x(\mathbf{r})|$ and $|E_y(\mathbf{r})|$ along $\mathbf{r} = \hat{\mathbf{z}}z$, $z \in [-8\lambda_0, 8\lambda_0]$, 
computed using the TS-VIE-GSTC solver for both the monoanisotropic (``M") and bianisotropic (``B") realizations, compared against the analytical 
solution.}
\label{Fig:Pol_Rot}
\end{figure}

The metasurface is designed to rotate the electric field of a normally incident plane wave by an angle $\theta_{\mathrm{p}}$ upon transmission. The prescribed field transformation specifies fields on the incident side, $\mathbf{E}_2(\mathbf{r})$, and the transmitted side, $\mathbf{E}_1(\mathbf{r})$, as
\begin{equation}
\begin{aligned}
\mathbf{E}_2(\mathbf{r}) &= \hat{\mathbf{x}} e^{-jk_0 z}\\
\mathbf{E}_1(\mathbf{r}) &= \hat{\mathbf{x}}\cos\theta_{\mathrm{p}}\, e^{-jk_0 z} 
+ \hat{\mathbf{y}} \sin\theta_{\mathrm{p}}\, e^{-jk_0 z}.
\end{aligned}
\end{equation}
For the monoanisotropic realization, strict Lorentz reciprocity is not enforced. The resulting susceptibility tensors exhibit gyrotropic behavior due to the asymmetric off-diagonal components in 
$\bar{\bar{\chi}}_{\mathrm{ee}}(\mathbf{r})$ and $\bar{\bar{\chi}}_{\mathrm{mm}}(\mathbf{r})$, which enable the required polarization rotation. The nonzero susceptibility tensors are
\begin{equation}
\begin{aligned}
     \chi_{\mathrm{ee}}^{xy}&=-{2j(1-\cos{\theta_{\mathrm{p}}})}/({k_0\sin{\theta_{\mathrm{p}}}})\\
     \chi_{\mathrm{ee}}^{yx}&={2j\sin{\theta_{\mathrm{p}}}}/[{k_0(\cos{\theta_{\mathrm{p}}}+1)}]\\
       \chi_{\mathrm{mm}}^{xy}&={-2j\sin{\theta_{\mathrm{p}}}}/[{ k_0(\cos{\theta_{\mathrm{p}}}+1)}]\\
     \chi_{\mathrm{mm}}^{yx}&=-{2j(\cos{\theta_{\mathrm{p}}}-1)}/({k_0\sin{\theta_{\mathrm{p}}}}).
\end{aligned}    
\end{equation}
For the bianisotropic realization, reciprocal gyrotropy is achieved via symmetric electric--magnetic coupling tensors satisfying $\bar{\bar{\chi}}_{\mathrm{me}}(\mathbf{r}) = -\bar{\bar{\chi}}_{\mathrm{em}}^T(\mathbf{r})$. The nonzero susceptibility tensors are
\begin{equation}
 \begin{aligned}
     &\chi_{\mathrm{em}}^{xx}=-{2j(\cos{\theta_{\mathrm{p}}}-1)}/({k_0\sin{\theta_{\mathrm{p}}}})\\
     &\chi_{\mathrm{em}}^{yy}={2j\sin{\theta_{\mathrm{p}}}}/[{k_0(\cos{\theta_{\mathrm{p}}}+1)}]\\
     &\chi_{\mathrm{me}}^{xx}=-{2j\sin{\theta_{\mathrm{p}}}}/[{k_0 (\cos{\theta_{\mathrm{p}}}+1)}]\\
     &\chi_{\mathrm{me}}^{yy}={2j(\cos{\theta_{\mathrm{p}}}-1)}/({k_0 \sin{\theta_{\mathrm{p}}}}).
\end{aligned}   
\end{equation}
The rotation angle and the Gaussian beam waist radius are set to $\theta_{\mathrm{p}} = 30^\circ$ and $w_b = 2\lambda_0$, respectively. In all simulations of this example, \eqref{final_mom} is solved without a preconditioner, and $\epsilon = 10^{-3}$.
\paragraph*{Case 1: Planar Metasurface}
Fig.~\ref{Fig:Pol_Rot}(a)--(b) illustrate $|E_x(\mathbf{r})|$ and $|E_y(\mathbf{r})|$ on the $yz$-plane computed using the TS-VIE-GSTC solver for the monoanisotropic polarization rotator with $\tau = \lambda_0/30$. The results demonstrate the gradual polarization conversion of the incident beam as it propagates along the $z$-axis.

To illustrate the additional design degrees of freedom offered by bianisotropic coupling, one representative bianisotropic case is considered at $\tau = \lambda_0/250$. Fig.~\ref{Fig:Pol_Rot}(c) compares $|E_x(\mathbf{r})|$ and $|E_y(\mathbf{r})|$ along the line $\mathbf{r} = \hat{\mathbf{z}}z$, with $z \in [-8\lambda_0, 8\lambda_0]$, computed by the TS-VIE-GSTC solver for both the monoanisotropic and bianisotropic realizations, against the analytical solution. Excellent agreement is observed between numerical and analytical results for both realizations, with only minor reflections visible.

To quantify the accuracy, $\mathrm{err}_{\ell_2}$ is computed along the line $\mathbf{r}_p =\hat{\mathbf{z}}[z_0 + (p-1)\Delta z]$ with $z_0 = 0.5308\lambda_0$, $\Delta z = 0.0053\lambda_0$, and $N_p =1401$. For the bianisotropic case, an error of $\mathrm{err}_{\ell_2} = 7.89 \times 10^{-3} $ is obtained with $N_{\mathrm{it}} = 67$. For the monoanisotropic case, the error and iteration count are evaluated for $\tau \in \{\lambda_0/30, \lambda_0/60,\lambda_0/90\}$, yielding $N_{\mathrm{it}} \in \{15, 19, 21\}$ and $\mathrm{err}_{\ell_2} \in \{7.65, 7.67, 7.64\} \times 10^{-3}$, respectively. The iteration counts grow slowly as $\tau$ decreases while the error remains stable, demonstrating the robustness of the proposed TS-VIE-GSTC solver across $\tau$.
\paragraph*{Case 2: Curved Metasurface}

The monoanisotropic polarization rotator is realized on a spherical cap of radius $R_{\mathrm{c}} = 5\lambda_0$, obtained by displacing each node of the planar disk of radius $R_{\mathrm{d}} = 4\lambda_0$ along $\hat{\mathbf{z}}$ by $ h(r) = \sqrt{R_{\mathrm{c}}^{2}-r^{2}} - \sqrt{R_{\mathrm{c}}^{2}-R_{\mathrm{d}}^{2}}$, where $r = \sqrt{x^2+y^2}$, so that the rim stays in place and the apex rises by $2\lambda_0$. The mesh therefore has the same $N_\perp$ and $N_\parallel$ as the planar disk, while its edges are stretched toward the rim, resulting in $l_{\mathrm{avg}} \approx \lambda_0/9$. The susceptibility tensors are synthesized for the curved geometry following~\cite{CelisSierra2026_ST}, so that they vary over $S$ with its local orientation. The sheet thickness is set to $\tau = \lambda_0/30$.

\begin{figure}
\centering
\subfigure[]{\includegraphics[width=0.45\linewidth]{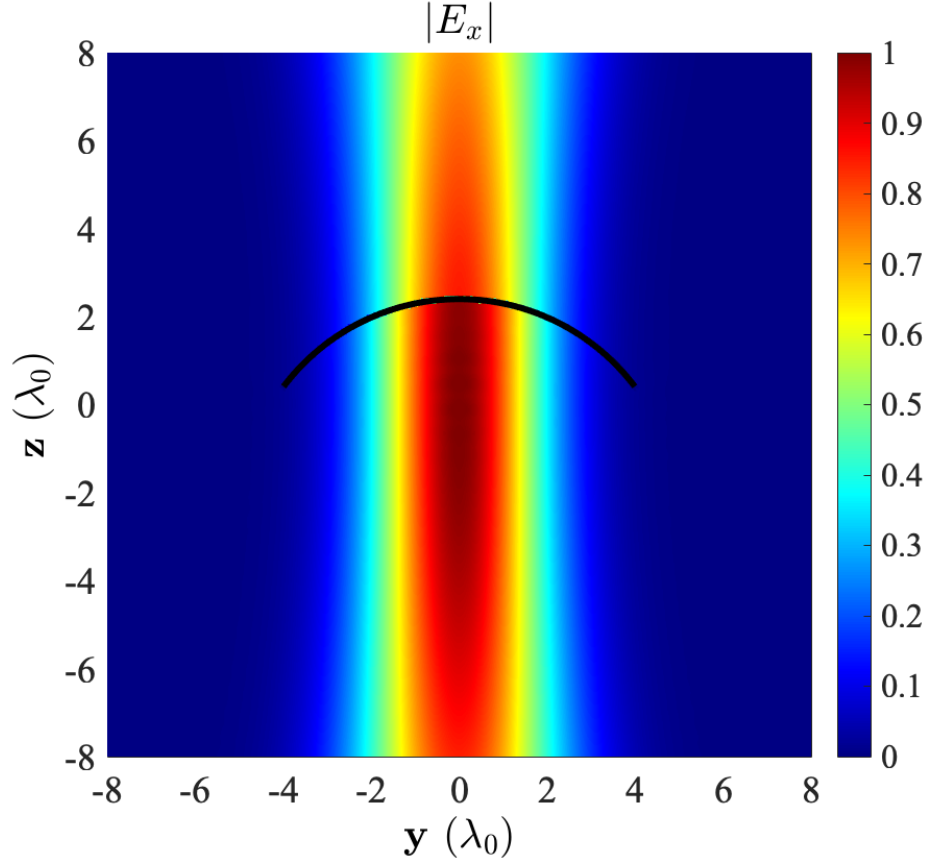}}\hfill
\subfigure[]{\includegraphics[width=0.45\linewidth]{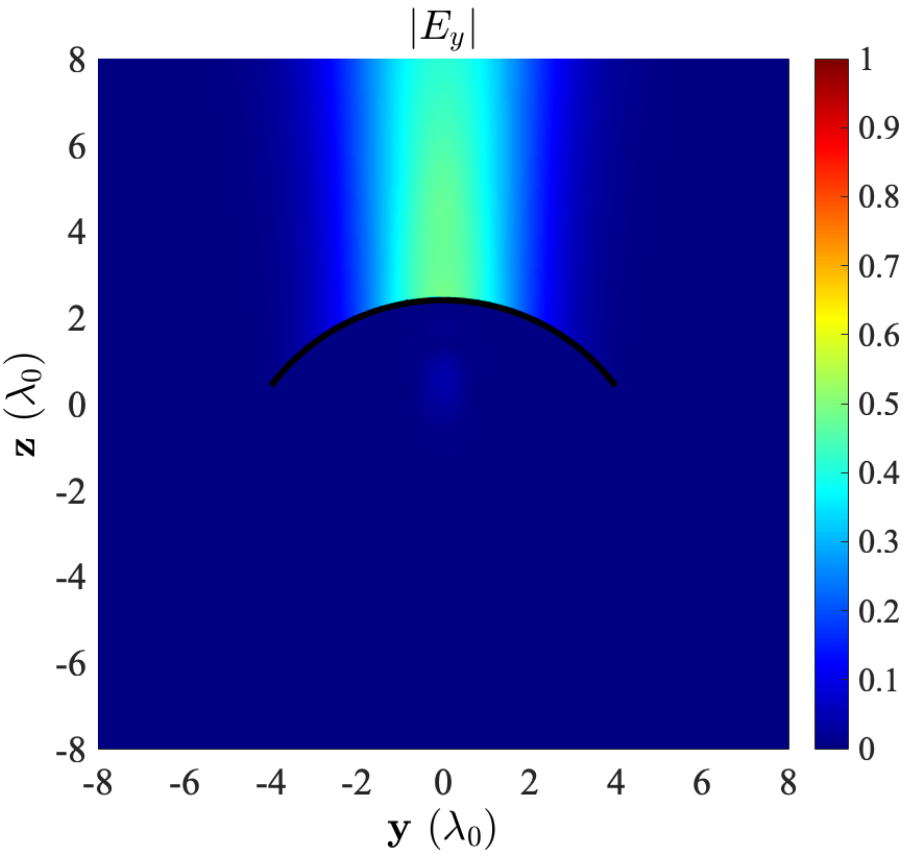}}\\
\subfigure[]{\includegraphics[width=0.6\linewidth]{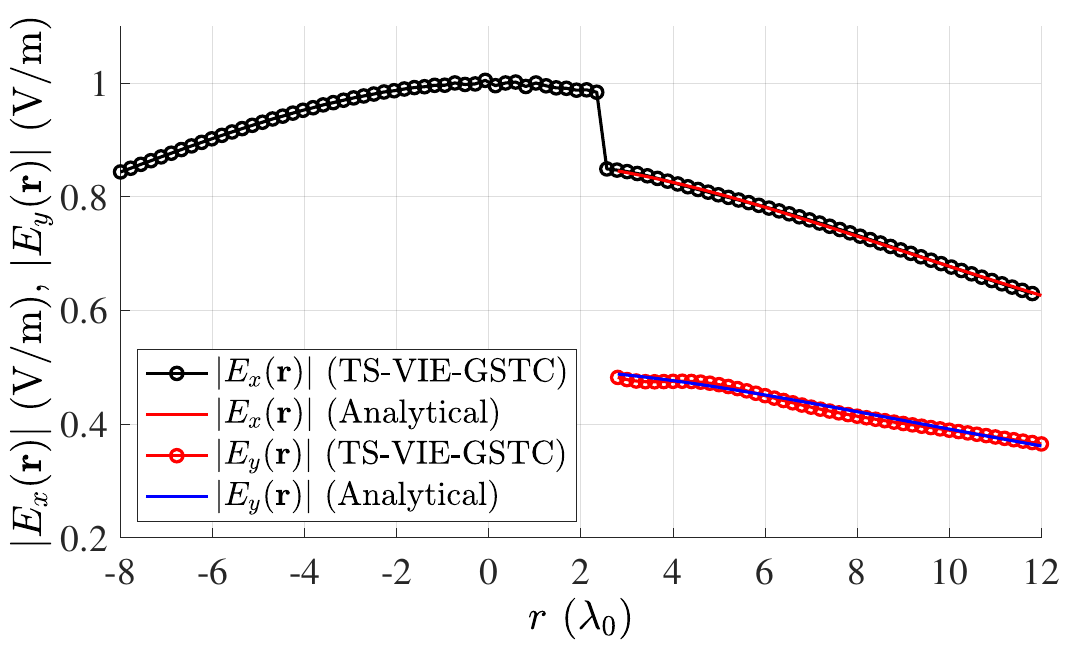}}
\caption{Curved polarization rotator.  
(a) $|E_x(\mathbf{r})|$ and (b) $|E_y(\mathbf{r})|$ on the $yz$-plane computed using the TS-VIE-GSTC solver for the monoanisotropic realization for $\tau=\lambda_0/30$. (c) $|E_x(\mathbf{r})|$ and $|E_y(\mathbf{r})|$ along $\mathbf{r} = \hat{\mathbf{z}}z$, $z \in [-8\lambda_0, 12\lambda_0]$, computed using the TS-VIE-GSTC solver compared against the analytical solution of the planar rotator.}
\label{Fig:CPol_Rot}
\end{figure}

Fig.~\ref{Fig:CPol_Rot}(a)--(b) present $|E_x(\mathbf{r})|$ and $|E_y(\mathbf{r})|$ on the $yz$-plane computed using the TS-VIE-GSTC solver. The polarization conversion is preserved on the curved sheet, for which the $\mathcal{K}[\bar{\bar{\beta}}_{i}\cdot \mathbf{X}](\mathbf{r})$ contributions are nonzero. Fig.~\ref{Fig:CPol_Rot}(c) compares $|E_x(\mathbf{r})|$ and $|E_y(\mathbf{r})|$ along the line $\mathbf{r} = \hat{\mathbf{z}}z$, $z \in [-8\lambda_0, 12\lambda_0]$, computed by the TS-VIE-GSTC solver against the analytical solution of the planar rotator, which is used as the reference since no analytical solution is available for the curved one.

To quantify the accuracy, $\mathrm{err}_{\ell_2}$ is computed along the line $\mathbf{r}_p = \hat{\mathbf{z}}[z_0 + (p-1)\Delta z]$ with $z_0 = 2.79\lambda_0$, $\Delta z = 0.0040\lambda_0$, and $N_p = 2301$, yielding $\mathrm{err}_{\ell_2} = 8.54\times 10^{-3}$ with $N_{\mathrm{it}} = 29$. Although the planar solution is not exact for the curved sheet, the error is comparable to that of the planar case. This is because the field along the axis is determined mainly by the region near the apex, where the cap is nearly perpendicular to the beam axis and the beam is concentrated. The iteration count increases moderately, from $N_{\mathrm{it}}=15$ to $N_{\mathrm{it}}=29$.

\subsubsection{Perfect Reflection}\label{sec:PEC}
Perfect reflection under normal incidence is realized by enforcing the following prescribed field transformation that specifies fields on the incident side, $\mathbf{E}_2(\mathbf{r})$, and the transmitted side, $\mathbf{E}_1(\mathbf{r})$, as
\begin{equation}
\begin{aligned}
\mathbf{E}_2(\mathbf{r}) & = \hat{\mathbf{x}} (e^{-jk_0 z} -e^{jk_0 z})\\
\mathbf{E}_1(\mathbf{r}) & = \mathbf{0}.
\end{aligned}
\end{equation}
This transformation corresponds to total reflection, where the transmitted field vanishes and the reflected field equals the phase-reversed incident field.

\begin{figure}[h]
\centering
\subfigure[]{\includegraphics[width=0.45\linewidth]{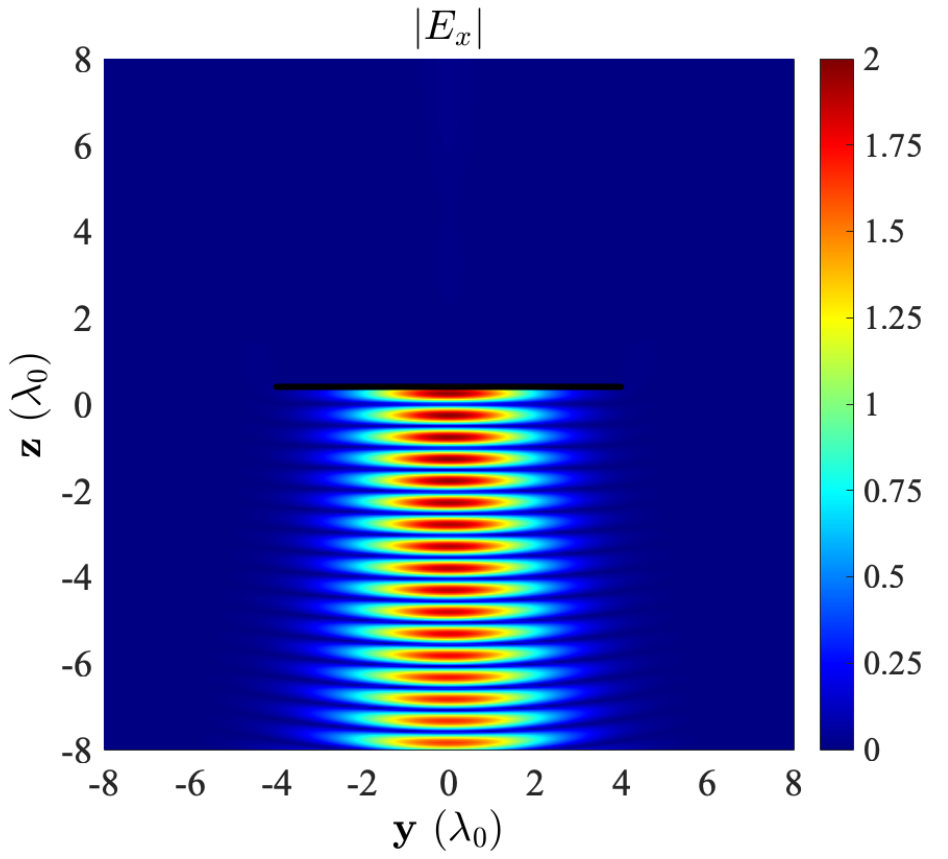}}\\
\subfigure[]{\includegraphics[width=0.6\linewidth]{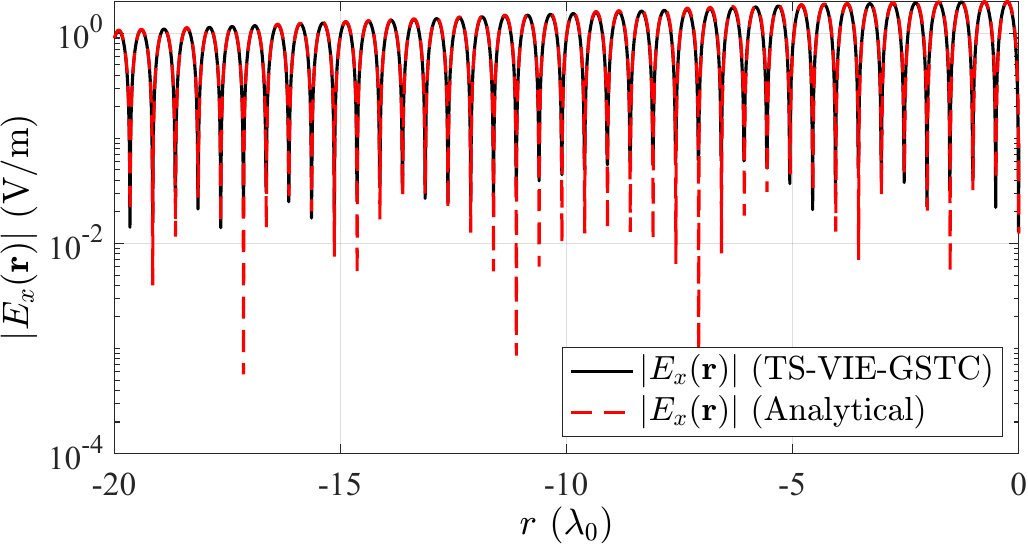}}
\caption{Perfect reflector. (a) $|E_x(\mathbf{r})|$ on the $yz$-plane computed using the TS-VIE-GSTC solver for $\tau =\lambda_0/30$. (b) $|E_x(\mathbf{r})|$ along $\mathbf{r} = \hat{\mathbf{z}}z$, $z \in [-20\lambda_0, 0]$, computed using the TS-VIE-GSTC solver, compared against the analytical solution.}
\label{Fig:PEC}
\end{figure}

For this transformation, the susceptibility tensors are synthesized assuming full bianisotropy while enforcing Lorentz reciprocity. The nonzero susceptibility components are
\begin{equation}
\label{eq:tensor_ref}
\begin{aligned}
    \chi_{\mathrm{ee}}^{xx}&={4 j(e^{2 j k_0 z_0}+1)}/[{k_0(e^{2 j k_0 z_0}-1)}]\\
     \chi_{\mathrm{em}}^{xy}&={2j}/{k_0}\\
      \chi_{\mathrm{me}}^{yx}&=-{2j}/{k_0}
\end{aligned}    
\end{equation}
where $z_0=0.2\,\mathrm{cm}$ is the z-coordinate of the metasurface center, coinciding with the mid-surface $S$. The Gaussian beam waist radius is set to $w_b = 2\lambda_0$. No preconditioner is used, so that the effect of the large susceptibility values on the convergence can be observed directly (Section~\ref{sec:comments}), and $\epsilon = 10^{-3}$. Fig.~\ref{Fig:PEC}(a) presents $|E_x(\mathbf{r})|$ on the $yz$-plane computed using the TS-VIE-GSTC solver for $\tau = \lambda_0/30$. The field distribution demonstrates complete reflection, with negligible transmission beyond the metasurface.

To quantify the reflection behavior, Fig.~\ref{Fig:PEC}(b) compares $|E_x(\mathbf{r})|$ along the line
$\mathbf{r} = \hat{\mathbf{z}}z$, $z \in [-20\lambda_0,0]$, computed by the TS-VIE-GSTC solver against the analytical solution. 
Good agreement is observed between numerical and analytical results across the observation interval. 

To quantify the accuracy, $\mathrm{err}_{\ell_2}$ is computed along the line $\mathbf{r}_p=\hat{\mathbf{z}}[z_0 + (p-1)\Delta z]$ with $z_0 = -24\lambda_0$, $\Delta z = 0.008\lambda_0$, and $N_p = 3000$. For $\tau \in \{\lambda_0/30, \lambda_0/60, \lambda_0/90\}$, the iteration counts are $N_{\mathrm{it}} \in \{234, 285, 328\}$ and the corresponding errors are $\mathrm{err}_{\ell_2} \in \{1.26, 1.30, 1.33\} \times 10^{-2}$. These results confirm that the proposed TS-VIE-GSTC solver accurately models perfect reflection. However, compared to the polarization rotation example, the iteration counts and errors are significantly higher, which is attributed to the large susceptibility values in~\eqref{eq:tensor_ref} that arise from the requirement to completely suppress transmission, leading to a more ill-conditioned system matrix.

\subsubsection{Beam Steering With Spatially Varying Susceptibilities}\label{sec:beamforming}
The metasurface is designed to steer a normally incident $x$-polarized plane wave to the angle $\theta_{\mathrm{d}}$ with respect to the $z$-axis upon transmission. The prescribed field transformation specifies fields on the incident side, $\mathbf{E}_2(\mathbf{r})$, and the transmitted side, $\mathbf{E}_1(\mathbf{r})$, as
\begin{equation}
\begin{aligned}
\mathbf{E}_{2}(\mathbf{r}) &= \hat{\mathbf{x}}e^{-j k_0 z} \\
\mathbf{E}_{1}(\mathbf{r}) &= \hat{\mathbf{x}}e^{-j k_0 (y \sin \theta_{\mathrm{d}} + z \cos \theta_{\mathrm{d}})}.
\end{aligned}
\end{equation}
Since the tangential wavenumbers of the incident and transmitted fields differ, the susceptibilities required to realize this transformation depend on position along $S$. The nonzero susceptibility components are
\begin{equation}
\label{eq:chi_bs}
\!\!\!\begin{aligned}
\chi^{xx}_{\mathrm{ee}}(\mathbf{r}) &= -\frac{j}{k_0}\left[(1 - \cos\theta_{\mathrm{d}})
  + j(1 + \cos\theta_{\mathrm{d}})\tan\Phi_{\mathrm{d}}(\mathbf{r})\right]\\
\chi^{yx}_{\mathrm{me}}(\mathbf{r}) &= \frac{2}{k_0}\tan\Phi_{\mathrm{d}}(\mathbf{r})
\end{aligned}
\end{equation}
where $\Phi_{\mathrm{d}}(\mathbf{r}) = (k_0/2)[z(\cos\theta_{\mathrm{d}} - 1) + y\sin\theta_{\mathrm{d}}]$. The steering angle and the Gaussian beam waist radius are set to $\theta_{\mathrm{d}} = 30^\circ$ and $w_b = 2\lambda_0$, respectively. Since $\tan\Phi_{\mathrm{d}}(\mathbf{r})$ is unbounded where $\Phi_{\mathrm{d}}(\mathbf{r}) = \pi/2 + n\pi$ for integer $n$, which occurs along four lines of constant $y$ on $S$, the susceptibilities reach large values near these lines. They remain finite in the discretized problem, since the tensors are evaluated at the element centroids (Section~\ref{sec:comments}). Since such large values are expected to increase the iteration count (Section~\ref{sec:PEC}), \eqref{final_mom} is left-preconditioned with the near-field preconditioner with $R_{\mathrm{pre}} = 0.3\lambda_0$, and $\epsilon = 10^{-5}$.

\begin{figure}[h]
\centering
\subfigure[]{\includegraphics[width=0.45\columnwidth]{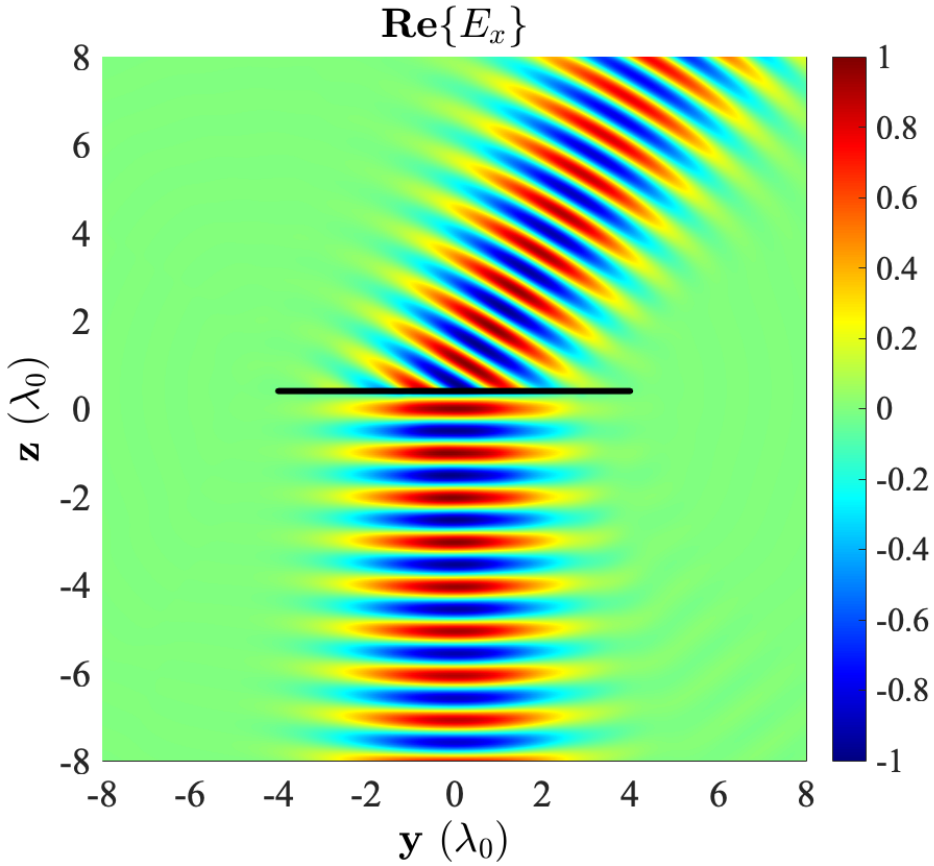}}\hfill
\subfigure[]{\includegraphics[width=0.45\columnwidth]{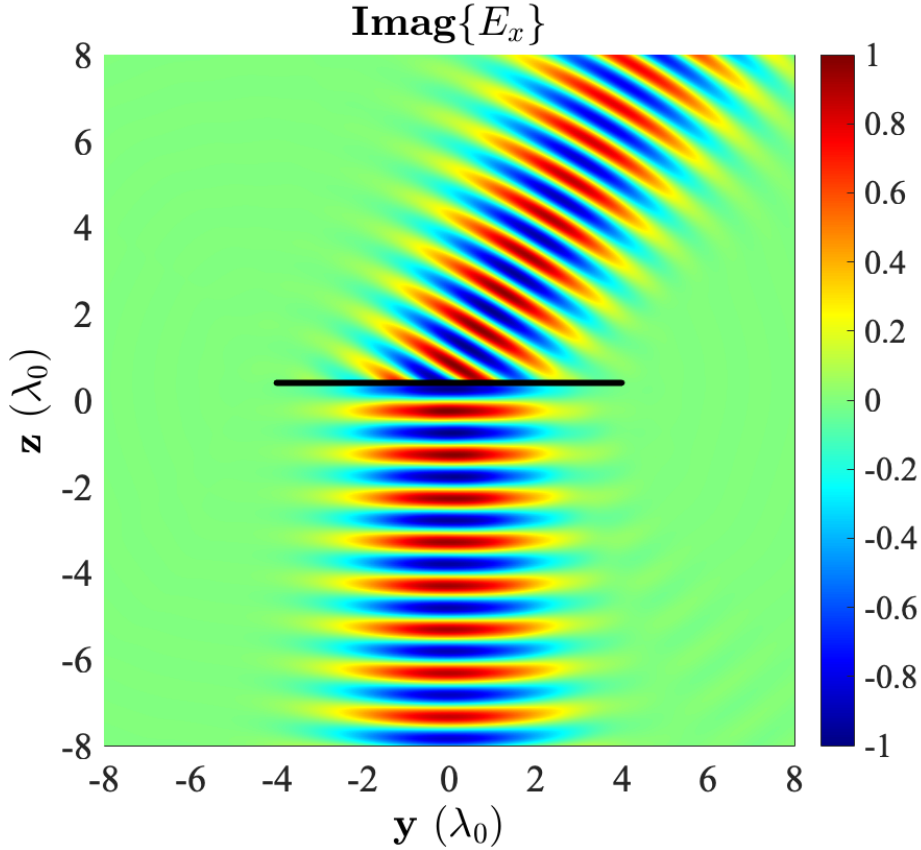}}
\caption{Beam-steering metasurface, $\theta_{\mathrm{d}} = 30^\circ$. (a) $\mathrm{Re}\{E_x(\mathbf{r})\}$ and (b) $\mathrm{Im}\{E_x(\mathbf{r})\}$ on the $yz$-plane computed using the TS-VIE-GSTC solver for $\tau=\lambda_0/30$.}
\label{Fig:BS}
\end{figure}

Fig.~\ref{Fig:BS}(a)--(b) present $\mathrm{Re}\{E_x(\mathbf{r})\}$ and $\mathrm{Im}\{E_x(\mathbf{r})\}$ on the $yz$-plane computed using the TS-VIE-GSTC solver for $\tau = \lambda_0/30$. The phase fronts of the transmitted field are tilted by $30^\circ$ with respect to those of the incident field, as prescribed, showing that the solver captures the spatially varying response of the sheet. Convergence is reached in $N_{\mathrm{it}} = 25$ iterations.

\subsubsection{Multi-Directional Attenuator}

A metasurface is designed to simultaneously attenuate incident $\hat{\mathbf{x}}$-polarized plane waves arriving from incidence angles $\theta^{\mathrm{inc}} \in \{0^\circ, 22.5^\circ, -22.5^\circ\}$. For each $\theta^{\mathrm{inc}}$, the prescribed field transformation specifies fields on the incident side, $\mathbf{E}_2(\mathbf{r})$, and the transmitted side, $\mathbf{E}_1(\mathbf{r})$, as 
\begin{equation}
\begin{aligned}
 \mathbf{E}_{2}(\mathbf{r}) &= \hat{\mathbf{x}} e^{-jk_{0}[y\sin \theta^{\mathrm{inc}}+z\cos\theta^{\mathrm{inc}}]} \\
\mathbf{E}_{1}(\mathbf{r}) &= \hat{\mathbf{x}} A e^{-jk_{0}[y\sin \theta^{\mathrm{inc}}+z\cos\theta^{\mathrm{inc}}]}.
\end{aligned}
\end{equation}
where $A$ is the attenuation factor.

The susceptibility tensors are synthesized assuming full bianisotropy while enforcing Lorentz reciprocity for the three values of $\theta^{\mathrm{inc}}$. The nonzero susceptibility components are
\begin{equation}
\begin{aligned}
     \chi^{xx}_{\mathrm{ee}}&={4 jA_{-}}/({k_0 A_{+}})\\
     \chi^{xy}_{\mathrm{em}}&=-{2jA_{-}  }/({k_0A_{+}})\\
     \chi^{yx}_{\mathrm{me}}&={2jA_{-}  }/({k_0 A_{+}})\\
     \chi^{zz}_{\mathrm{mm}}&=-{4 jA_{-}}/[{k_0A_{+}(\cos 22.5^{\circ}+1)}].
\end{aligned}    
\end{equation}
where $A_{\pm}=A\pm1$. The attenuation factor and the Gaussian beam waist radius are set to $A=0.5$ and $w_b = 2\lambda_0$, respectively. For all three cases, \eqref{final_mom} is left-preconditioned with the near-field preconditioner with $R_{\mathrm{pre}} = 0.3\lambda_0$, and $\epsilon = 10^{-5}$.

\begin{figure}[h]
\centering
\subfigure[]{\includegraphics[width=0.45\columnwidth]{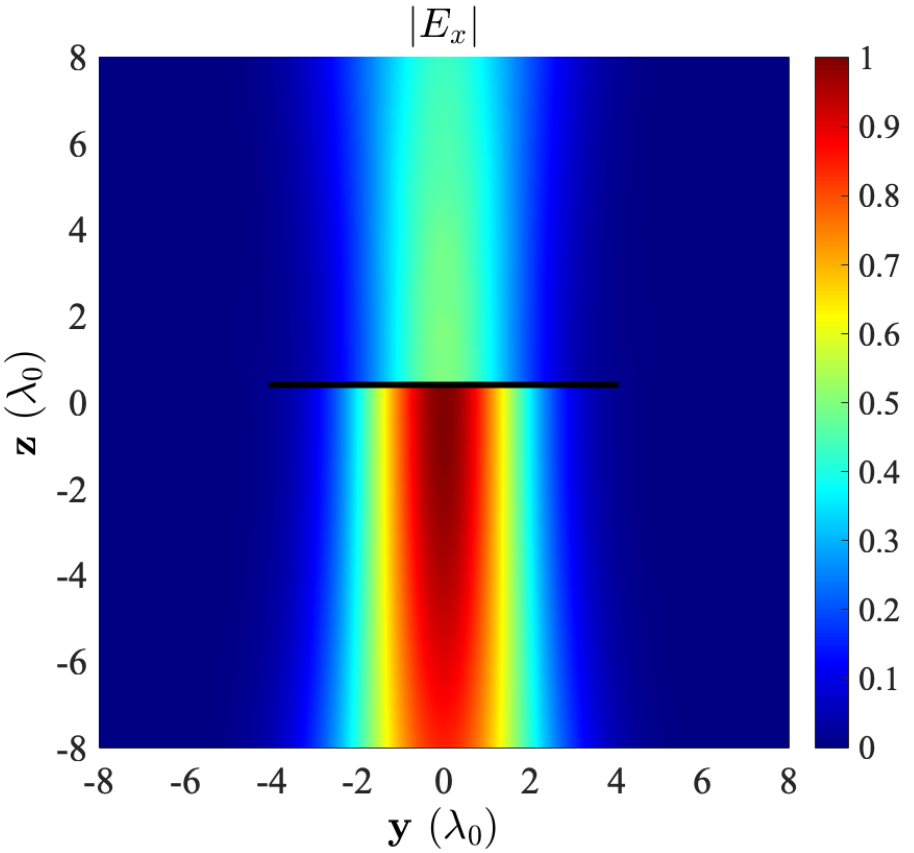}}\\
\subfigure[]{\includegraphics[width=0.6\columnwidth]{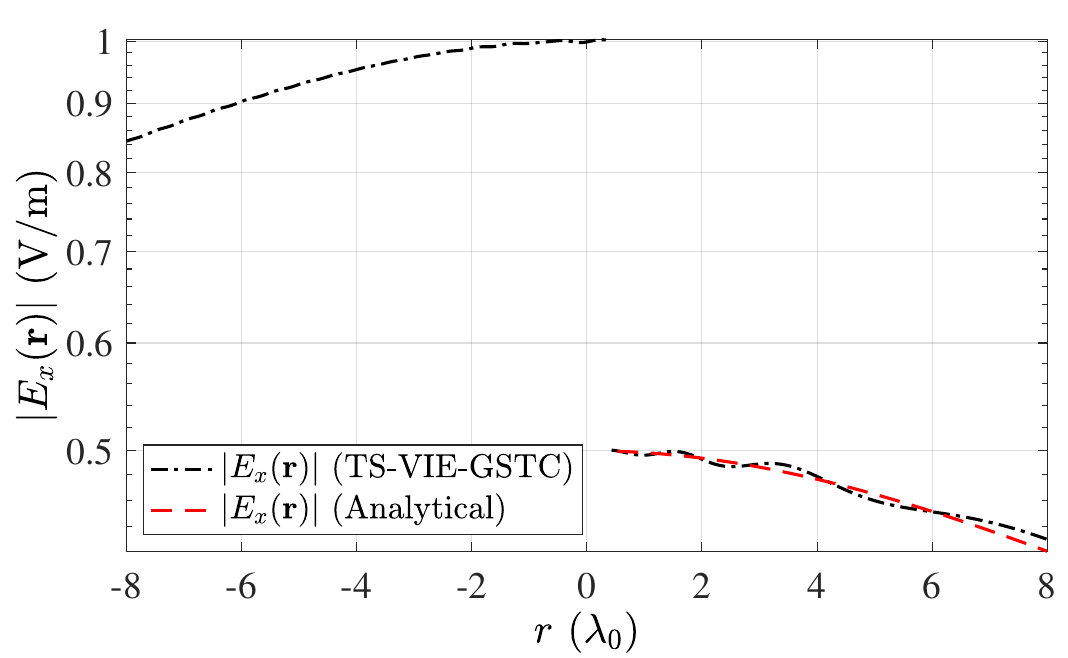}}
\caption{Multi-directional attenuator, $\theta^{\mathrm{inc}} = 0^\circ$. (a) $|E_{x}(\mathbf{r})|$ on the $yz$-plane computed using the TS-VIE-GSTC solver for $\tau=\lambda_0/30$. (b) $|E_x(\mathbf{r})|$ along $\mathbf{r} = \hat{\mathbf{z}}z$, $z \in [-8\lambda_0, 8\lambda_0]$, computed using the TS-VIE-GSTC solver, compared against the analytical solution.}
\label{Fig:Att_0}
\end{figure}

\paragraph*{Case 1: Normal Incidence ($\theta^{\mathrm{inc}}=0^\circ$)}
Fig.~\ref{Fig:Att_0}(a) presents $|E_x(\mathbf{r})|$ on the $yz$-plane computed using the TS-VIE-GSTC solver for $\tau = \lambda_0/30$. The field profile demonstrates the intended attenuation on the transmission side. Fig.~\ref{Fig:Att_0}(b) compares $|E_x(\mathbf{r})|$ along $\mathbf{r}=\hat{\mathbf{z}}z$, $z\in [-8\lambda_0, 8\lambda_0]$, computed using the TS-VIE-GSTC solver against the analytical solution. 

To quantify the accuracy, $\mathrm{err}_{\ell_2}$ is computed along the line $\mathbf{r}_p = \hat{\mathbf{z}}[z_0+(p-1) \Delta z]$ with $z_0=0.5308 \lambda_0$, $\Delta z = 0.0053 \lambda_0$, and $N_p = 1401$. For $\tau \in \{\lambda_0/30, \lambda_0/60, \lambda_0/90\}$, the iteration counts are $N_{\mathrm{it}} \in \{10, 10, 11\}$ and the corresponding errors are $\mathrm{err}_{\ell_2} \in \{1.29, 1.31, 1.33\} \times 10^{-2}$. The iteration count is practically independent of $\tau$, indicating that the convergence is insensitive to the sheet thickness, while the error remains nearly constant, demonstrating stable accuracy.

\paragraph*{Case 2: Oblique Incidence ($\theta^{\mathrm{inc}}=22.5^\circ$)}

\begin{figure}[h]
\centering
\subfigure[]{\includegraphics[width=0.45\columnwidth]{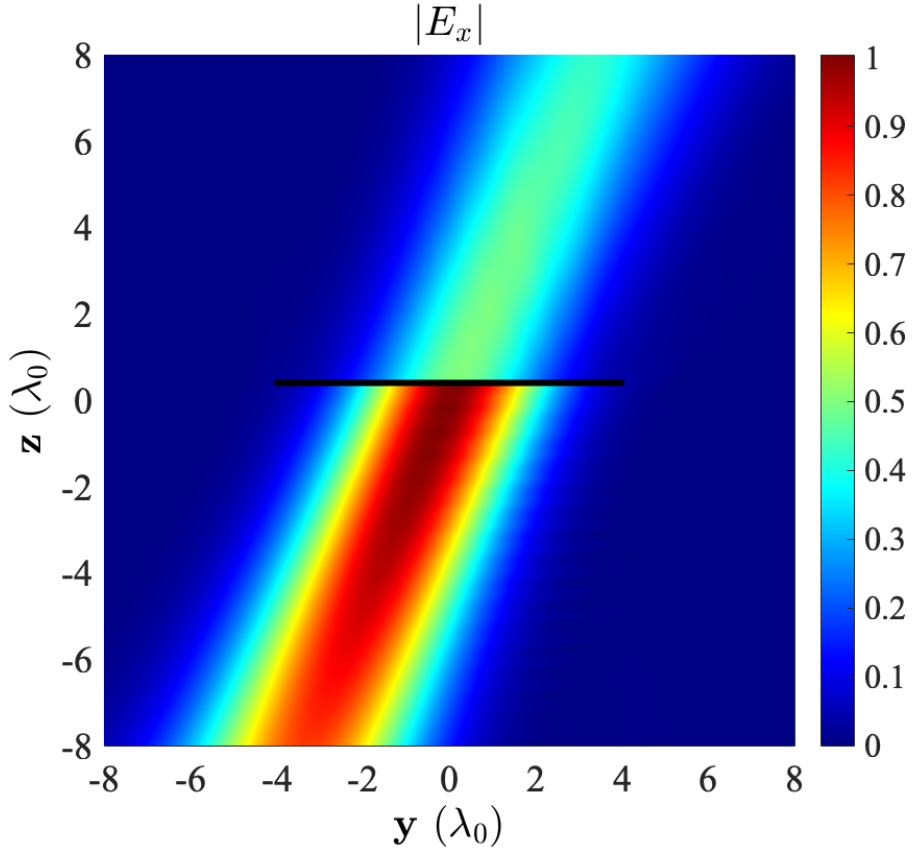}}\\
\subfigure[]{\includegraphics[width=0.6\columnwidth]{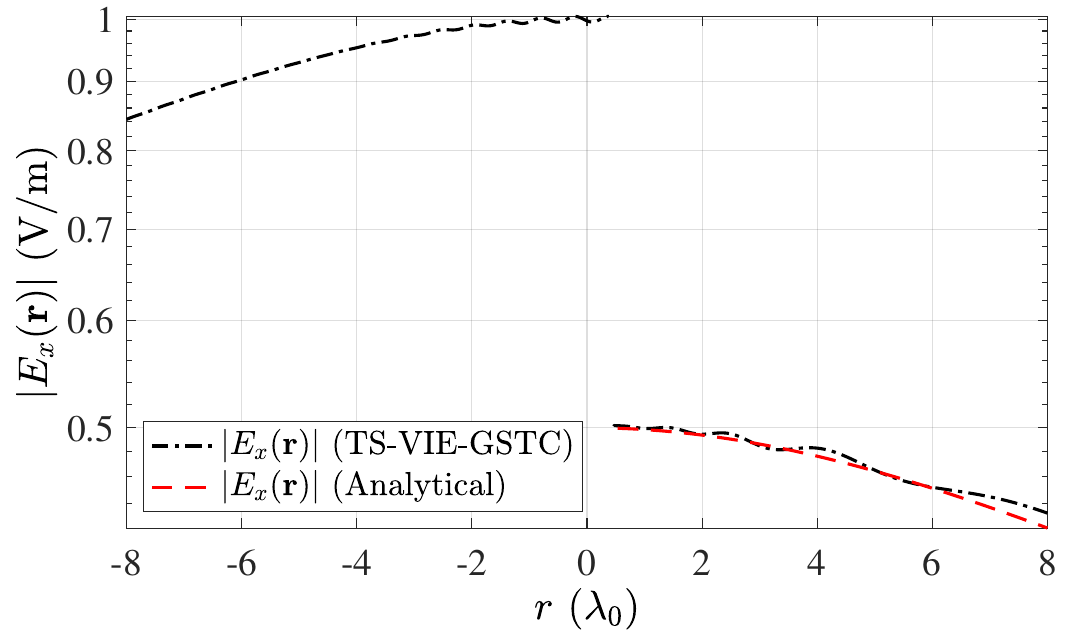}}
\caption{Multi-directional attenuator, $\theta^{\mathrm{inc}} = 22.5^\circ$. (a) $|E_x(\mathbf{r})|$ on the $yz$-plane computed using the TS-VIE-GSTC solver for $\tau = \lambda_0/30$. (b) $|E_x(\mathbf{r})|$ along $\mathbf{r} = r[\hat{\mathbf{y}}\sin
\theta^{\mathrm{inc}}+\hat{\mathbf{z}}\cos\theta^{\mathrm{inc}}]$, $r \in [-8\lambda_0, 8\lambda_0]$, computed using the TS-VIE-GSTC solver, compared against the analytical solution.}
\label{Fig:Att_P}
\end{figure}

Fig.~\ref{Fig:Att_P}(a) presents $|E_x(\mathbf{r})|$ on the $yz$-plane computed using the TS-VIE-GSTC solver for $\tau = \lambda_0/30$. The transmitted beam maintains the prescribed attenuation. Fig.~\ref{Fig:Att_P}(b) compares $|E_x(\mathbf{r})|$ along $\mathbf{r}=r[\hat{\mathbf{y}} \sin\theta^{\mathrm{inc}}+\hat{\mathbf{z}} \cos \theta^{\mathrm{inc}}]$, $r\in [-8\lambda_0, 8\lambda_0]$, computed using the TS-VIE-GSTC solver against the analytical solution. 

To quantify the accuracy, $\mathrm{err}_{\ell_2}$ is computed along the line $\mathbf{r}_p =(r_0+[p-1]\Delta r)[\hat{\mathbf{y}} \sin\theta^{\mathrm{inc}}+\hat{\mathbf{z}} \cos \theta^{\mathrm{inc}}]$ with $r_0=0.5308 \lambda_0$, $\Delta r = 0.0053 \lambda_0$, and $N_p = 1401$. For $\tau \in \{\lambda_0/30, \lambda_0/60, \lambda_0/90\}$, the iteration counts are $N_{\mathrm{it}} \in \{11, 11, 12\}$ and the corresponding errors are $\mathrm{err}_{\ell_2} \in \{1.05,1.04, 1.04\} \times 10^{-2}$.

\begin{figure}[t!]
\centering
\subfigure[]{\includegraphics[width=0.45\columnwidth]{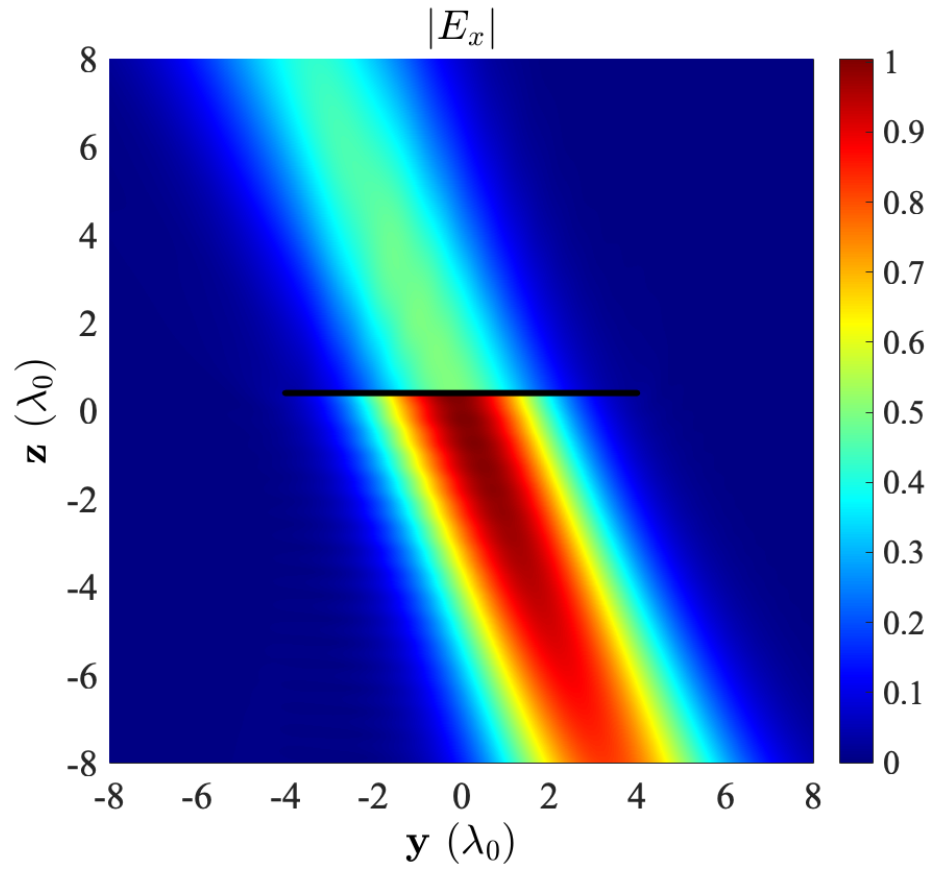}}\\
\subfigure[]{\includegraphics[width=0.6\columnwidth]{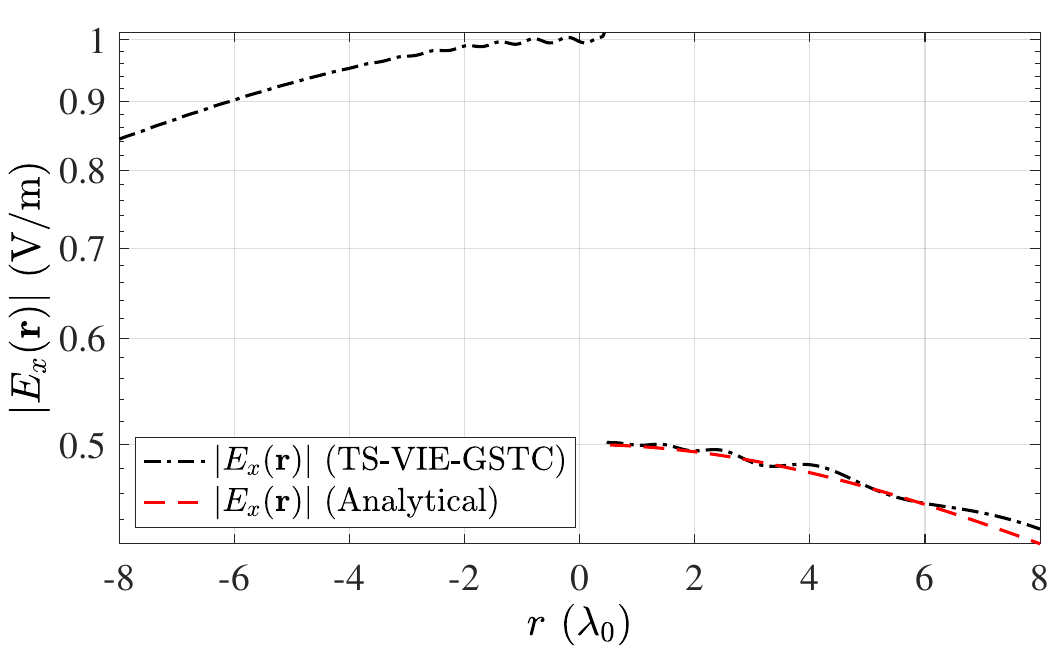}}
\caption{Multi-directional attenuator, $\theta^{\mathrm{inc}} = -22.5^\circ$. (a) $|E_x(\mathbf{r})|$ on the $yz$-plane computed using the TS-VIE-GSTC solver for $\tau = \lambda_0/30$. (b) $|E_x(\mathbf{r})|$ along $\mathbf{r} = r[\hat{\mathbf{y}}\sin\theta^{\mathrm{inc}}+\hat{\mathbf{z}}\cos\theta^{\mathrm{inc}}]$, $r \in [-8\lambda_0, 8\lambda_0]$, computed using the TS-VIE-GSTC solver, compared against the analytical solution.}
\label{Fig:Att_M}
\end{figure}

\paragraph*{Case 3: Oblique Incidence ($\theta^{\mathrm{inc}}=-22.5^\circ$)}
Fig.~\ref{Fig:Att_M}(a) presents $|E_x(\mathbf{r})|$ on the $yz$-plane computed using the TS-VIE-GSTC solver for $\tau = \lambda_0/30$. As expected from the symmetry of the metasurface and its susceptibility tensors with respect to the $xz$-plane, the field distribution mirrors that of the $22.5^\circ$ case. Fig.~\ref{Fig:Att_M}(b) compares $|E_x(\mathbf{r})|$ along $\mathbf{r}=r[\hat{\mathbf{y}} \sin\theta^{\mathrm{inc}}+\hat{\mathbf{z}} \cos \theta^{\mathrm{inc}}]$, $r\in [-8\lambda_0, 8\lambda_0]$, computed using the TS-VIE-GSTC solver against the analytical solution. 

To quantify the accuracy, $\mathrm{err}_{\ell_2}$ is computed along the line $\mathbf{r}_p =(r_0+[p-1]\Delta r)[\hat{\mathbf{y}} \sin\theta^{\mathrm{inc}}+\hat{\mathbf{z}} \cos \theta^{\mathrm{inc}}]$ with $r_0=0.5308 \lambda_0$, $\Delta r = 0.0053 \lambda_0$, and $N_p=1401$. For $\tau \in \{\lambda_0/30, \lambda_0/60, \lambda_0/90\}$, the iteration counts are $N_{\mathrm{it}} \in \{11, 11, 12\}$ and the corresponding errors are $\mathrm{err}_{\ell_2} \in \{1.04, 1.04, 1.04\} \times 10^{-2}$, which are nearly identical to those of the $22.5^\circ$ case. 

The slight discrepancies between the two oblique-incidence cases are attributed to the nonsymmetric discretization. Overall, the results confirm that the proposed TS-VIE-GSTC solver accurately models multi-directional attenuation while maintaining numerical robustness across varying incidence angles and sheet thicknesses.

\subsubsection{Oblique Phase-Shift Transformation}
In this final numerical example, a metasurface is designed to impose a prescribed phase shift on an obliquely incident wave. Two synthesis realizations are compared: a bianisotropic realization operating for three incidence directions $\{-\theta^{\mathrm{inc}}, 0^\circ, \theta^{\mathrm{inc}}\}$, and a monoanisotropic realization restricted to a single incidence direction. For a given $\theta^{\mathrm{inc}}$, the prescribed field transformation specifies fields on the incident side, $\mathbf{E}_2(\mathbf{r})$, and the transmitted side, $\mathbf{E}_1(\mathbf{r})$, as
\begin{equation}
 \begin{aligned}
 \mathbf{E}_{2}(\mathbf{r}) &=\hat{\mathbf{x}}e^{-jk_{0}[y\sin \theta^{\mathrm{inc}} +z\cos \theta^{\mathrm{inc}}]} \\
\mathbf{E}_{1}(\mathbf{r}) &= \hat{\mathbf{x}}e^{-jk_{0}[y\sin \theta^{\mathrm{inc}}+z\cos \theta^{\mathrm{inc}}]+j\varphi^{\mathrm{PS}}}.
\end{aligned}   
\end{equation}
where $\varphi^{\mathrm{PS}}$ denotes the prescribed phase shift. For the bianisotropic realization, the nonzero susceptibility components are:
\begin{equation}\label{eq:PS_bi}
\begin{aligned}
     \chi^{xx}_{\mathrm{ee}}&={4jF_{-}^{\mathrm{PS}}}/({ k_0F_{+}^{\mathrm{PS}}})\\
     \chi^{xy}_{\mathrm{em}}&=-{2jF_{-}^{\mathrm{PS}}  }/({k_0  F_{+}^{\mathrm{PS}}})\\
     \chi^{yx}_{\mathrm{me}}&={2jF_{-}^{\mathrm{PS}}  }/({k_0  F_{+}^{\mathrm{PS}}})\\
     \chi^{zz}_{\mathrm{mm}}&=-{4jF_{-}^{\mathrm{PS}} }/[F_{+}^{\mathrm{PS}}{k_0    (\cos \theta^{\mathrm{inc}} + 1)}]
\end{aligned}     
\end{equation}
where $F_{\pm}^{\mathrm{PS}} = e^{j\varphi^{\mathrm{PS}}} \pm 1$.

For the monoanisotropic realization, which is restricted to a single angle of incidence and employs only two degrees of freedom, the nonzero susceptibility components are:
\begin{equation}
\begin{aligned}
\chi^{xx}_{\mathrm{ee}} & ={2 j\cos \theta^{\mathrm{inc}}F_{-}^{\mathrm{PS}}}/({ k_0F_{+}^{\mathrm{PS}}}) \\
\chi^{yy}_{\mathrm{mm}} & ={2jF_{-}^{\mathrm{PS}}}/(F_{+}^{\mathrm{PS}}{ k_0 \cos \theta^{\mathrm{inc}})}.
\end{aligned}
\end{equation}

The Gaussian beam waist radius is set to $w_b = \lambda_0$, and the beam is incident on the metasurface with $\tau = \lambda_0/250$ at $\theta^{\mathrm{inc}} = 22.5^\circ$. The prescribed phase shift is set to $\varphi^{\mathrm{PS}} = 22.5^\circ$. For both realizations, \eqref{final_mom} is left-preconditioned with the near-field preconditioner with $R_{\mathrm{pre}} = 0.6\lambda_0$, and $\epsilon = 10^{-5}$.

\begin{figure}[h]
\centering
\subfigure[]{\includegraphics[width=0.6\columnwidth]{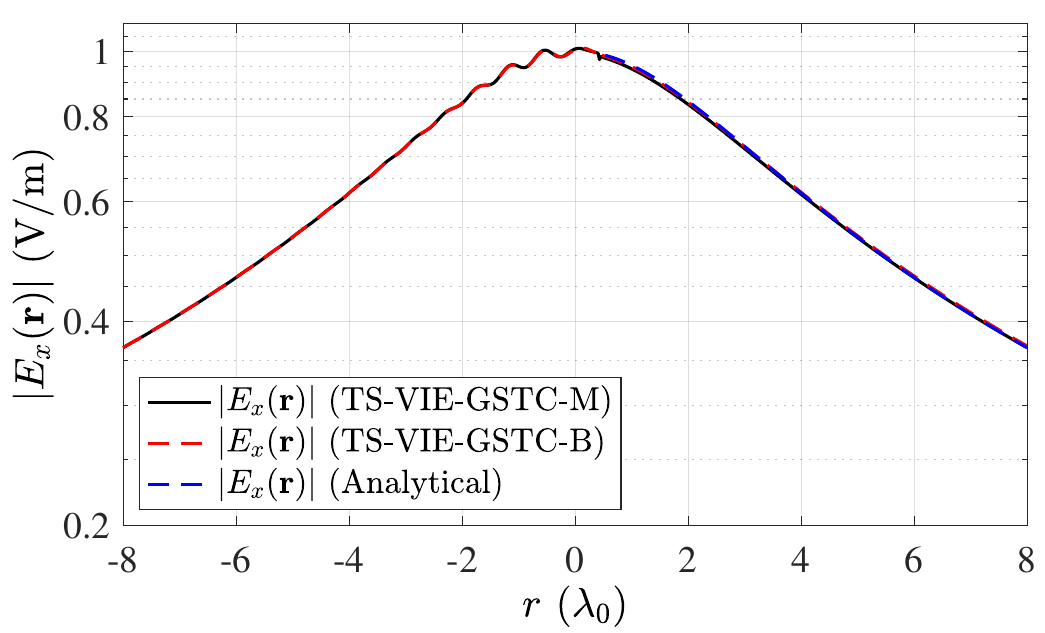}}\\
\subfigure[]{\includegraphics[width=0.6\columnwidth]{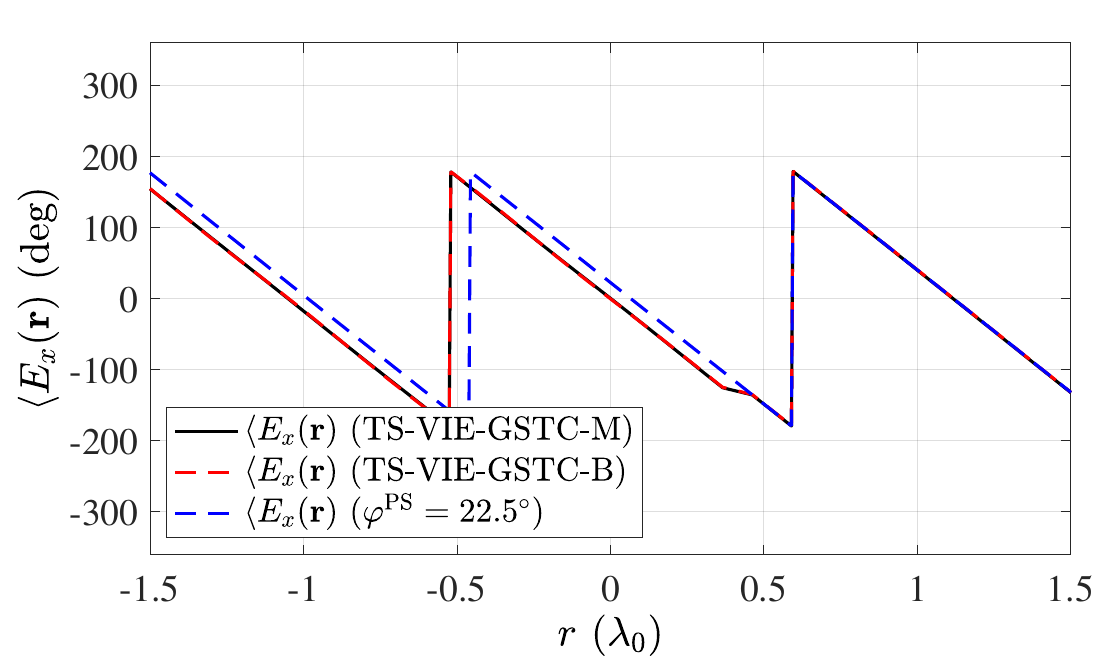}}
\caption{Oblique phase-shift transformation, $\theta^{\mathrm{inc}} = 22.5^\circ$. (a) $|E_x(\mathbf{r})|$ and (b) $\angle E_x(\mathbf{r})$ (in degrees) along $\mathbf{r} = r[\hat{\mathbf{y}}\sin\theta^{\mathrm{inc}}+\hat{\mathbf{z}}\cos\theta^{\mathrm{inc}}]$, computed using the TS-VIE-GSTC solver for both the monoanisotropic (``M'') and bianisotropic (``B'') realizations, compared against the analytical solution. The amplitude is evaluated for $r \in [-8\lambda_0, 8\lambda_0]$ and the phase for $r \in [-1.5\lambda_0, 1.5\lambda_0]$.}
\label{Fig:PS}
\end{figure}

Fig.~\ref{Fig:PS}(a)--(b) compare $|E_x(\mathbf{r})|$ and $\angle E_x(\mathbf{r})$ along the line $\mathbf{r}=r[\hat{\mathbf{y}} \sin\theta^{\mathrm{inc}}+\hat{\mathbf{z}} \cos \theta^{\mathrm{inc}}]$, computed using the TS-VIE-GSTC solver for both the monoanisotropic and bianisotropic realizations against the analytical solution, for $r\in [-8\lambda_0, 8\lambda_0]$ and $r\in [-1.5\lambda_0, 1.5\lambda_0]$, respectively.

To quantify the accuracy, $\mathrm{err}_{\ell_2}$ is computed along the line $\mathbf{r}_p =(r_0+[p-1]\Delta r)[\hat{\mathbf{y}} \sin\theta^{\mathrm{inc}}+\hat{\mathbf{z}} \cos \theta^{\mathrm{inc}}]$ with $r_0=0.5308 \lambda_0$, $\Delta r = 0.0053 \lambda_0$, and $N_p=1401$. For the bianisotropic and monoanisotropic realizations, the iteration counts are $N_{\mathrm{it}} \in \{151, 52\}$ and the corresponding errors are $\mathrm{err}_{\ell_2} \in \{5.98, 10.12\} \times 10^{-3}$, respectively. For the bianisotropic realization, the same quantities are also evaluated at the other two design angles, $\theta^{\mathrm{inc}} = 0^\circ$ and $\theta^{\mathrm{inc}} =-22.5^\circ$, yielding $N_{\mathrm{it}} \in \{75, 178\}$ and $\mathrm{err}_{\ell_2} \in \{5.21, 5.99\} \times 10^{-3}$, respectively, confirming that the metasurface operates at all three angles. The $\theta^{\mathrm{inc}} = -22.5^\circ$ case, which requires the largest number of iterations among the preconditioned simulations, involves $2(N_\parallel + N_\perp) = 59\,172$ unknowns and takes $41\,\mathrm{min}$ in total on $56$ cores with a peak memory of $72\,\mathrm{GB}$.

This example illustrates the trade-off between the two realizations: the monoanisotropic one, restricted to a single incidence direction, converges in roughly three times fewer iterations, whereas the bianisotropic one operates at all three design angles with a lower error at a higher iteration cost. Furthermore, the results validate the observation in~\cite{Achouri_Angular_Scattering} that oblique phase-shift transformations can be achieved using only tangential susceptibility components when restricted to a single incidence direction.

\paragraph*{Comparison with the SIE-GSTC Solver}
To illustrate the effect of the normal susceptibility component, the TS-VIE-GSTC solver is compared with the SIE-GSTC solver of~\cite{CelisSierra2026_ST}, which enforces only the tangential susceptibility components ($\chi^{zz}_{\mathrm{mm}}$ is neglected). The metasurface is modeled as an equivalent thin cylindrical sheet of radius $1\,\mathrm{cm}$ centered at $(0,0,0.2\,\mathrm{cm})$, and the bianisotropic realization~\eqref{eq:PS_bi} is used. The same mesh, with $l_{\mathrm{avg}} = \lambda_0/10$, is used by both solvers, resulting in $N_\perp = 2\,970$ triangles and $N_\parallel = 4\,518$ edges for the TS-VIE-GSTC solver. The TS-VIE-GSTC and SIE-GSTC systems have $14\,976$ and $8\,784$ unknowns, respectively. The incidence angle, the phase shift, the Gaussian beam waist radius, and the sheet thickness are set to $\theta^{\mathrm{inc}} = 30^\circ$, $\varphi^{\mathrm{PS}} = 45^\circ$, $w_b = \lambda_0$, and $\tau = \lambda_0/250$, respectively. The TS-VIE-GSTC system is solved using the near-field preconditioner with $R_{\mathrm{pre}} = 0.6\lambda_0$, and the SIE-GSTC system without a preconditioner. For both solvers, $\epsilon = 10^{-5}$.

\begin{figure}[h]
\centering
\subfigure[]{\includegraphics[width=0.45\linewidth]{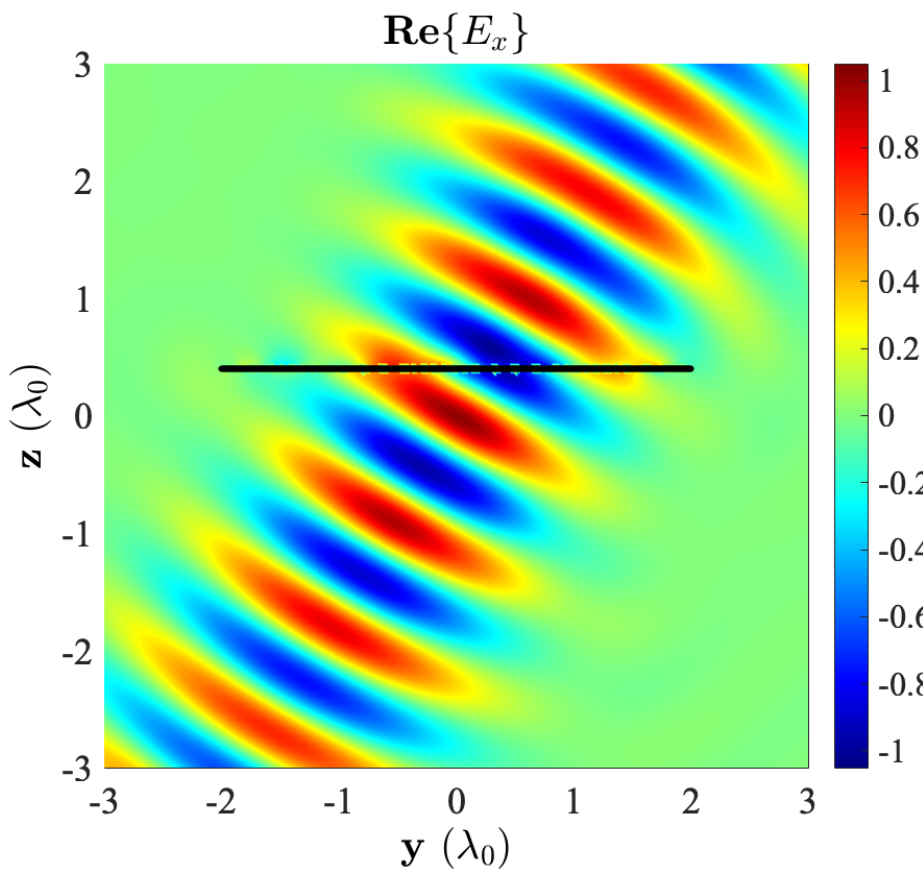}}\hfill
\subfigure[]{\includegraphics[width=0.45\linewidth]{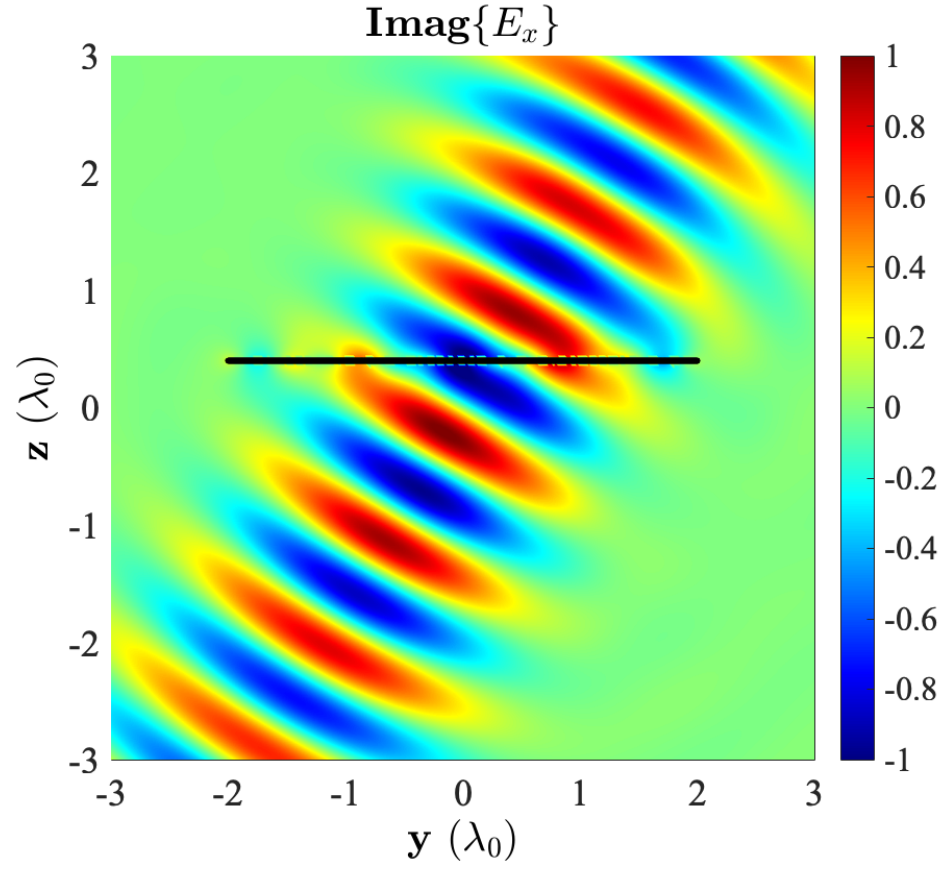}}\\
\subfigure[]{\includegraphics[width=0.6\linewidth]{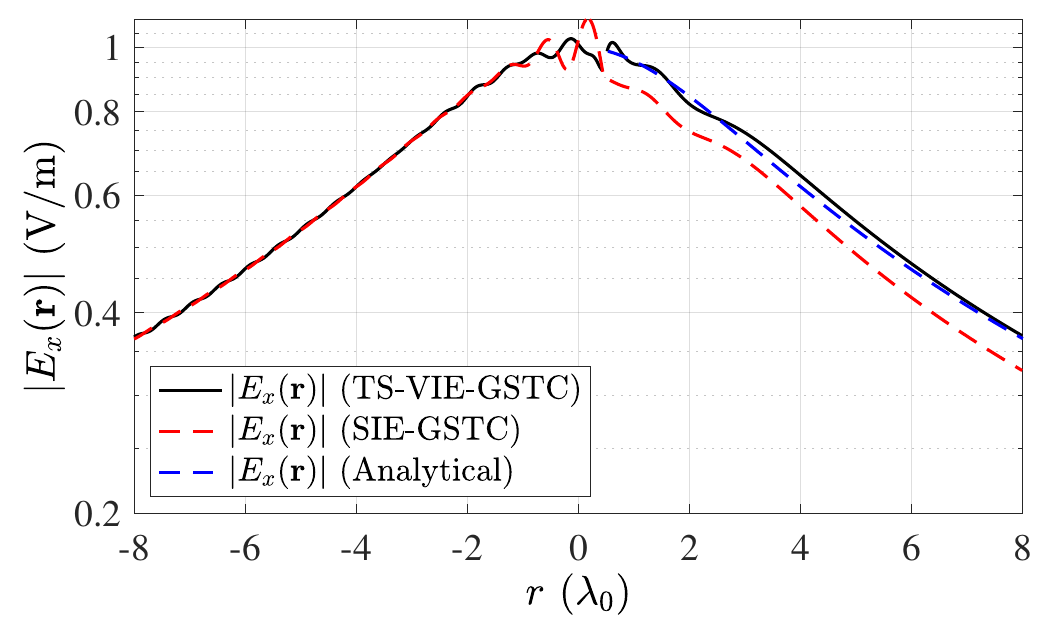}}
\caption{Oblique phase-shift transformation, $\theta^{\mathrm{inc}}=30^{\circ}$, $\varphi^{\mathrm{PS}} = 45^\circ$. (a) $\mathrm{Re}\{E_x(\mathbf{r})\}$ and (b) $\mathrm{Im}\{E_x(\mathbf{r})\}$ on the $yz$-plane computed using the TS-VIE-GSTC solver for $\tau=\lambda_0/250$. (c) $|E_x(\mathbf{r})|$ along $\mathbf{r} = r[\hat{\mathbf{y}}\sin\theta^{\mathrm{inc}}+\hat{\mathbf{z}}\cos\theta^{\mathrm{inc}}]$, $r \in [-8\lambda_0, 8\lambda_0]$, computed using the TS-VIE-GSTC and SIE-GSTC solvers, compared against the analytical solution.}
\label{Fig:PS_P}
\end{figure}

Fig.~\ref{Fig:PS_P}(a)--(b) present $\mathrm{Re}\{E_x(\mathbf{r})\}$ and $\mathrm{Im}\{E_x(\mathbf{r})\}$ on the $yz$-plane computed using the TS-VIE-GSTC solver. Fig.~\ref{Fig:PS_P}(c) compares $|E_x(\mathbf{r})|$ along the line $\mathbf{r} = r[\hat{\mathbf{y}}\sin\theta^{\mathrm{inc}}+\hat{\mathbf{z}}\cos\theta^{\mathrm{inc}}]$, $r \in [-8\lambda_0, 8\lambda_0]$, computed using the TS-VIE-GSTC and SIE-GSTC solvers against the analytical solution. The TS-VIE-GSTC result follows the analytical solution, whereas the SIE-GSTC result deviates from it on the transmission side, which is attributed to the neglected $\chi^{zz}_{\mathrm{mm}}$.

To quantify the accuracy, $\mathrm{err}_{\ell_2}$ is computed along the line $\mathbf{r}_p =(r_0+[p-1]\Delta r)[\hat{\mathbf{y}} \sin\theta^{\mathrm{inc}}+\hat{\mathbf{z}} \cos \theta^{\mathrm{inc}}]$ with $r_0=0.5308 \lambda_0$, $\Delta r = 0.0053 \lambda_0$, and $N_p=1401$. For the TS-VIE-GSTC and SIE-GSTC solvers, the iteration counts are $N_{\mathrm{it}} \in \{33, 128\}$ and the corresponding errors are $\mathrm{err}_{\ell_2} \in \{3.09, 11.34\} \times 10^{-2}$, respectively. The TS-VIE-GSTC solver thus reduces the error by a factor of 3.67 by enforcing the normal susceptibility component that the SIE-GSTC formulation cannot represent.

\section{Conclusions}\label{sec:Conclusions}

A TS-VIE-GSTC solver is presented for the simulation of 3D bianisotropic metasurfaces. The formulation represents the metasurface as an equivalent TS and rigorously reduces the governing VIEs to surface integrals by invoking the thin-sheet approximation, treating tangential and normal flux density components as distinct sets of unknowns. A central contribution of this work is the comprehensive extension of thin-sheet reduction techniques to bianisotropic media, which requires rigorous treatment of volume bound-charge contributions arising from tensor-vector interactions, an aspect not addressed in existing anisotropic or bi-isotropic TS formulations. The constitutive tensors of the equivalent TS are derived directly from the GSTC susceptibility tensors, enabling rigorous enforcement of the GSTCs, including normal field interactions, within a surface-integral based framework while retaining the flux-based VIE character of the formulation.

The accuracy and robustness of the proposed solver are demonstrated through a systematic set of numerical examples, including a parametric study of a monoisotropic spherical shell against the analytical Mie-series solution and five canonical bianisotropic wave transformations: polarization rotation on planar and curved sheets, perfect reflection, beam steering with spatially varying susceptibilities, multi-directional attenuation, and oblique phase-shift transformation. Good agreement is observed between numerical and analytical results, and the iteration count remains constant or grows moderately as the sheet thickness decreases, whereas it increases with the material contrast and with the magnitude of the susceptibilities. A comparison with an SIE-GSTC solver that enforces only the tangential susceptibility components further shows that accounting for the normal susceptibility component significantly reduces the error for the oblique phase-shift transformation.

The proposed TS-VIE-GSTC framework provides a rigorous computational platform for modeling bianisotropic metasurfaces in electrically large open-domain environments. Future work will focus on a rigorous analysis of the thin-sheet reduced operators, including alternative testing schemes for the normal contributions, more effective preconditioning strategies for configurations with large susceptibility values, matrix compression and acceleration techniques to enable simulation of electrically large metasurfaces, and automated tensor-optimization techniques for efficient synthesis of complex electromagnetic transformations.

\newpage\clearpage

\begin{appendices}\label{sec:matvec_entries}
\section{}

\subsection{Entries of the Matrix and the Right-hand Side Vector}\label{sec:matrix_entries}

Throughout this appendix, the inner product between two vectors $\mathbf{a}(\mathbf{r})$ and $\mathbf{b}(\mathbf{r})$ is defined as
\begin{equation*}
    \Big\langle \mathbf{a}(\mathbf{r}),\mathbf{b}(\mathbf{r})  \Big\rangle = \int_{D} \mathbf{a}(\mathbf{r}) \cdot \mathbf{b}(\mathbf{r})\,ds
\end{equation*}
where $D$ is the intersection of the supports of $\mathbf{a}(\mathbf{r})$ and $\mathbf{b}(\mathbf{r})$.

The system in~\eqref{final_mom} can be written in the expanded form as
\begin{align}
\underbrace{{\begin{bmatrix}
\bar{\bar{Z}}_{11}&\bar{\bar{Z}}_{12}&\bar{\bar{Z}}_{13}&\bar{\bar{Z}}_{14}\\
\bar{\bar{Z}}_{21}&\bar{\bar{Z}}_{22}&\bar{\bar{Z}}_{23}&\bar{\bar{Z}}_{24}\\
\bar{\bar{Z}}_{31}&\bar{\bar{Z}}_{32}&\bar{\bar{Z}}_{33}&\bar{\bar{Z}}_{34}\\
\bar{\bar{Z}}_{41}&\bar{\bar{Z}}_{42}&\bar{\bar{Z}}_{43}&\bar{\bar{Z}}_{44}
\end{bmatrix}}}_{\displaystyle\bar{\bar{Z}}}\cdot
\underbrace{{\begin{bmatrix}
\bar{I}_{\mathrm{D},\parallel}\\
\bar{I}_{\mathrm{D},\perp}\\
\bar{I}_{\mathrm{B},\parallel}\\
\bar{I}_{\mathrm{B},\perp}\\
\end{bmatrix}}}_{\displaystyle \bar{I}}=
\underbrace{{\begin{bmatrix}
\bar{V}_{\mathrm{E},\parallel}\\
\bar{V}_{\mathrm{E},\perp}\\
\bar{V}_{\mathrm{H}, \parallel}\\
\bar{V}_{\mathrm{H},\perp}\\
\end{bmatrix}}}_{\displaystyle \bar{V}}.\label{eqmom_equation}
\end{align}
The entries of each block matrix $\bar{\bar{Z}}_{kl}$, $k,l\in \{1,2,3,4\}$ are
\begin{equation*}
\{\bar{\bar{Z}}_{11}\}_{mn} = \Big\langle\mathbf{f}_m(\mathbf{r}),\bar{\bar{\alpha}}_1\cdot\mathbf{f}_n(\mathbf{r}) -\frac{1}{\varepsilon_0}\mathcal{L}_{A}[\bar{\bar{\beta}}_1\cdot\mathbf{f}_n](\mathbf{r})-\frac{1}{\varepsilon_0}\mathcal{T}_{\phi}[\bar{\bar{\beta}}_1\cdot\mathbf{f}_n](\mathbf{r})+j\omega\,\mathcal{K}[\bar{\bar{\beta}}_4\cdot\mathbf{f}_n](\mathbf{r}) \Big\rangle
\end{equation*}
\begin{equation*}
\{\bar{\bar{Z}}_{12}\}_{mq} = \Big\langle\mathbf{f}_m(\mathbf{r}),\bar{\bar{\alpha}}_1\cdot\hat{\mathbf{n}}(\mathbf{r})p_q(\mathbf{r}) -\frac{1}{\varepsilon_0}\mathcal{L}_{A}[\bar{\bar{\beta}}_1\cdot\hat{\mathbf{n}}p_q](\mathbf{r})-\frac{1}{\varepsilon_0}\mathcal{N}_{\phi}[\bar{\bar{\beta}}_1\cdot\hat{\mathbf{n}}p_q](\mathbf{r})+j\omega\,\mathcal{K}[\bar{\bar{\beta}}_4\cdot\hat{\mathbf{n}}p_q](\mathbf{r}) \Big\rangle
\end{equation*}
\begin{equation*}
\{\bar{\bar{Z}}_{13}\}_{mn} = \Big\langle\mathbf{f}_m(\mathbf{r}),\bar{\bar{\alpha}}_2\cdot\mathbf{f}_n(\mathbf{r}) -\frac{1}{\varepsilon_0}\mathcal{L}_{A}[\bar{\bar{\beta}}_2\cdot\mathbf{f}_n](\mathbf{r})-\frac{1}{\varepsilon_0}\mathcal{T}_{\phi}[\bar{\bar{\beta}}_2\cdot\mathbf{f}_n](\mathbf{r})+j\omega\,\mathcal{K}[\bar{\bar{\beta}}_3\cdot\mathbf{f}_n](\mathbf{r}) \Big\rangle
\end{equation*}
\begin{equation*}
\{\bar{\bar{Z}}_{14}\}_{mq} = \Big\langle\mathbf{f}_m(\mathbf{r}),\bar{\bar{\alpha}}_2\cdot\hat{\mathbf{n}}(\mathbf{r})p_q(\mathbf{r}) -\frac{1}{\varepsilon_0}\mathcal{L}_{A}[\bar{\bar{\beta}}_2\cdot\hat{\mathbf{n}}p_q](\mathbf{r})-\frac{1}{\varepsilon_0}\mathcal{N}_{\phi}[\bar{\bar{\beta}}_2\cdot\hat{\mathbf{n}}p_q](\mathbf{r})+j\omega\,\mathcal{K}[\bar{\bar{\beta}}_3\cdot\hat{\mathbf{n}}p_q](\mathbf{r}) \Big\rangle
\end{equation*}
\begin{equation*}
\{\bar{\bar{Z}}_{21}\}_{hn} = \Big\langle\hat{\mathbf{n}}(\mathbf{r})p_h(\mathbf{r}),\bar{\bar{\alpha}}_1\cdot\mathbf{f}_n(\mathbf{r}) -\frac{1}{\varepsilon_0}\mathcal{L}_{A}[\bar{\bar{\beta}}_1\cdot\mathbf{f}_n](\mathbf{r})-\frac{1}{\varepsilon_0}\mathcal{T}_{\phi}[\bar{\bar{\beta}}_1\cdot\mathbf{f}_n](\mathbf{r})+j\omega\,\mathcal{K}[\bar{\bar{\beta}}_4\cdot\mathbf{f}_n](\mathbf{r}) \Big\rangle
\end{equation*}
\begin{equation*}
\{\bar{\bar{Z}}_{22}\}_{hq} = \Big\langle\hat{\mathbf{n}}(\mathbf{r})p_h(\mathbf{r}),\bar{\bar{\alpha}}_1\cdot\hat{\mathbf{n}}(\mathbf{r})p_q(\mathbf{r}) -\frac{1}{\varepsilon_0}\mathcal{L}_{A}[\bar{\bar{\beta}}_1\cdot\hat{\mathbf{n}}p_q](\mathbf{r})-\frac{1}{\varepsilon_0}\mathcal{N}_{\phi}[\bar{\bar{\beta}}_1\cdot\hat{\mathbf{n}}p_q](\mathbf{r})+j\omega\,\mathcal{K}[\bar{\bar{\beta}}_4\cdot\hat{\mathbf{n}}p_q](\mathbf{r}) \Big\rangle
\end{equation*}
\begin{equation*}
\{\bar{\bar{Z}}_{23}\}_{hn} = \Big\langle\hat{\mathbf{n}}(\mathbf{r})p_h(\mathbf{r}),\bar{\bar{\alpha}}_2\cdot\mathbf{f}_n(\mathbf{r}) -\frac{1}{\varepsilon_0}\mathcal{L}_{A}[\bar{\bar{\beta}}_2\cdot\mathbf{f}_n](\mathbf{r})-\frac{1}{\varepsilon_0}\mathcal{T}_{\phi}[\bar{\bar{\beta}}_2\cdot\mathbf{f}_n](\mathbf{r})+j\omega\,\mathcal{K}[\bar{\bar{\beta}}_3\cdot\mathbf{f}_n](\mathbf{r}) \Big\rangle
\end{equation*}
\begin{equation*}
\{\bar{\bar{Z}}_{24}\}_{hq} = \Big\langle\hat{\mathbf{n}}(\mathbf{r})p_h(\mathbf{r}),\bar{\bar{\alpha}}_2\cdot\hat{\mathbf{n}}(\mathbf{r})p_q(\mathbf{r}) -\frac{1}{\varepsilon_0}\mathcal{L}_{A}[\bar{\bar{\beta}}_2\cdot\hat{\mathbf{n}}p_q](\mathbf{r})-\frac{1}{\varepsilon_0}\mathcal{N}_{\phi}[\bar{\bar{\beta}}_2\cdot\hat{\mathbf{n}}p_q](\mathbf{r})+j\omega\,\mathcal{K}[\bar{\bar{\beta}}_3\cdot\hat{\mathbf{n}}p_q](\mathbf{r}) \Big\rangle
\end{equation*}
\begin{equation*}
\{\bar{\bar{Z}}_{31}\}_{mn}= \Big\langle\mathbf{f}_m(\mathbf{r}),\bar{\bar{\alpha}}_3\cdot\mathbf{f}_n(\mathbf{r}) -\frac{1}{\mu_0}\mathcal{L}_{A}[\bar{\bar{\beta}}_4\cdot\mathbf{f}_n](\mathbf{r})-\frac{1}{\mu_0}\mathcal{T}_{\phi}[\bar{\bar{\beta}}_4\cdot\mathbf{f}_n](\mathbf{r})-j\omega\,\mathcal{K}[\bar{\bar{\beta}}_1\cdot\mathbf{f}_n](\mathbf{r}) \Big\rangle
\end{equation*}
\begin{equation*}
\{\bar{\bar{Z}}_{32}\}_{mq} = \Big\langle\mathbf{f}_m(\mathbf{r}),\bar{\bar{\alpha}}_3\cdot\hat{\mathbf{n}}(\mathbf{r})p_q(\mathbf{r}) -\frac{1}{\mu_0}\mathcal{L}_{A}[\bar{\bar{\beta}}_4\cdot\hat{\mathbf{n}}p_q](\mathbf{r})-\frac{1}{\mu_0}\mathcal{N}_{\phi}[\bar{\bar{\beta}}_4\cdot\hat{\mathbf{n}}p_q](\mathbf{r})-j\omega\,\mathcal{K}[\bar{\bar{\beta}}_1\cdot\hat{\mathbf{n}}p_q](\mathbf{r}) \Big\rangle
\end{equation*}
\begin{equation*}
\{\bar{\bar{Z}}_{33}\}_{mn} = \Big\langle\mathbf{f}_m(\mathbf{r}),\bar{\bar{\alpha}}_4\cdot\mathbf{f}_n(\mathbf{r}) -\frac{1}{\mu_0}\mathcal{L}_{A}[\bar{\bar{\beta}}_3\cdot\mathbf{f}_n](\mathbf{r})-\frac{1}{\mu_0}\mathcal{T}_{\phi}[\bar{\bar{\beta}}_3\cdot\mathbf{f}_n](\mathbf{r})-j\omega\,\mathcal{K}[\bar{\bar{\beta}}_2\cdot\mathbf{f}_n](\mathbf{r}) \Big\rangle
\end{equation*}
\begin{equation*}
\{\bar{\bar{Z}}_{34}\}_{mq} = \Big\langle\mathbf{f}_m(\mathbf{r}),\bar{\bar{\alpha}}_4\cdot\hat{\mathbf{n}}(\mathbf{r})p_q(\mathbf{r}) -\frac{1}{\mu_0}\mathcal{L}_{A}[\bar{\bar{\beta}}_3\cdot\hat{\mathbf{n}}p_q](\mathbf{r})-\frac{1}{\mu_0}\mathcal{N}_{\phi}[\bar{\bar{\beta}}_3\cdot\hat{\mathbf{n}}p_q](\mathbf{r})-j\omega\,\mathcal{K}[\bar{\bar{\beta}}_2\cdot\hat{\mathbf{n}}p_q](\mathbf{r}) \Big\rangle
\end{equation*}
\begin{equation*}
\{\bar{\bar{Z}}_{41}\}_{hn} = \Big\langle\hat{\mathbf{n}}(\mathbf{r})p_h(\mathbf{r}),\bar{\bar{\alpha}}_3\cdot\mathbf{f}_n(\mathbf{r}) -\frac{1}{\mu_0}\mathcal{L}_{A}[\bar{\bar{\beta}}_4\cdot\mathbf{f}_n](\mathbf{r})-\frac{1}{\mu_0}\mathcal{T}_{\phi}[\bar{\bar{\beta}}_4\cdot\mathbf{f}_n](\mathbf{r})-j\omega\,\mathcal{K}[\bar{\bar{\beta}}_1\cdot\mathbf{f}_n](\mathbf{r}) \Big\rangle
\end{equation*}
\begin{equation*}
\{\bar{\bar{Z}}_{42}\}_{hq} = \Big\langle\hat{\mathbf{n}}(\mathbf{r})p_h(\mathbf{r}),\bar{\bar{\alpha}}_3\cdot\hat{\mathbf{n}}(\mathbf{r})p_q(\mathbf{r}) -\frac{1}{\mu_0}\mathcal{L}_{A}[\bar{\bar{\beta}}_4\cdot\hat{\mathbf{n}}p_q](\mathbf{r})-\frac{1}{\mu_0}\mathcal{N}_{\phi}[\bar{\bar{\beta}}_4\cdot\hat{\mathbf{n}}p_q](\mathbf{r})-j\omega\,\mathcal{K}[\bar{\bar{\beta}}_1\cdot\hat{\mathbf{n}}p_q](\mathbf{r}) \Big\rangle
\end{equation*}
\begin{equation*}
\{\bar{\bar{Z}}_{43}\}_{hn} = \Big\langle\hat{\mathbf{n}}(\mathbf{r})p_h(\mathbf{r}),\bar{\bar{\alpha}}_4\cdot\mathbf{f}_n(\mathbf{r}) -\frac{1}{\mu_0}\mathcal{L}_{A}[\bar{\bar{\beta}}_3\cdot\mathbf{f}_n](\mathbf{r})-\frac{1}{\mu_0}\mathcal{T}_{\phi}[\bar{\bar{\beta}}_3\cdot\mathbf{f}_n](\mathbf{r})-j\omega\,\mathcal{K}[\bar{\bar{\beta}}_2\cdot\mathbf{f}_n](\mathbf{r}) \Big\rangle
\end{equation*}
\begin{equation*}
\{\bar{\bar{Z}}_{44}\}_{hq} = \Big\langle\hat{\mathbf{n}}(\mathbf{r})p_h(\mathbf{r}),\bar{\bar{\alpha}}_4\cdot\hat{\mathbf{n}}(\mathbf{r})p_q(\mathbf{r}) -\frac{1}{\mu_0}\mathcal{L}_{A}[\bar{\bar{\beta}}_3\cdot\hat{\mathbf{n}}p_q](\mathbf{r})-\frac{1}{\mu_0}\mathcal{N}_{\phi}[\bar{\bar{\beta}}_3\cdot\hat{\mathbf{n}}p_q](\mathbf{r})-j\omega\,\mathcal{K}[\bar{\bar{\beta}}_2\cdot\hat{\mathbf{n}}p_q](\mathbf{r}) \Big\rangle
\end{equation*}
for $m,n = 1,\ldots,N_{\parallel}$, $h,q = 1,\ldots,N_{\perp}$. The entries of the excitation vectors $\{\bar{V}_{\mathrm{E},\parallel}\}_m$, $\{\bar{V}_{\mathrm{E},\perp}\}_h$, $\{\bar{V}_{\mathrm{H},\parallel}\}_m$, and $\{\bar{V}_{\mathrm{H},\perp}\}_h$ are
\begin{align*}
   \{\bar{V}_{\mathrm{E},\parallel}\}_m&=\Big\langle\mathbf{f}_m(\mathbf{r}),  \mathbf{E}^{\mathrm{inc}}(\mathbf{r})\Big\rangle, \quad
    \{\bar{V}_{\mathrm{E},\perp} \}_h=\Big\langle\hat{\mathbf{n}}(\mathbf{r})p_h(\mathbf{r}),  \mathbf{E}^{\mathrm{inc}}(\mathbf{r})\Big\rangle\\
      \{\bar{V}_{\mathrm{H},\parallel} \}_m&=\Big\langle\mathbf{f}_m(\mathbf{r}),  
     \mathbf{H}^{\mathrm{inc}}(\mathbf{r})\Big\rangle,\quad
    \{\bar{V}_{\mathrm{H},\perp} \}_h=\Big\langle\hat{\mathbf{n}}(\mathbf{r})p_h(\mathbf{r}),  \mathbf{H}^{\mathrm{inc}}(\mathbf{r})\Big\rangle
\end{align*}
for $m=1,2,\dots,N_{\parallel}$ and $h=1,2,\dots,N_{\perp}$.

\subsection{Singularity Treatment}\label{sec:inner_product}
The entries of the block matrices in~\eqref{eqmom_equation} require the computation of six distinct operator-basis function combinations: $\mathcal{L}_A[\bar{\bar{\beta}}_{i}\cdot\mathbf{f}_n](\mathbf{r})$, $\mathcal{L}_A[\bar{\bar{\beta}}_{i}\cdot\hat{\mathbf{n}}{p}_q](\mathbf{r})$, $\mathcal{T}_\phi[\bar{\bar{\beta}}_{i}\cdot
\mathbf{f}_n](\mathbf{r})$, $\mathcal{N}_{\phi}[\bar{\bar{\beta}}_{i}\cdot\hat{\mathbf{n}}{p}_q](\mathbf{r})$, $\mathcal{K}[\bar{\bar{\beta}}_{i}\cdot\mathbf{f}_n](\mathbf{r})$, and $\mathcal{K}[\bar{\bar{\beta}}_{i}\cdot\hat{\mathbf{n}}p_q](\mathbf{r})$. In all cases, $i\in\{1,2,3,4\}$ is determined by the index of the matrix block. 

Considering $\bar{\bar{\beta}}_{i} \cdot \mathbf{f}_n(\mathbf{r}^{\prime})=\bar{\bar{\beta}}_{i,\parallel} \cdot \mathbf{f}_n(\mathbf{r}^{\prime})$ in~\eqref{eq:la_reduced}, ${\mathcal{L}}_{A}[\bar{\bar{\beta}}_{i}\cdot \mathbf{f}_n](\mathbf{r})$ is expressed as 
\begin{equation}
\label{eq:la_rwg}
\mathcal{L}_A[\bar{\bar{\beta}}_{i} \cdot \mathbf{f}_n](\mathbf{r})=\tau k_0^2 \int_S G(\mathbf{r}, \mathbf{r}^{\prime})\, \bar{\bar{\beta}}_{i,\parallel} \cdot \mathbf{f}_n(\mathbf{r}^{\prime})\,d s^{\prime}.
\end{equation}
The tensor-vector product $\bar{\bar{\beta}}_{i,\parallel} \cdot \mathbf{f}_n(\mathbf{r}^{\prime})$ does not alter the singularity of $G(\mathbf{r},\mathbf{r}^{\prime})$. Similarly, ${\mathcal{L}_{A}}[\bar{\bar{\beta}}_{i}\cdot \hat{\mathbf{n}}p_q](\mathbf{r})$ is expressed as
\begin{equation}
\label{eq:la_p}
\mathcal{L}_{A}[\bar{\bar{\beta}}_{i}\cdot \hat{\mathbf{n}}p_q](\mathbf{r})=\tau k_0^2\int_S G(\mathbf{r},\mathbf{r}^{\prime})\,\beta_{i,\perp} \hat{\mathbf{n}}(\mathbf{r}^{\prime})p_q(\mathbf{r}^{\prime})\,d s^{\prime}.
\end{equation}
The vector $\beta_{i,\perp} \hat{\mathbf{n}}(\mathbf{r}^{\prime})p_q(\mathbf{r}^{\prime})$ does not alter the singularity order of $G(\mathbf{r},\mathbf{r}^{\prime})$. For both ${\mathcal{L}}_{A}[\bar{\bar{\beta}}_{i}\cdot \mathbf{f}_n](\mathbf{r})$ and ${\mathcal{L}_{A}}[\bar{\bar{\beta}}_{i}\cdot \hat{\mathbf{n}}p_q](\mathbf{r})$, the singularity extraction techniques described in~\cite{wilton_potential_1984, graglia_numerical_1993,CFIE_Pasi} are used.

Using~\eqref{eq:t_phi_reduced}, $\mathcal{T}_{\phi}[\bar{\bar{\beta}}_{i}\cdot \mathbf{f}_n](\mathbf{r})$ is expressed
\begin{equation}
\begin{aligned}
 &\mathcal{T}_{\phi}[\bar{\bar{\beta}}_{i}\cdot\mathbf{f}_n](\mathbf{r}) = -\tau\int_{C} \nabla G(\mathbf{r},\mathbf{r}^{\prime})\,
        \hat{\mathbf{t}}(\mathbf{r}^{\prime})\cdot[\bar{\bar{\beta}}_{i,\parallel}
        \cdot\mathbf{f}_n(\mathbf{r}^{\prime})]\,dl^{\prime} \\
 &+\frac{\tau}{2}\int_{S}  \nabla G(\mathbf{r},\mathbf{r}^{\prime}+\frac{\tau\hat{\mathbf{n}}(\mathbf{r}^{\prime})}{2}) \beta_{i,\bot}\nabla^{\prime}_{\parallel}\cdot\mathbf{f}_n(\mathbf{r}^{\prime})\,  ds^{\prime}+\frac{\tau}{2}\int_{S}  \nabla G(\mathbf{r},\mathbf{r}^{\prime}-\frac{\tau\hat{\mathbf{n}}(\mathbf{r}^{\prime})}{2}) \beta_{i,\bot}\nabla^{\prime}_{\parallel}\cdot\mathbf{f}_n(\mathbf{r}^{\prime})\,ds^{\prime}\\
\label{eq:t_rwg}    &+\tau\int_{S}\nabla G(\mathbf{r},\mathbf{r}^{\prime})\,
        [\bar{\bar{\beta}}_{i,\parallel}:\nabla^{\prime}_{\parallel}
        \mathbf{f}_n(\mathbf{r}^{\prime})]\,ds^{\prime} -\tau\int_{S}\nabla G(\mathbf{r},\mathbf{r}^{\prime})\,
        \beta_{i,\perp}\,\nabla^{\prime}_{\parallel}\cdot\mathbf{f}_n(\mathbf{r}^{\prime})\,ds^{\prime}.
\end{aligned}
\end{equation}
In~\eqref{eq:t_rwg}, the double inner product $\bar{\bar{\beta}}_{i,\parallel}:\nabla'_{\parallel}\mathbf{f}_n(\mathbf{r}^{\prime})$ is computed using
\begin{equation}
\bar{\bar{\beta}}_{i,\parallel}: \nabla_{\parallel} \mathbf{f}_n(\mathbf{r})= \pm \frac{l_n}{2 A_n^{\pm}} \operatorname{tr}(\bar{\bar{\beta}}_{i,\parallel})    
\end{equation}
where $\operatorname{tr}(\bar{\bar{\beta}}_{i,\parallel})$ denotes the trace of tensor $\bar{\bar{\beta}}_{i,\parallel}$, $l_n$ is the length of the $n$-th edge, and $A_n^{\pm}$ are the areas of the two triangular elements associated with the $n$-th edge. The tensor-vector product $\bar{\bar{\beta}}_{i,\parallel}\cdot\mathbf{f}_n(\mathbf{r}^{\prime})$, the divergence $\nabla'_{\parallel}\cdot\mathbf{f}_n(\mathbf{r}^{\prime})$, and the double inner product $\bar{\bar{\beta}}_{i,\parallel}:\nabla'_{\parallel}\mathbf{f}_n(\mathbf{r}^{\prime})$ do not alter the singularity of $\nabla G(\mathbf{r},\mathbf{r}^{\prime})$.

Using~\eqref{eq:n_phi_reduced}, $\mathcal{N}_{\phi}[\bar{\bar{\beta}}_{i}\cdot \hat{\mathbf{n}}p_q](\mathbf{r})$ is expressed as
\begin{equation}
\label{eq:n_p}
\begin{aligned}
\mathcal{N}_{\phi}[\bar{\bar{\beta}}_{i}\cdot\hat{\mathbf{n}}p_q](\mathbf{r})
    &= -\int_{S}\nabla G(\mathbf{r},
        \mathbf{r}^{\prime}+\frac{\tau\hat{\mathbf{n}}(\mathbf{r}^{\prime})}{2})\beta_{i,\perp}p_q(\mathbf{r}^{\prime})\,ds^{\prime}\\
    &+\int_{S}\nabla G(\mathbf{r},
        \mathbf{r}^{\prime}-\frac{\tau\hat{\mathbf{n}}(\mathbf{r}^{\prime})}{2})\beta_{i,\perp}p_q(\mathbf{r}^{\prime})\,ds^{\prime}.
\end{aligned}
\end{equation}
Clearly, $\beta_{i,\perp}p_q(\mathbf{r}^{\prime})$ does not alter the singularity of $\nabla G[\mathbf{r},\mathbf{r}^{\prime}\pm(\tau/2)\hat{\mathbf{n}}(\mathbf{r}^{\prime})]$ as $\tau \to 0$.

Considering $\bar{\bar{\beta}}_{i} \cdot \mathbf{f}_n(\mathbf{r}^{\prime})=\bar{\bar{\beta}}_{i,\parallel} \cdot \mathbf{f}_n(\mathbf{r}^{\prime})$ and $\bar{\bar{\beta}}_{i}\cdot\hat{\mathbf{n}}(\mathbf{r}^{\prime})p_q(\mathbf{r}^{\prime}) = \beta_{i,\perp}\hat{\mathbf{n}}(\mathbf{r}^{\prime})$ in~\eqref{eq:k_reduced}, $\mathcal{K}[\bar{\bar{\beta}}_{i}\cdot\mathbf{f}_n](\mathbf{r})$ and $\mathcal{K}[\bar{\bar{\beta}}_{i}\cdot\hat{\mathbf{n}}p_q](\mathbf{r})$ are expressed as
\begin{equation}
\label{eq:k_rwg}
\mathcal{K}[\bar{\bar{\beta}}_{i}\cdot\mathbf{f}_n](\mathbf{r})= \tau\int_{S}\nabla G(\mathbf{r},\mathbf{r}^{\prime})\times
\bar{\bar{\beta}}_{i,\parallel}\cdot\mathbf{f}_n(\mathbf{r}^{\prime})\,ds^{\prime}
\end{equation}
\begin{equation}
\label{eq:k_p}
\mathcal{K}[\bar{\bar{\beta}}_{i}\cdot\hat{\mathbf{n}}p_q](\mathbf{r})= \tau\int_{S}\nabla G(\mathbf{r},\mathbf{r}^{\prime})\times\beta_{i,\perp}\hat{\mathbf{n}}(\mathbf{r}^{\prime})\,ds^{\prime}.
\end{equation}
The tensor-vector product $\bar{\bar{\beta}}_{i,\parallel} \cdot \mathbf{f}_n(\mathbf{r}^{\prime})$ in~\eqref{eq:k_rwg} and the vector $\beta_{i,\perp} \hat{\mathbf{n}}(\mathbf{r}^{\prime})$ in~\eqref{eq:k_p} do not alter the singularity order of $\nabla G(\mathbf{r},\mathbf{r}^{\prime})$. For $\mathcal{T}_{\phi}[\bar{\bar{\beta}}_{i}\cdot \mathbf{f}_n](\mathbf{r})$, $\mathcal{N}_{\phi}[\bar{\bar{\beta}}_{i}\cdot \hat{\mathbf{n}}p_q](\mathbf{r})$, $\mathcal{K}[\bar{\bar{\beta}}_{i}\cdot\mathbf{f}_n](\mathbf{r})$, and $\mathcal{K}[\bar{\bar{\beta}}_{i}\cdot\hat{\mathbf{n}}p_q](\mathbf{r})$, the singularity extraction techniques described in~\cite{CFIE_Pasi,hodges_evaluation_1997} are used. 

\subsection{Edge Integrals and Spatially Varying Tensors}\label{sec:inhomogeneous}
After discretization, $\bar{\bar{\beta}}_i$ is evaluated at the centroid of each triangular element and assumed constant over it, making the sheet piecewise homogeneous. The derivation of Section~\ref{sec:TS-VIEs}, which assumes $\nabla\cdot\bar{\bar{\beta}}_i = 0$ within $V$, then applies to each element separately: in~\eqref{eq:la_rwg}--\eqref{eq:k_p}, $S$ and $C$ become the surface and the boundary of each element in the support of the basis function.

In $\mathcal{L}_A[\bar{\bar{\beta}}_{i}\cdot\mathbf{f}_n](\mathbf{r})$, $\mathcal{L}_A[\bar{\bar{\beta}}_{i}\cdot\hat{\mathbf{n}}p_q](\mathbf{r})$, $\mathcal{K}[\bar{\bar{\beta}}_{i}\cdot\mathbf{f}_n](\mathbf{r})$, $\mathcal{K}[\bar{\bar{\beta}}_{i}\cdot\hat{\mathbf{n}}p_q](\mathbf{r})$, $\mathcal{N}_{\phi}[\bar{\bar{\beta}}_{i}\cdot\hat{\mathbf{n}}p_q](\mathbf{r})$, and the surface integrals of $\mathcal{T}_{\phi}[\bar{\bar{\beta}}_{i}\cdot\mathbf{f}_n](\mathbf{r})$, the tensors appear only as multiplicative factors of the basis functions or of their element-wise divergence and gradient. These terms are evaluated with the tensor of each element without additional terms. In particular, a discontinuity of $\beta_{i,\perp}$ across an element edge does not produce a line charge, since the line integral over $C$ in~\eqref{eq:t_rwg} involves only $\bar{\bar{\beta}}_{i,\parallel}$. This line integral, on the other hand, is evaluated over the boundaries $\partial T_n^{\pm}$ of the two triangular elements $T_n^{\pm}$ supporting the RWG function $\mathbf{f}_n(\mathbf{r})$, i.e.,
\begin{equation}
\int_{C}(\cdot)\,dl^{\prime} = \oint_{\partial T_n^{+}}(\cdot)\,dl^{\prime} + \oint_{\partial T_n^{-}}(\cdot)\,dl^{\prime}
\end{equation}
where each boundary is traversed with the in-plane outward normal $\hat{\mathbf{t}}(\mathbf{r})$ and the tensor of its own element. These boundaries comprise the four outer edges of $T_n^{\pm}$, each integrated once, and the edge shared by $T_n^{+}$ and $T_n^{-}$, integrated once from each side with opposite normals and, in general, different tensors. If $\mathbf{f}_n(\mathbf{r})$ is a half RWG function, $C$ reduces to the boundary of its single element. When $\bar{\bar{\beta}}_{i,\parallel}$ is isotropic and identical on both elements, $\hat{\mathbf{t}}(\mathbf{r})\cdot[\bar{\bar{\beta}}_{i,\parallel}\cdot\mathbf{f}_n(\mathbf{r})]$ vanishes on the outer edges and is continuous across the shared edge, so that only the edges on the boundary of $S$ contribute, as in~\cite{TS}. When $\bar{\bar{\beta}}_{i,\parallel}$ is anisotropic or differs between $T_n^{+}$ and $T_n^{-}$, this quantity is neither zero nor continuous, and all edges contribute.
\end{appendices}
\bibliographystyle{unsrt}
\bibliography{References_HB_V1}

\newpage\clearpage

\section*{Figures}

\newpage

\newpage

\newpage

\newpage

\newpage

\newpage

\newpage

\newpage

\newpage

\newpage

\newpage

\newpage

\newpage

\end{document}